\documentclass[useAMS,usenatbib]{mn2e}
\usepackage[T1]{fontenc}
\usepackage{aecompl}
\usepackage{graphicx}
\usepackage{array}
\usepackage{lscape} 
\usepackage{txfonts}
\usepackage{longtable}
\usepackage{color}
\usepackage{ulem}    
\usepackage{times}
\usepackage{subfigure}
\usepackage{booktabs}
\newcommand{\temp}{$T_{\rm e}$}
\newcommand{\efftemp}{$T_{\rm eff}$}

\newcommand\ion[2]{#1$\;${\scshape{#2}}}%                       % ion, i.e., CII = \ion{C}{ii}

\newcommand{\aboii}{$\mbox{O}^{+}$}
\newcommand{\aboiii}{$\mbox{O}^{++}$}
\newcommand{\abnii}{$\mbox{N}^{+}$}

\newcommand{\abneiii}{$\mbox{Ne}^{++}$}

\newcommand{\abnev}{$\mbox{Ne}^{+4}$}
\newcommand{\absii}{$\mbox{S}^{+}$}
\newcommand{\absiii}{$\mbox{S}^{++}$}

\newcommand{\abariii}{$\mbox{Ar}^{++}$}

\newcommand{\abclii}{$\mbox{Cl}^{+}$}
\newcommand{\abcliii}{$\mbox{Cl}^{++}$}
\newcommand{\abcliv}{$\mbox{Cl}^{+3}$}

\newcommand{\gio}{$\mbox{O}^{++}$/($\mbox{O}^{+}$+$\mbox{O}^{++}$)}
\newcommand{\gihe}{$\mbox{He}^{++}$/($\mbox{He}^{+}$+$\mbox{He}^{++}$)}
\newcommand{\inthe}{$I$(\ion{He}{ii})/$I$(H$\beta$)}
\newcommand{\icfm}{ICF$_{\rm m}$}
\newcommand{\icff}{ICF$_{\rm f}$}
\newcommand*{\TabV}[1]{%
  \setlength{\fboxrule}{0pt}\fbox{\hspace*{-\fboxsep}#1\hspace*{-\fboxsep}}}
\newcommand{\apj}{ApJ}
\newcommand{\apjs}{ApJS}

\newcommand{\aap}{A{\&}A}
\newcommand{\aaps}{A{\&}AS}
\newcommand{\aj}{AJ}

\newcommand{\mnras}{MNRAS}

\newcommand{\pasp}{PASP}
\newcommand{\pasa}{PASA}
\newcommand{\rmaa}{RMxAA}

\newcommand{\jpb}{JPhB}

\newcommand{\ssr}{{\it Space Science Reviews}}
\title[ICFs in PNe]{Ionization Correction Factors for Planetary Nebulae : \\
I- Using optical spectra}
\author[G. Delgado-Inglada, C. Morisset, and G. Stasi\'nska]
{Gloria Delgado-Inglada$^{1}$\thanks{E-mail: gloria.delgado.inglada@gmail.com (GDI); 
Chris.Morisset@gmail.com (CM); grazyna.stasinska@obspm.fr (GS)}, 
Christophe Morisset$^{1}$\footnotemark[1], and Gra\.zyna Stasi\'nska$^{2}$\footnotemark[1]\\
$^{1}$Instituto de Astronom{\'\i}a, Universidad Nacional Aut\'onoma de M\'exico, Apdo. Postal 70264, M\'ex. D.F., 04510, Mexico\\
$^{2}$LUTH, Observatoire de Paris, CNRS, Universit\'e Paris Diderot; Place Jules Janssen 92190 Meudon, France}

\begin{document}

\date{Accepted 2014 February 19. Received 2014 February 6; in original form 2013 December 16.}
\pagerange{\pageref{firstpage}--\pageref{lastpage}} \pubyear{2014}
\maketitle

\label{firstpage}

\begin{abstract}
We compute a large grid of photoionization models that covers a wide range of physical parameters and is representative of most of the observed PNe.  
Using this grid, we derive new formulae for the ionization correction factors (ICFs) of He, O, N, Ne, S, Ar, Cl, and C. Analytical expressions to 
estimate the uncertainties arising from our ICFs are also provided. This should be useful since these uncertainties are usually not considered 
when estimating the error bars in element abundances. 

Our ICFs are valid over a variety of assumptions such as the input metallicities, the spectral energy distribution of the ionizing source, the gas 
distribution, or the presence of dust grains. Besides, the ICFs are adequate both for large aperture observations and for pencil-beam observations 
in the central zones of the nebulae.

We test our ICFs on a large sample of observed PNe that extends as far as possible in ionization, central star temperature, and metallicity, by checking 
that the Ne/O, S/O, Ar/O, and Cl/O ratios show no trend with the degree of ionization. 

Our ICFs lead to significant differences in the derived abundance ratios as compared with previous determinations, 
especially for N/O, Ne/O, and Ar/O.

\end{abstract}


%%%%%%%%%%%%%%%%%%%%%%%%%%%%%%%%%%%%%%%%%%%%%%%%%%%%%%%%%%%%%%%%%%%%%%%%%%
\section{Introduction}
\label{sec:intro}

Planetary nebulae (PNe) are powerful tools to study chemical evolution in galaxies. The spectra of PNe have bright emission lines that allow us 
to calculate physical conditions and ionic abundances in the ionized gas. The abundance of sulfur, neon, argon, 
and chlorine in PNe reflect the abundances of the interstellar medium at the time the progenitor stars were born. Therefore, these 
abundances can be used as tracers of interstellar abundances. On the other hand, helium, nitrogen, carbon, and oxygen abundances change
as a consequence of nucleosynthesis processes occurring in the star before and during the PN phase. The analysis of these abundances can 
give us clues on the efficiencies of nucleosynthesis in low and intermediate mass stars. 

The total abundance of one particular element is obtained by adding up the ionic abundances of all the ions present in a nebula. 
However, not all the ions are observed, either because the lines are emitted in a different spectral range 
from the one observed or because the lines are too weak to be detected. Thus, one must estimate the contribution of unobserved ions 
to the total abundance using ionization correction factors (ICFs). First ICFs were proposed by \citet{Peimbert_69} and 
they were based on ionization potential similarities. However these ICFs should be taken with caution since 
the ionization structures do not depend only on ionization potentials \citep[see, e.g.,][]{Stasinska_02}. 
Alternatively, ICFs may be derived from photoionization models, as those derived, e.g., by \citet{Stasinska_78}; Kingsburgh \& Barlow (1994, 
hereafter KB94); and \citet{Kwitter_01}. They are more reliable but so far have been based on a small number of models. 

In this paper we derive new recipes for the ICFs of He, O, N, Ne, S, Ar, Cl, and C using a large grid of models that spans over a wide range of 
parameters and is representative of most of the observed PNe. We also provide analytical formulae to estimate the error bars 
associated with our ICFs. Our motivations are two. First, some of the commonly adopted ICFs do not correct properly for unobserved ions. 
Second, the errors in ICFs are usually not considered when estimating errors in element abundances. Our results should be useful for empirical 
abundance studies. In particular, an accurate determination of the element abundances and their associated uncertainties is essential in the 
study of abundance gradients in galaxies. 

We examine the accuracy of our ICFs by applying them to several families of photoionization models computed from different 
assumptions. 
The studied performed by \citet{Alexander_97} is the only one in the literature investigating the errors arising from the 
use of ICFs in PNe. Our sample of models cover a wider range of physical parameters and we include several assumptions not 
considered in \citet{Alexander_97}, such as the non-constant density distribution, the combination of two models, or the inclusion 
of dust grains in the nebulae. We also test our ICFs on a large sample of observed PNe, and compare the computed abundances with the 
ones obtained with other ICFs (mainly the ones by KB94). 

The analysis here is restricted to ions with emission lines in the optical range, ICFs for ultraviolet and infrared observations will be 
discussed in a forthcoming paper. 
%%%%%%%%%%%%%%%%%%%%%%%%%%%%%%%%%%%%%%%%%%%%%%%%%%%%%%%%%%%%%%%%%%%%%%%%%%
\section{Photoionization models}
\label{sec:mod}

The ionization structure of a nebula depends principally on the spectral energy distribution shape of the ionizing spectrum, 
in this case essentially determined by the effective temperature (\efftemp) of the ionizing star. For a given \efftemp, 
it also depends on the ratio of the flux of ionizing photons reaching a given point in the nebula and the local electron density. 
The larger this ratio, the higher the ionization state. Finally, it also depends on the electron temperature of the nebular gas, 
since recombination is more efficient at low temperature. Photoionization codes allow one to compute the detailed ionization 
structure of the various elements present in a model nebula, by taking into account all the processes that govern ionization 
and recombination (i.e. mostly photoionization, radiative and di-electronic recombination, and charge exchange), as well 
as all the heating and cooling processes that determine the electron temperature.

For our study, we have computed a grid of spherical photoionization models using Cloudy c10.01 \citep{Ferland_98}.  
The free parameters are the effective temperature (\efftemp) and the luminosity (L$_{ *}$) of the central star, 
the inner nebular radius (R$_{\rm in}$), and the hydrogen density of the gas (n$_{\rm H}$). These parameters 
take the values shown in Table~\ref{tab:1}. 

We use both blackbodies (B) and Rauch atmospheres (Rau) for the spectral energy distributions of the ionizing stars, 
and include constant (C) and gaussian (G) nebular density distributions. 
Rauch models are constructed with the atmospheres from \citet{Rauch_03} for \efftemp\ = 50000, 75000, 
100000, 125000, 150000, and 180000 K. We only use one value for the stellar surface gravity, $\log(g)=6$. 
This parameter varies from $\log(g)\sim4$ to  $\log(g)\sim6.5$ during the evolution of the PN but we do not expect 
this has an effect on our results. 

For radiation bounded (R) models the calculations are performed until the ionic fraction of H$^+$ falls below 
0.02. Matter bounded (M) models are simply obtained by reading out Cloudy results at 40\%, 60\%, and 80\% of 
the total gas mass. This percentage is given by the parameter $f_{\rm mass}$. Radiation bounded models 
have $f_{\rm mass}$ = 1 whereas matter bounded models have $0.4 < f_{\rm mass} < 1$.
In order to mimic some aspherical PNe we combine pairs of the above models (Co) with the same 
ionizing source, inner radius and gas density but different quantities of gas. 

The input abundances used for the solar (S) models are the default PN abundances in Cloudy except 
for those elements with abundances equal to 10$^{-20}$, for which we use their values in the Cloudy ISM set. The PN set
is based on the works by \citet{Aller_83} and \citet{Khromov_89} and is not necessarily representative of solar 
neighborhood abundances \citep[see, e.g,][]{Rodriguez_11, Rodriguez_12}. But, as we will show in 
Section~\ref{sec:effect}, our ICFs do not depend on the detailed abundances. The input abundances 
for the elements studied here are, in units of $12 + \log$(X/H), 11.00 (He), 8.89 (C), 8.26 (N), 8.64 (O), 8.04 (Ne), 
7.0 (S), 5.23 (Cl), and 6.43 (Ar).
To explore the effect of metallicity in our results we also compute high metallicity models (with twice solar abundances, H) 
and low metallicity models (with half solar abundances, L).  

Table~\ref{tab:2} describes the characteristics and names of the different families of models computed. The number 
in the last column refers to the final number of models of each family after applying all the selection criteria described 
below. 

\begin{table}
\centering
\caption{Input parameters for the models\label{tab:1}.}
\begin{tabular}{cccccc}
\hline \hline \\[-2ex]
\multicolumn{6}{c}{{\bf $T_{{\bf eff}}$} (10$^3$ K)} \\
[0.5ex] \hline \\[-1.8ex]
25  & 35  & 50  & 75 & 100 & 125 \\
150 & 180 & 210 & 240 & 270 & 300\\
[0.5ex] \hline \\[-1.8ex]
\multicolumn{6}{c}{{\bf $R_{{\bf in}}$} (cm)}\\
[0.5ex] \hline \\[-1.8ex]
$3\times10^{15}$ & 10$^{16}$ & $3\times10^{16}$ & 10$^{17}$ & $3\times10^{17}$ & 10$^{18}$ \\
$3\times10^{18}$ \\
[0.5ex] \hline \\[-1.8ex]
\multicolumn{6}{c}{{\bf $n_{{\bf H}}$} (cm$^{-3}$)}\\
[0.5ex] \hline \\[-1.8ex]
30 & 100 & 300 & 1000 & 3000 & 10000 \\
30000 & 100000 & 300000 \\
[0.5ex] \hline \\[-1.8ex]
\multicolumn{6}{c}{{\bf $L_{{\bf *}}$} (L$_\odot$)}\\
[0.5ex] \hline \\[-1.8ex]
200 & 1000 & 3000 & 5600 & 10000 & 17800 \\
\hline \hline \\[-2ex]
\end{tabular}
\end{table}

\begin{table*}
\caption{Families of photoionization models}             
\label{tab:2}      
\centering          
\begin{tabular}{llllllll}
\hline\hline       
Family name & Ionizing Source & Density law & Bounded & Metallicity & Dust & Combined Models & Number of models\\ 
 & & & & & & & after applying the filters\\
\hline                    
BCRS             & blackbody (B)   & constant (C) & radiation bounded (R) & solar (S) & no & no & 668\\
BCMS             & blackbody (B)   & constant (C) & matter bounded (M)    & solar (S) & no & no & 2286\\
RauCRS         & Rauch (Rau)         & constant (C) & radiation bounded (R) & solar (S) & no & no & 409\\
RauCMS         & Rauch (Rau)         & constant (C) & matter bounded (M) & solar (S) & no & no & 1410\\
BGRS             & blackbody (B)   & gaussian (G) & radiation bounded (R) & solar (S) & no & no & 725\\
BGMS             & blackbody (B)   & gaussian (G) & matter bounded (M) & solar (S) & no & no & 2323\\
BCRH             & blackbody (B)   & constant (C) & radiation bounded (R) & twice solar (H) & no & no & 697\\
BCMH             & blackbody (B)   & constant (C) & matter bounded (M)    & twice solar (H) & no & no & 2397\\
BCRL              & blackbody (B)   & constant (C) & radiation bounded (R) & half solar (L) & no & no & 637\\
BCML              & blackbody (B)   & constant (C) & matter bounded (M)    & half solar (L) & no & no & 2211\\
BCRSD           & blackbody (B)   & constant (C) & radiation bounded (R) & solar (S) & yes (D) & no & 730\\
BCMSD          & blackbody (B)   & constant (C) & matter bounded (M)    & solar (S) & yes (D)& no & 2436\\
BCRSCo         & blackbody (B)   & constant (C) & radiation bounded (R) & solar (S) & no & yes (Co) & 1000\\
BCMSCo        & blackbody (B)   & constant (C) & matter bounded (M)    & solar (S) & no & yes (Co) & 3635\\
\hline                  
\end{tabular}
\end{table*}

\subsection{Selection criteria for the models}

\begin{figure}
\includegraphics[width=\hsize,trim = 20 0 40 0,clip =yes]{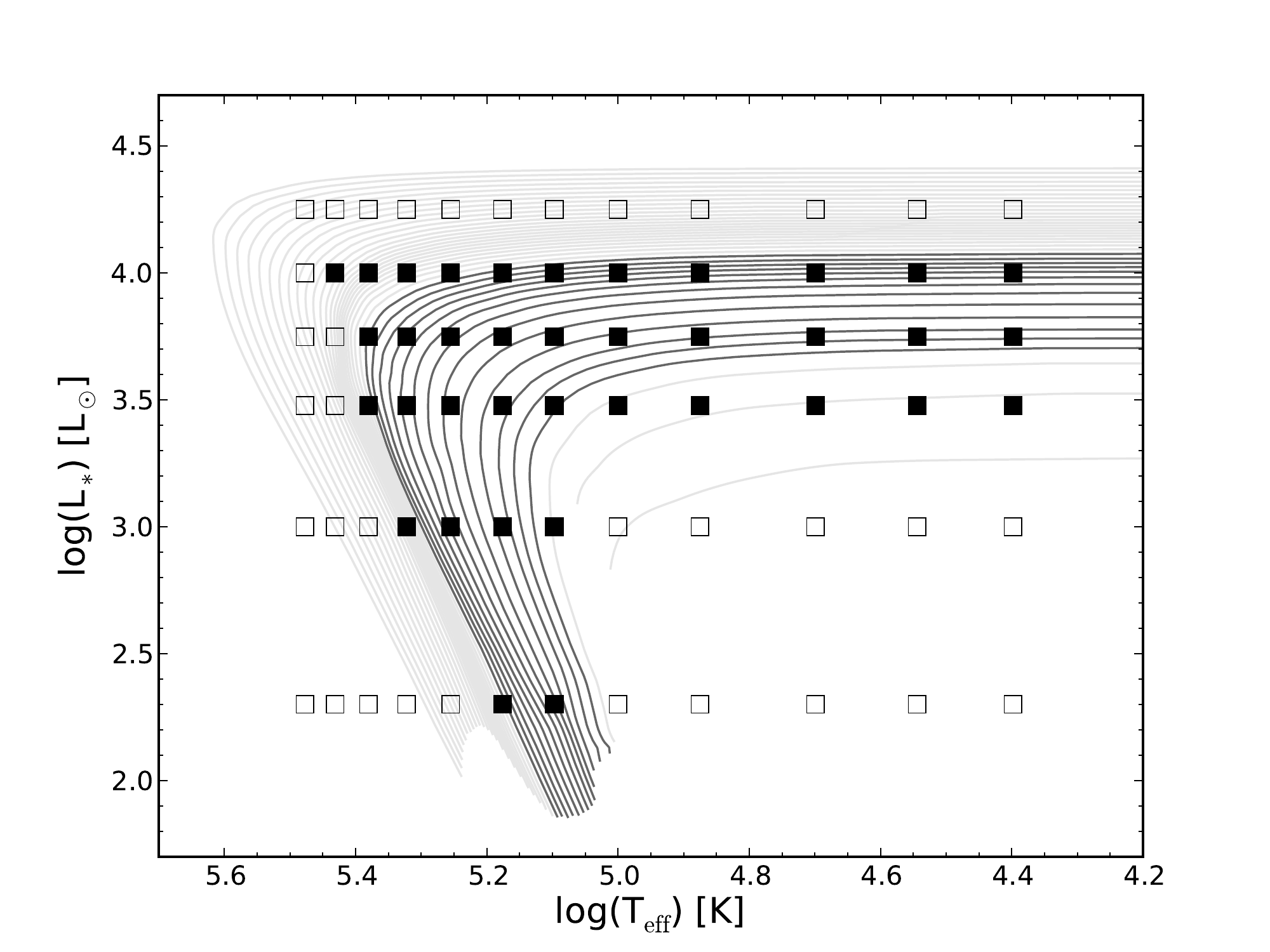}
\caption{Hertzsprung-Russel diagram for our blackbody models (squares). The evolutionary tracks interpolated from the 
\citet{Schoenberner_83} and \citet{Bloecker_95} tracks are overplotted. The tracks of PN central stars with masses in the 
range 0.57--0.70 M$_{\odot}$ are shown in dark grey whereas those with masses below 0.55 M$_{\odot}$ or above 
0.70 M$_{\odot}$ are shown in light grey. The filled squares represent the combinations of stellar temperatures and 
luminosities that are retained for the computation of the ICFs.\label{fig:mod_1}}
\end{figure}

We apply different filters to the initial grid of 22265 BCRS and BCMS models computed from all the combination of parameters showed 
in Table~\ref{tab:1}. We exclude those models with hydrogen masses above one solar mass since higher nebular masses are not 
observed \citep[e.g.,][]{Barlow_87, Gathier_87, Stasinska_91}.
We also exclude the combinations of \efftemp\ and L$_{{\bf *}}$ that fall outside the evolutionary tracks 
from \citet{Schoenberner_83} and \citet{Bloecker_95} of PN central stars with masses between $\sim0.58$ and $\sim0.70$ 
M$_{\odot}$ (see black symbols in Figure~\ref{fig:mod_1}). PNe with such central stars are indeed very rare \citep{Stasinska_97}. 

Another parameter that we use to trim our grid of models is the H$\beta$ surface brightness, S(H$\beta$). 
Based on typical values of S(H$\beta$) in observed samples of Galactic PNe \citep[e.g.][]{Tsamis_04, Liu_04}, 
we restrict $\log$ S(H$\beta$) values in our grid to the range $10^{-13}$ -- $10^{-11}$ erg s$^{-1}$ cm$^{-2}$ arcsec$^{-2}$. 

We also exclude the models having at the same time large (small) outer radius and high (low) electron density. To do this, 
we restrict our trimmed grid of models to $2\times10^{53}\leq n_{\rm H}R_{\rm out}^3\leq 3\times10^{56}$. This filter 
is based on Figure (18) from \citet{Marigo_01} which represents electron densities as a function of nebular radius 
for a sample of Galactic PNe together with their model predictions. 

\begin{figure}
\includegraphics[width=\hsize,trim = 25 0 45 0,clip =yes]{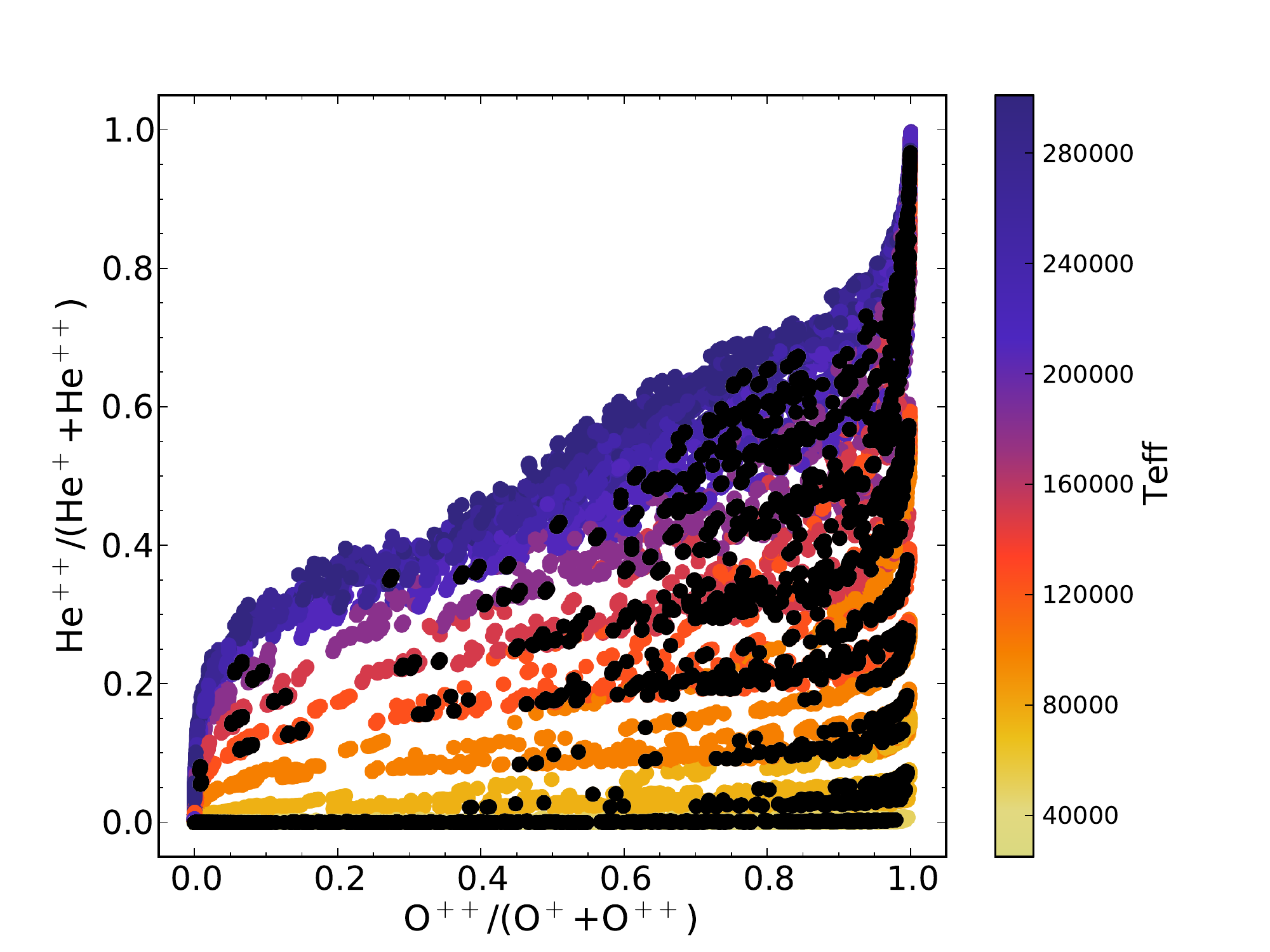}
\caption{Values of \gihe\ as a function of \gio\ for the initial grid of BCRS and BCMS models (in colors). Overplotted 
(in black) are the models resulting from the applied filters (see the text).The color bar located on the side runs from 
low to high values of the effective temperature.}\label{fig:mod_2}
\end{figure}

Figure~\ref{fig:mod_2} displays the values of \gihe\ as a function of \gio\ for the initial grid of $\sim$22300 BCRS 
and BCMS models (in different colors according to \efftemp). We also plot in the figure the grid of 2820 models 
resulting from the above filters (in black). This trimmed grid is used to derive the different 
ICF recipes and their associated uncertainties. In section~\ref{sec:effect} we discuss the validity of our computed ICFs 
for the models other than BCRS and BCMS. The comparison between the values of  \gihe\ as a function of \gio\ for the 
models and the sample of PNe will be presented later.

We include one more restriction when computing the analytical ICFs from our models. We only 
consider those models where the intensities of the common lines used to compute ionic abundances are above 
$\sim10^{-4}\times I({\rm H}\beta)$, otherwise the needed lines are probably not observed.

All the computed models are included in the Mexican Million Models database (3MdB, \citealt{Morisset_09} under 
the reference PNe\_2014. The 3MdB will be available for the community this year. In the meantime the grid is available 
from the authors upon request.

%%%%%%%%%%%%%%%%%%%%%%%%%%%%%%%%%%%%%%%%%%%%%%%%%%%%%%%%%%%%%%%%%%%%%%%%%%
\section{Observational sample}
\label{sec:sample}

Here we constitute a database of line intensities for observed PNe. First it will be used to check whether the range of 
observational ionization properties of our trimmed grid of models corresponds to the range observed in real nebulae. 
Also, it will be used to check whether our ICFs lead to a reasonable behavior of abundance ratios (such 
as Ne/O and Ar/O) with the degree of ionization, as done e.g. by \citet{TorresPeimbert_77}. Spectroscopic observations of 
PNe in the literature are numerous. We do not need to construct a complete sample, but rather to construct a 
sample that extends as far as possible in ionization, central star temperature, and metallicity. Therefore, we 
merge data from several sources, PNe from the Galactic disk, bulge as well as PNe from the Magellanic 
Clouds.  

The observational data for the Magellanic Clouds are taken from \citet{Leisy_06}. The data for the Galactic PNe belongs 
to several sources that contain large samples of objects: \citet{GR_09, GR_12, Gorny_09, Henry_10, Liu_04,  Milingo_10,
Tsamis_03, Wang_07, Wesson_04, Wesson_05}. We also include some Galactic PNe from the compilation by 
Delgado-Inglada \& Rodr\'iguez (2014, ApJ, accepted). In total, we have observations for 138 LMC PNe, 45 SMC PNe, and 
around 200 Galactic PNe. 

Note that the observations are not all equivalent, neither in terms of signal-to-noise nor in term of aperture size with 
respect to the total angular size of the nebulae. For example, observations of extragalactic PNe cover the entire 
nebular shells, while those of nearby nebulae are generally made with slits covering only a small portion 
of the objects (often the central or the brightest region). Only a few authors \citep{Liu_04, Tsamis_03} have obtained 
integrated spectra for nearby PNe by sweeping the slit across the object. The spectra from \citet{GR_09, GR_12, 
Liu_04, Tsamis_03, Wang_07, Wesson_04, Wesson_05} are the deepest ones and they possess the highest resolution. 

\begin{figure}
\centering
\includegraphics[width=9cm,trim = 0 0 0 0,clip =yes]{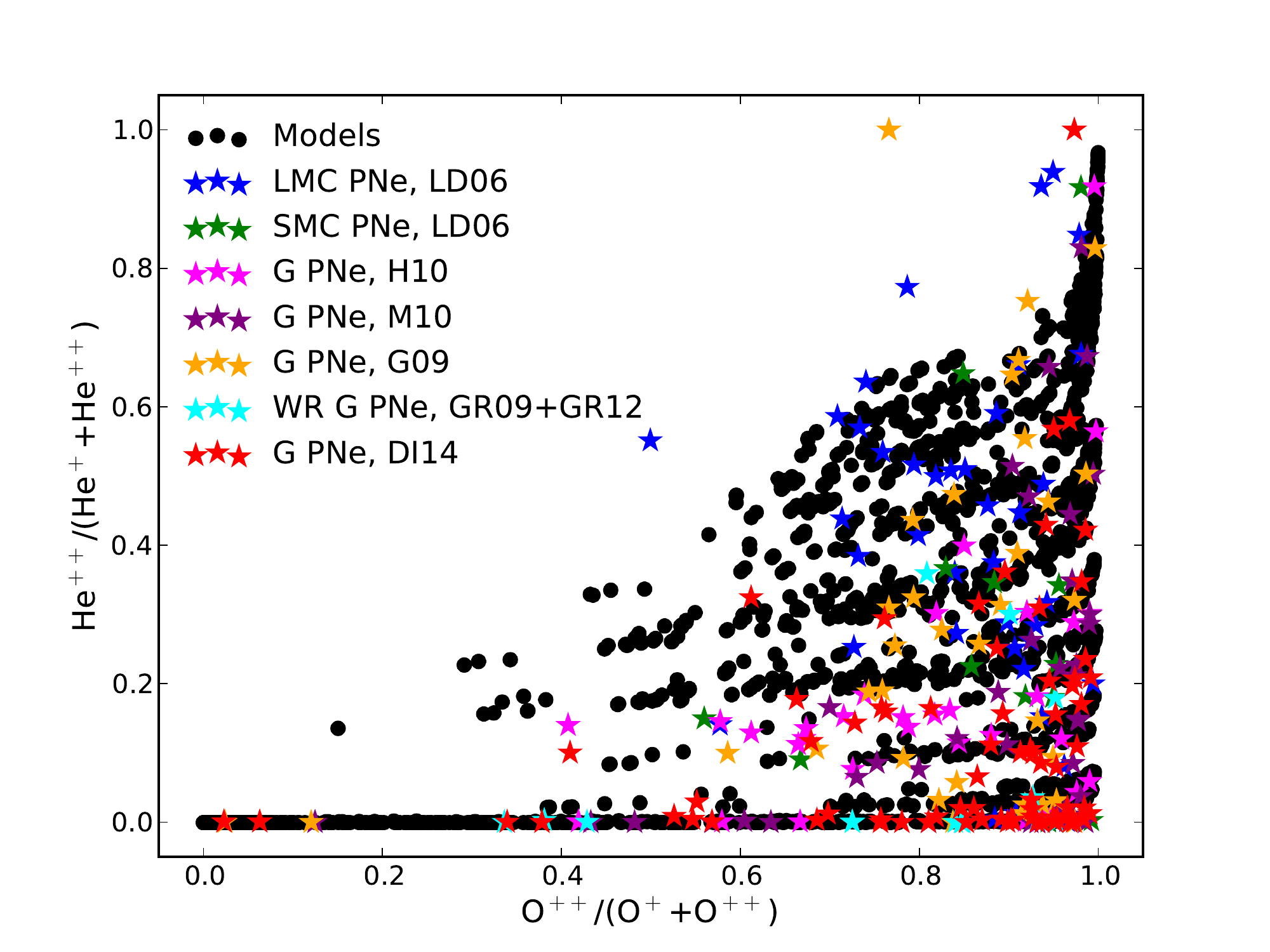}\\
\caption{Values of \gihe\ as a function of \gio\ for the whole sample of observed PNe (colored stars). The data are taken from: 
\citet{Leisy_06} (LD06), \citet{Henry_10} (H10), \citet{Milingo_10} (M10), \citet{Gorny_09} (G09), \citet{GR_09, GR_12} (GR09, GR12), 
and from the compilation by \citet{DI_14} (DI14). The black circles represent the values computed for our grid of blackbody models 
with constant density (the same as shown in Fig.~\ref{fig:mod_2}). \label{fig:obs_1}}
\end{figure}

Figure~\ref{fig:obs_1} shows the abundance ratios \gihe\ as a function of \gio\ for the observed PNe (stars) and for our 
BCRS and BCMS models (circles). It can be seen that the models cover almost the whole range of ionization, with the 
exception of a few cases that correspond to some of the lowest quality spectra. This figure confirms that our photoionization 
models are representative of most of the observed PNe. The calculation of the ionic abundances for the observational sample 
is explained in detail in Section~\ref{sec:ion_ab}.

%%%%%%%%%%%%%%%%%%%%%%%%%%%%%%%%%%%%%%%%%%%%%%%%%%%%%%%%%%%%%%%%%%%%%%%%%%
\section{Ionization correction factors\label{sec:icfs}}

\subsection{Methods and notations}
\label{sec:methods}

 Ionic abundances ratios are computed as the ratios of the intensities of lines  emitted by the parent ions divided by the ratios of 
 the corresponding emissivities of those lines (see e.g. Stasi\'nska 2006). 
 
The ICF for a given element, ICF(X), is the factor by which one should multiply 
the sum of all the observed ionic abundances ($\Sigma$X$^{+i}$/H$^+$) in order to obtain the total 
abundances of the element (X/H). 
 
Let us take the example of oxygen to define the quantities and notations that will be used throughout this paper. 
There are two ions which emit lines in the optical domain\footnote{The neutral oxygen 
line at 6300 \AA\ is observed in many PNe. However its abundance must not be included in formulae giving the total 
abundance of oxygen with respect to hydrogen since O$^0$ coexists with H$^0$ while only the abundance of H$^+$ 
is measured through the hydrogen recombination lines.}: O$^{+}$ and O$^{++}$.
The ionization correction factor to obtain the O/H ratio, better written as ICF(O$^{+}$+O$^{++}$) to specify from which ions it is obtained, is defined by:
 
 \begin{equation}
 \label{eq:OoH}
 \frac{\rm{O}}{\rm{H}}=\left(\frac{\rm{O^{+}}}{\rm{H}^{+}}+\frac{\rm{O^{++}}}{\rm{H}^{+}}\right)\rm{ICF}(\rm{O}^{+}+\rm{O}^{++})
 \end{equation}
 where 
 
 \begin{equation}
 \label{eq:icfmod}
\rm{ICF}(\rm O^{+}+\rm O^{++})=
\frac{\mathnormal{x}({\rm H^{+}})}{\mathnormal{x}({\rm O^{+}})+\mathnormal{x}({\rm O^{++}})}.
 \end{equation}
 The $x$ values are the relative ionic fractions weighted by the electron density $n_e$. For example, for O$^{++}$, 
 $x$(O$^{++}$) is defined as:
 \begin{equation}
 \label{eq:x}
 x({\rm O^{++}})= \frac {\int\!n({\rm O^{++}})n_e dV}{\int\!n({\rm O^{}})n_e dV}. 
\end{equation}
 
For each photoionization model, we can obtain ICF(O$^{+}$+O$^{++}$) from Eq.\ref{eq:icfmod}, since the ionic fractions are outputs of the models. The result will be called 
\icfm(O$^{+}$+O$^{++}$). 

We obtain an analytical fit for ICF(O$^{+}$+O$^{++}$), called \icff(O$^{+}$+O$^{++}$), by looking for correlations between the values of \icfm(O$^{+}$+O$^{++}$) and  ionic 
fractions given by the models for other, easily observed, ions. These ionic fractions are chosen to be
\begin{equation}
    \label{eq:upsilon}
\upsilon= \frac{\rm{He}^{++}}{(\rm{He}^{+}+\rm{He}^{++})}  
 \end{equation}
or 
\begin{equation}
    \label{eq:omega}
\omega= \frac{\rm{O}^{++}}{(\rm{O}^{+}+\rm{O}^{++})}.   
 \end{equation}

We have chose the $\upsilon$ and $\omega$ parameters because they are well-behaved functions varying between 0 and 1 that are reliably calculated from optical observations (the involved  lines are bright and easily detected, and the ratios are weak functions of the electron temperature). 
 
The quantity 
\begin{equation}
\label{eq:delta} 
\Delta/\rm{ICF}_{\rm f}({\rm{O^{+}}+\rm{O}^{++}})=\frac{\rm{ICF}_{\rm m}(\rm{O^{+}}+\rm{O}^{++})-\rm{ICF}_{\rm f}(\rm{O^{+}}+\rm{O}^{++})}{\rm{ICF}_{\rm f}(\rm{O^{+}}+\rm{O}^{++})}
\end{equation} 
 represents, for each model, the relative uncertainty of \icff(O$^{+}$+O$^{++}$).      
 
To provide a general expression for this uncertainty, we find an analytical fit to the upper and lower envelope of
$\Delta/\rm{ICF}_{\rm f}({\rm{O^{+}}+\rm{O}^{++}})$ as a function of $\upsilon$ or $\omega$. 
Note that the upper and lower envelopes are generally not symmetric. The fits are represented by the functions 
$\varepsilon^+$(O$^{+}$+O$^{++}$) and $\varepsilon^-$(O$^{+}$+O$^{++}$), for the upper and lower envelopes, 
respectively.

Thus, the computed abundance of oxygen, including the error bars,  can be written as: 

\begin{equation} 
    \label{eq:OoHcalc}
 \frac{\rm{O}}{\rm{H}}=\left(\frac{\rm{O^{+}}}{\rm{H}^{+}}+\frac{\rm{O^{++}}}{\rm{H}^{+}}\right)\rm{ICF}_{\rm f}(\rm{O}^{+}+\rm{O}^{++})\left(1\pm^{\varepsilon^+(\rm{O}^{+}+\rm{O}^{++})}_{\varepsilon^-(\rm{O}^{+}+\rm{O}^{++})}\right). 
\end{equation}
and the upper and lower error bars in $\log$(O/H), expressed in dex, as: 
\begin{equation} 
\log(1+\varepsilon^+)
\end{equation}
and 
\begin{equation} 
-\log(1-\varepsilon^-),
\end{equation}
respectively.

Note that our error bars are rather conservative. Users may wish to reduce them by a reasonable factor of their choice, however their trend with $\upsilon$ or $\omega$ should be maintained. For each element we define  regions in $\upsilon$ and $\omega$ where the error bars are so large that no meaningful abundance can be computed. The only reliable way to obtain the elemental abundances in such a case is to perform a well constrained detailed photoionization modelling of the object.

In the following subsections we deal with ICFs for He, O, N, Ne, S, Cl, Ar, and C. They are based on Figures~\ref{fig:he}--\ref{fig:c_1} in 
which the upper panels represents \icfm\ as a function of either $\upsilon$ or $\omega$. The errors are estimated using the lower panels 
which represent the values of $\Delta$/\icff\ as a function of $\upsilon$ or $\omega$. The errors in dex are showed in the right axis. The 
color bars located at the side of each panel run, in most of the cases, from low to high values of \efftemp\ (upper panels) or $f_{\rm mass}$ 
(lower panels). Table~\ref{tab:3} compiles all the ICFs we derived and the analytical expressions to estimate the associated error bars.

All the concepts and methods defined above can be readily generalized to the case where the abundances are computed with respect to O instead of H.

In this section we restrict ourselves to the BCRS and BCMS models, the other families of models will be discussed in Section~\ref{sec:effect}. 

\subsection{Helium}

Many observed PNe show \ion{He}{ii} lines in their spectra implying that the abundance of neutral helium is 
probably negligible for them. From our models, the contribution of He$^{0}$ to the total abundance of helium can 
be more than 20\% only in those objects with no contribution of He$^{++}$, 
i.e, $\upsilon$ = 0. 

The derivation of an ICF for neutral helium based on other ions is not recommend since the relative populations of 
helium ions depend essentially on the effective temperature of the central star, whereas the ones of metal 
ions also depend on the ionization parameter. The ICFs proposed by \citet{Peimbert_92, Zhang_03} and 
\citet{Peimbert_69} lead to helium abundances that show a hint of a trend with the degree of ionization. 

\begin{figure}
\subfigure{\includegraphics[width=\hsize,trim = 20 10 50 0,clip =yes]{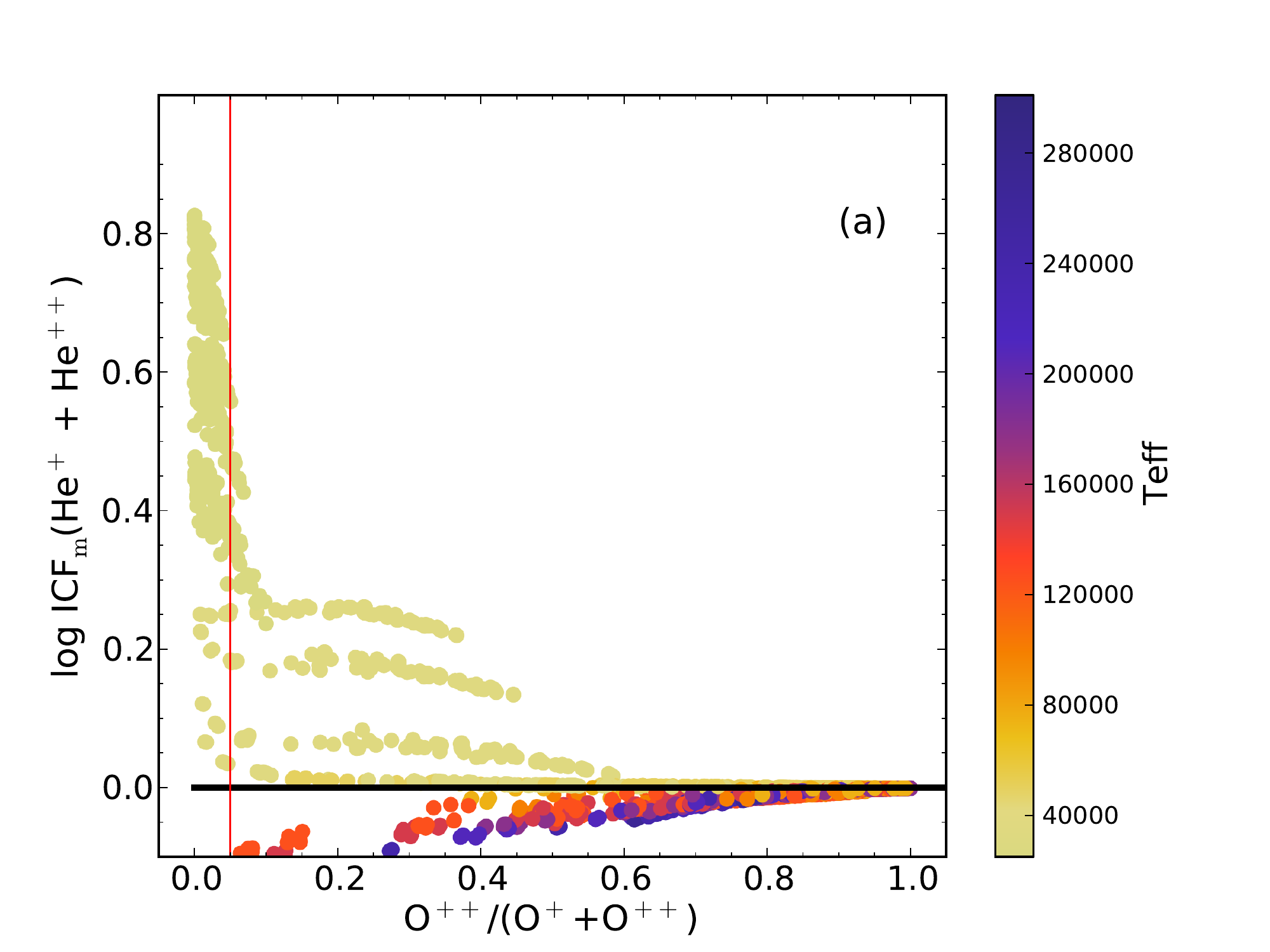}}
\subfigure{\includegraphics[width=\hsize,trim = 25 0 60 10,clip =yes]{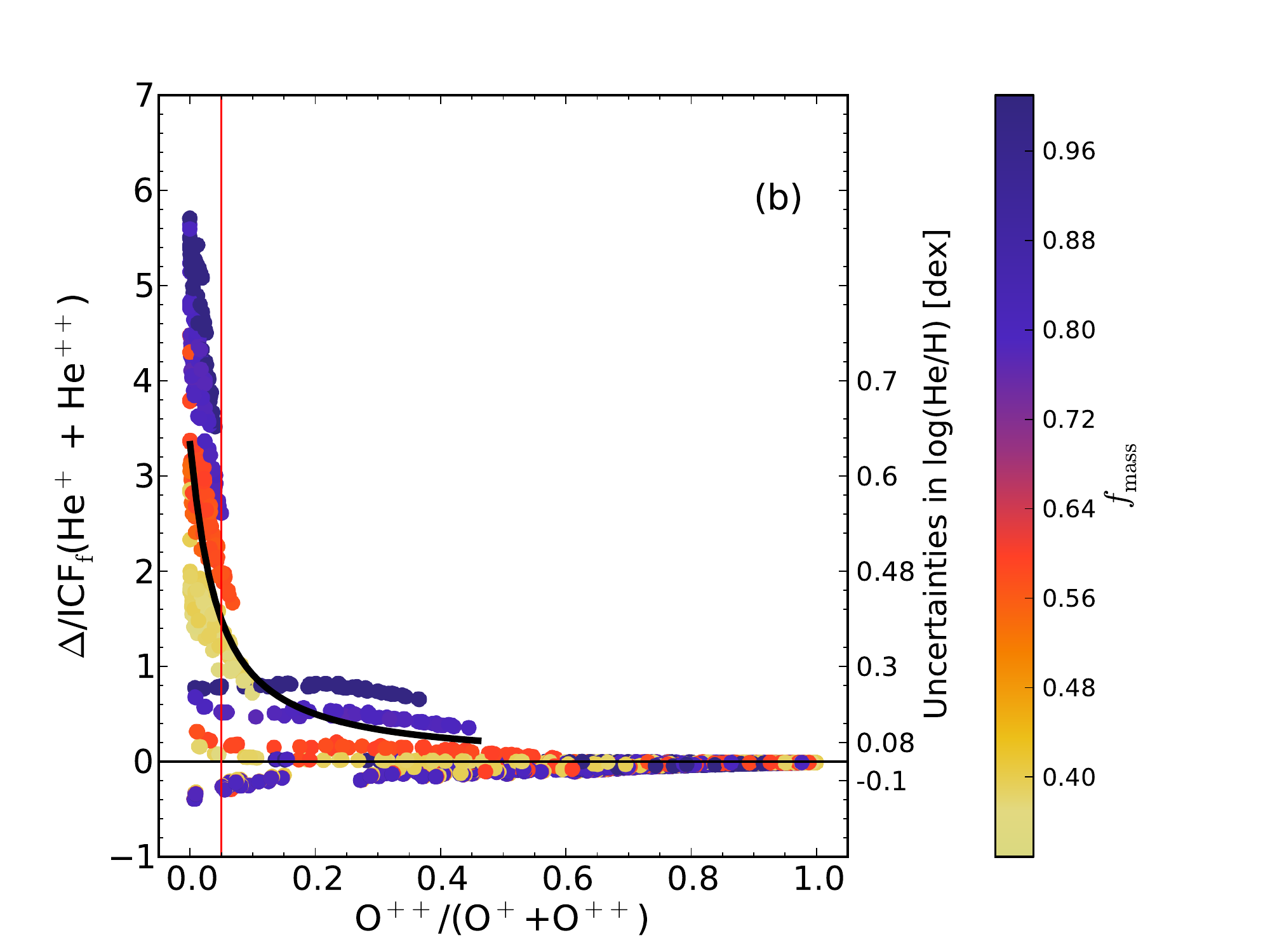}}
\caption{(a) Values of \icfm(He$^{+}$ + He$^{++}$) as a function of $\omega$ for our BCRS and BCMS models. 
The thick line represents \icff(He$^{+}$ + He$^{++}$). The color bar located on the side runs from low to high 
values of \efftemp. (b) Values of $\Delta$/\icff(He$^{+}$ + He$^{++}$) as a function of $\omega$ for our 
BCRS and BCMS models. The thick line represents the values of $\varepsilon^+$ for the models with $\omega$ $<$ 0.5. 
The thin horizontal line represents \icfm(He$^{+}$ + He$^{++}$) = \icff(He$^{+}$ + He$^{++}$). 
The color bar located on the side runs from low to high values of $f_{\rm mass}$. The vertical line in 
both panels delimits the value of $\omega$ (0.05) above which our ICF is valid.\label{fig:he}}
\end{figure}

We suggest to calculate He/H simply by adding He$^{+}$/H$^+$ and He$^{++}$/H$^+$ abundances: 
\begin{equation}
\label{eq:he_1}
\mbox{ICF}_{\rm f}({\rm He}^{+}+{\rm He}^{++})=1,
\end{equation}
(see Fig.~\ref{fig:he}a) and integrate the correction for neutral helium in the error bar. As expected, 
Figure~\ref{fig:he}b shows that the correction for neutral helium is important 
only for the models having ionizing sources with \efftemp~$\lesssim50000$ K. The highest corrections are 
found for radiation bounded PNe (dark purple dots with $f_{\rm mass} = 1$) having \inthe~$<0.02$. 
The value of $\varepsilon^+$ is estimated from Figure~\ref{fig:he}b as: 

\begin{equation}
\label{eq:he_2}
\varepsilon^{+}({\rm He}^{+}+{\rm He}^{++})=\frac{0.1}{0.03+\omega^{1.1}}, 
\end{equation}
represented by a thick solid line. 

Note that Equations~(\ref{eq:he_1}) and (\ref{eq:he_2}) provide estimates of helium abundances for objects 
with $\omega \lesssim 0.45$, but for detailed abundance studies tailored photoionization models should be constructed. 

\subsection{Oxygen}

From optical observations we can derive O$^{+}$ and O$^{++}$ abundances through
several bright lines, such as [\ion{O}{ii}] $\lambda\lambda$3727, 29 and 
[\ion{O}{iii}] $\lambda$4959, $\lambda$5007. These two ions are expected to be the 
major contributors to total oxygen abundances in low ionization PNe. In PNe of higher 
ionization, the contribution of other ions may be significant. 

\begin{figure}
\subfigure{\includegraphics[width=\hsize,trim = 20 10 50 0,clip =yes]{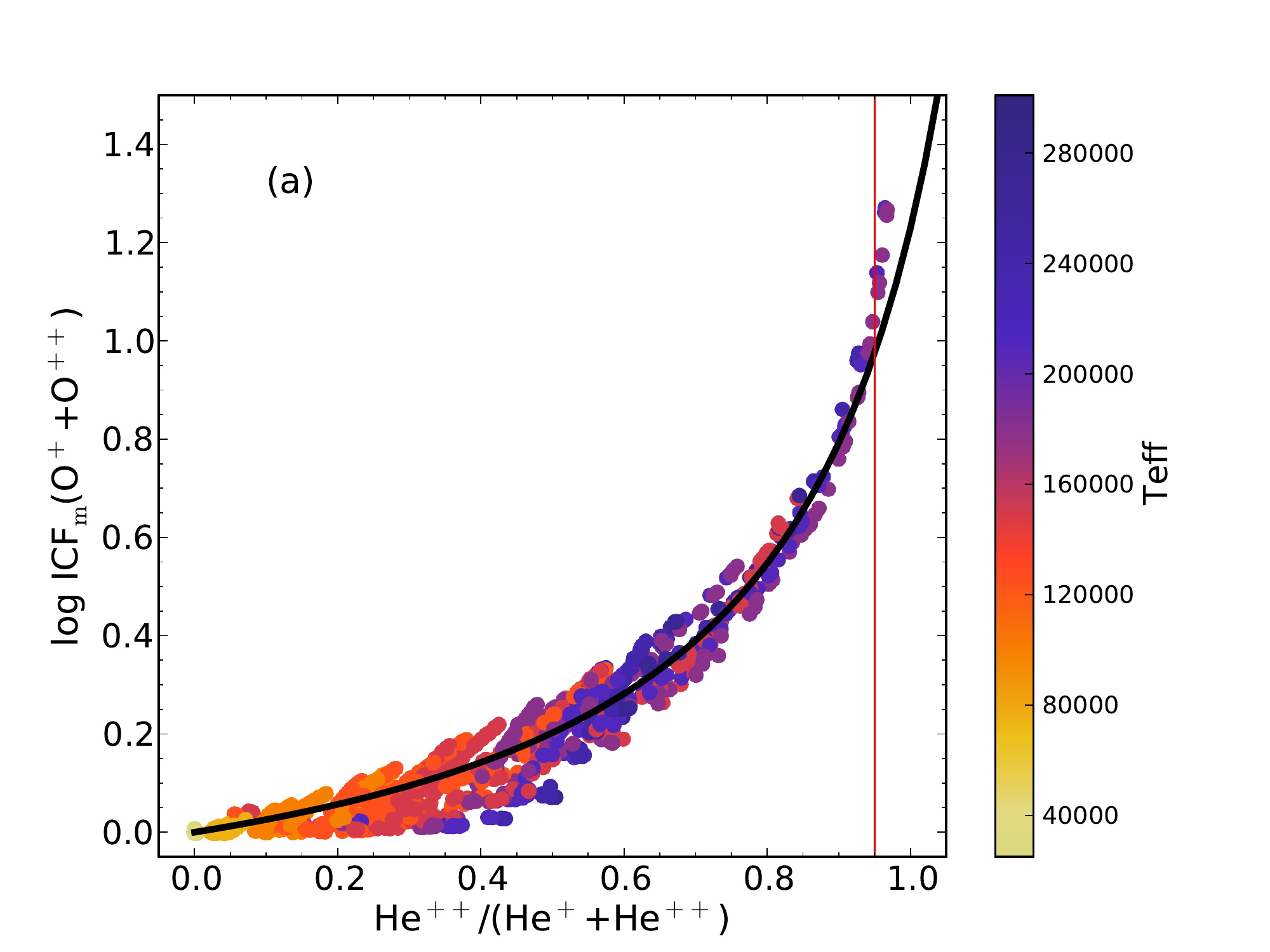}}
\subfigure{\includegraphics[width=\hsize,trim = 20 0 50 10,clip =yes]{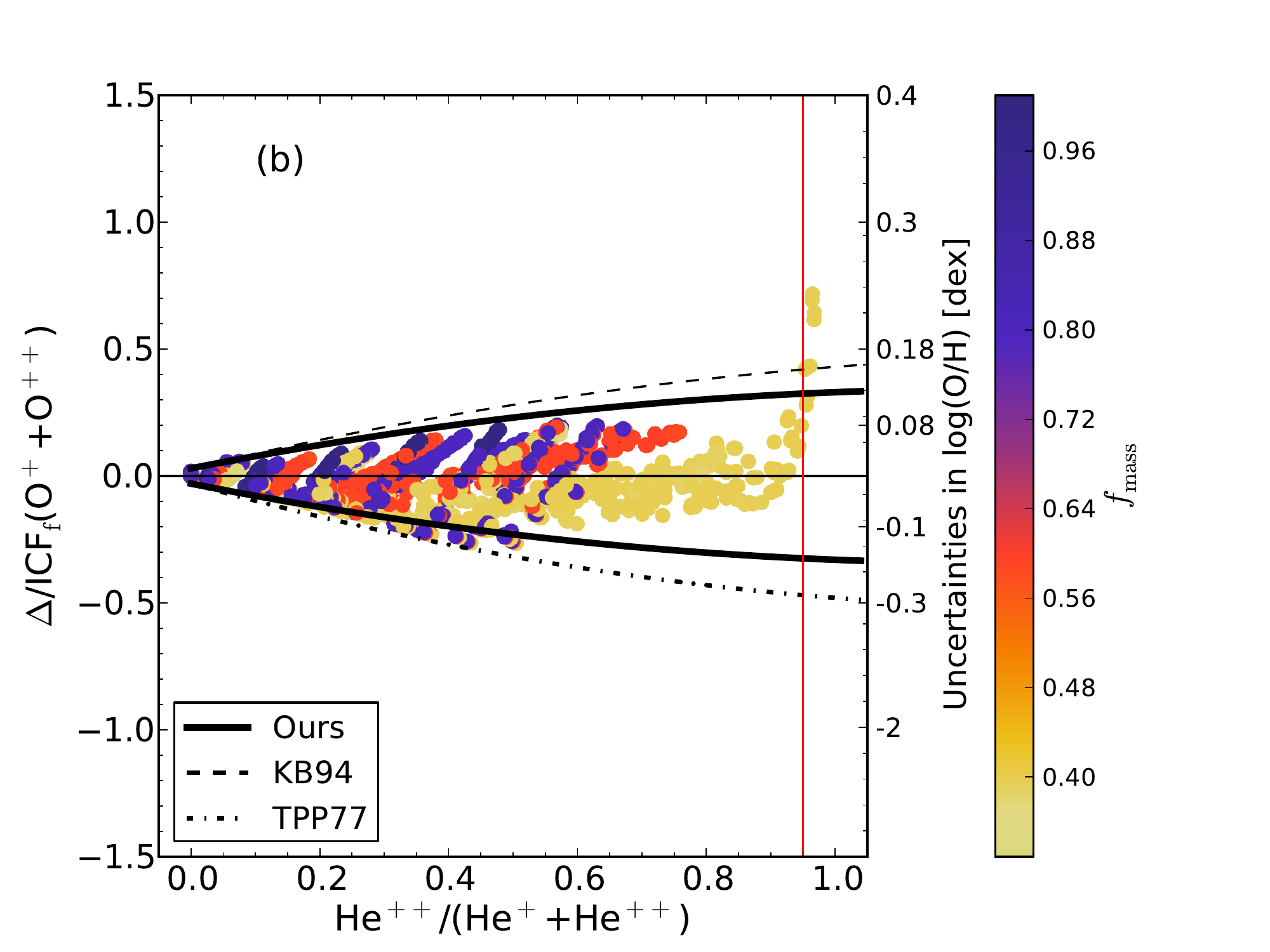}}
\caption{(a) Values of ICF$_{\rm m}$(O$^{+}$ + O$^{++}$) as a function of $\upsilon$ for our 
BCRS and BCMS models. The solid line is ICF$_{\rm f}$(O$^{+}$ + O$^{++}$) from equation~(\ref{eq:o_1}). 
The color bar located on the 
side runs from low to high values of \efftemp. (b) Values of $\Delta$/\icff(O$^{+}$ + O$^{++}$) as a function 
of $\upsilon$ for our BCRS and BCMS models. The thick solid lines are $\varepsilon^{+}$ and $\varepsilon^{-}$ 
and the thin solid line represents \icfm(O$^{+}$ + O$^{++}$) = \icff(O$^{+}$ + O$^{++}$). For comparison purposes 
we show the uncertainties associated with the ICFs from KB94 (dashed line) and \citet{TorresPeimbert_77} 
(dash-dotted line). The color bar located 
on the side runs from low to high values of $f_{\rm mass}$. The vertical line in 
both panels delimits the value of $\upsilon$ (0.95) below which our ICF is valid.\label{fig:o}}
\end{figure}

Figure~\ref{fig:o}a displays the values of ICF$_{\rm m}({\rm O}^{+}+{\rm O}^{++})$ as 
a function of $\upsilon$ for our models. 
From this figure we propose:

\begin{equation}
\label{eq:o_1}
\log \mbox{ICF}_{\rm f}({\rm O}^{+}+{\rm O}^{++})=\frac{0.08\upsilon+0.006\upsilon^2}{0.34-0.27\upsilon}.
\end{equation}
This function is represented by a solid line in the figure. It does not fit the models with $\upsilon$ larger than 0.95. 
However in that case, the ICF becomes very large and the oxygen abundance very uncertain. As a matter of fact 
we do not recommend to compute oxygen abundances in such cases (unless a well constrained tailored photoionization 
model is computed). 

Figure~\ref{fig:o}b shows $\Delta$/\icff(O$^{+}$ + O$^{++}$) as a function 
of $\upsilon$. Note that radiation bounded models (dark purple dots with $f_{\rm mass} = 1$)
do not reach the highest values of $\upsilon$. 
Only the matter bounded models in which we trimmed away around 40\%--60\% (yellow dots with $f_{\rm mass} \sim 0.4$--0.6) 
of the nebular mass have $\upsilon$ values beyond 0.6. This 
suggests that PNe with $\upsilon$ above 0.6 are probably matter bounded nebulae.

The values of $\varepsilon^+$  and $\varepsilon^-$ can be estimated as: 
\begin{equation}
\label{eq:o_2}
\varepsilon({\rm O}^{+}+{\rm O}^{++}) =  0.03 + 0.5\upsilon - 0.2\upsilon^2.
\end{equation}
The uncertainties associated with the ICF for oxygen (between 0.1 and 0.15 dex) can constitute a significant part of 
the total error budget on the total abundance of this element.   

For comparison we also plot in Figure~\ref{fig:o}b the error bars associated with the two ICFs most 
frequently adopted in the literature, those proposed by KB94 (dashed line) and \citet{TorresPeimbert_77} (dash-dotted line). 
In the first case, upper error bars are somewhat higher than the ones associated 
with our ICF whereas lower error bars are equal to ours. As for the ICF from \citet{TorresPeimbert_77},  there are no upper 
error bars since this ICF always overestimates oxygen abundances. The lower error bars are somewhat higher than ours.  

\subsection{Nitrogen}

In the optical range, the only nitrogen lines to be commonly detected are the ones emitted by N$^{+}$. 
This ion is often found in a small proportion in PNe, since its low ionization potential (29.6 eV) can be 
easily reached, and thus, using a correct ICF is crucial to obtain reliable values of total nitrogen abundances. 
We find that the contribution of N$^{+}$ to the total abundance of nitrogen can be less than 20\% in objects with 
$\omega\gtrsim0.6$. 

\begin{figure}
\subfigure{\includegraphics[width=\hsize,trim = 20 10 50 0,clip =yes]{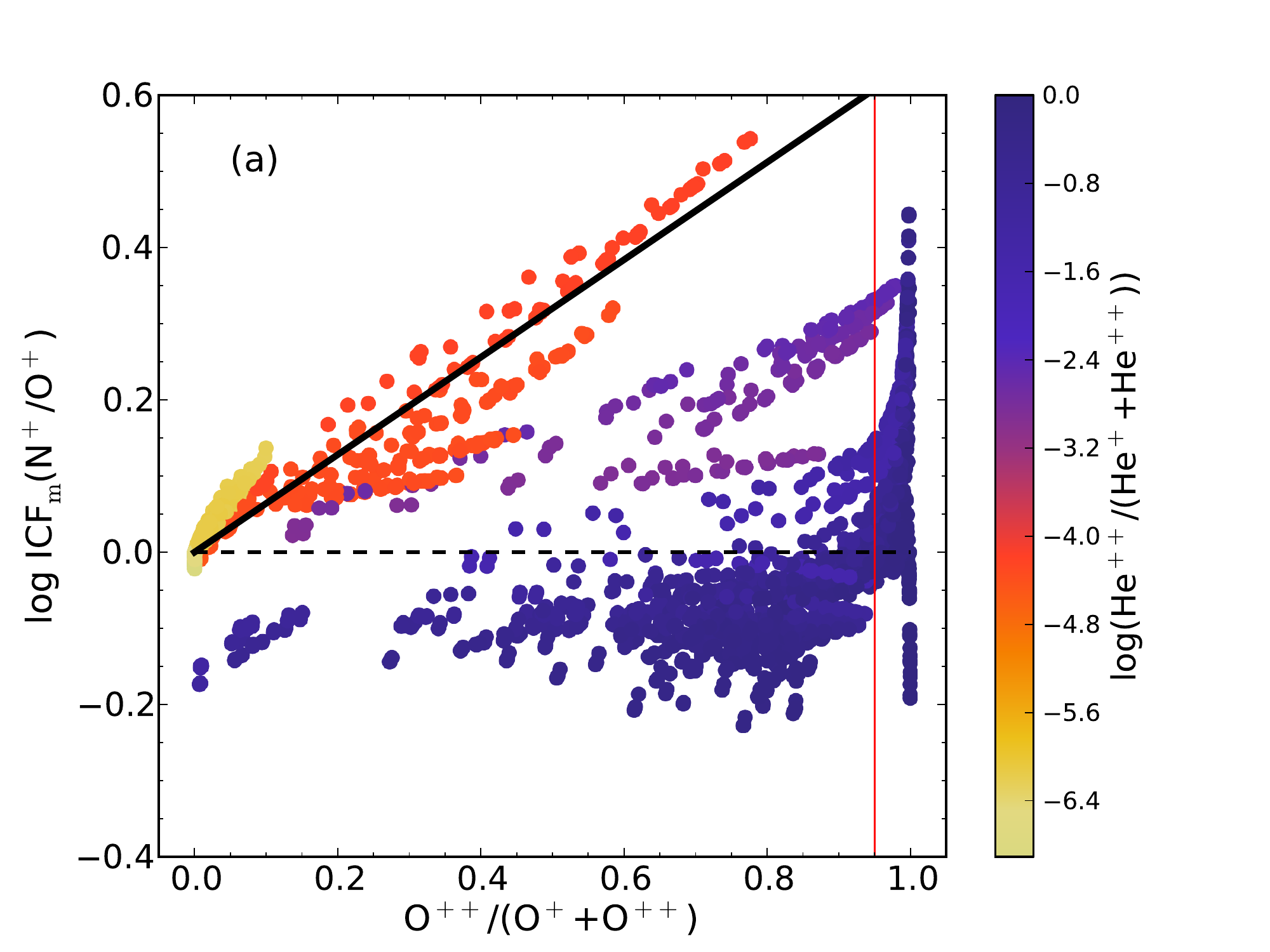}}
\subfigure{\includegraphics[width=\hsize,trim = 20 0 50 10,clip =yes]{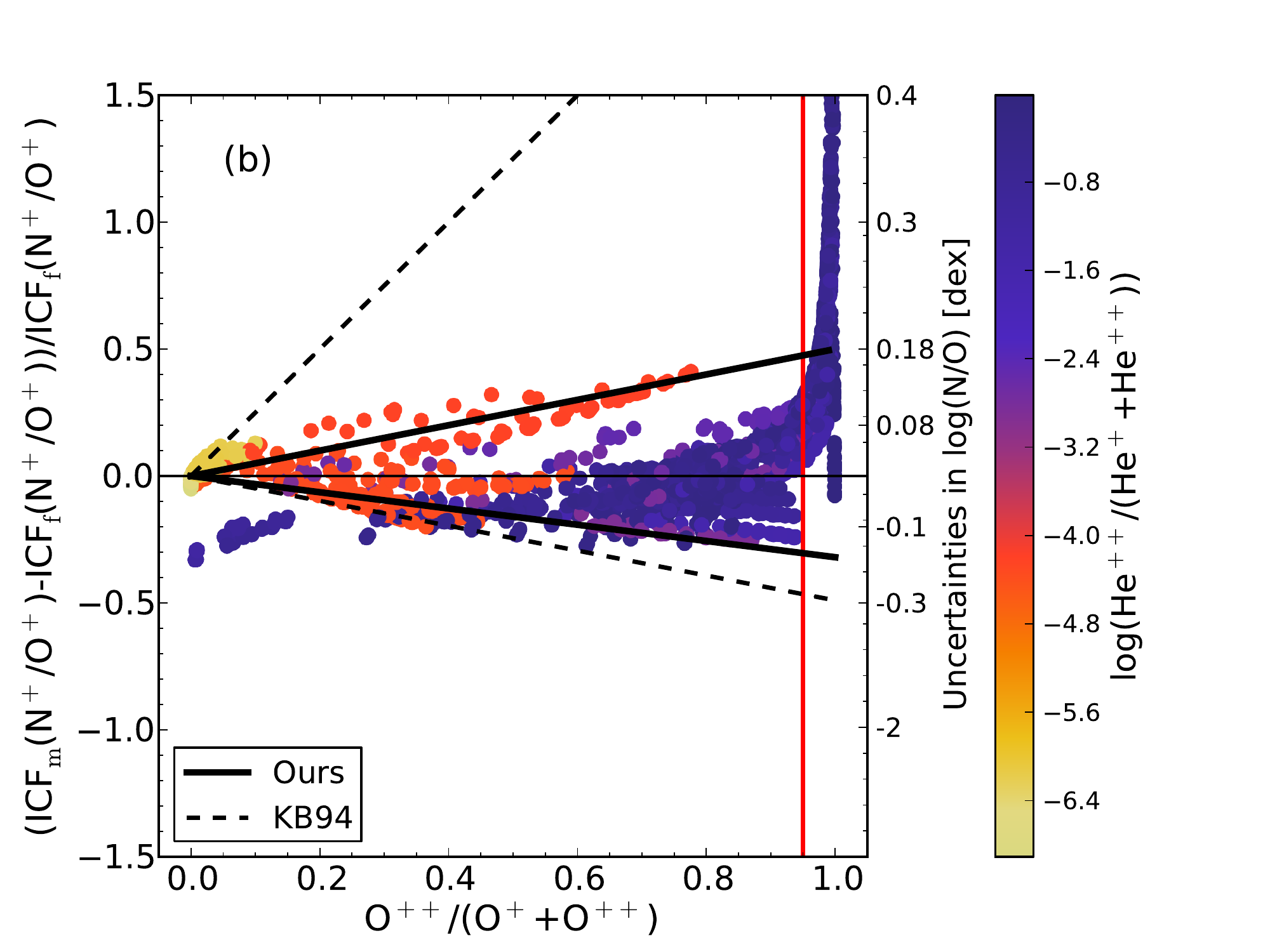}}
\caption{(a) Values of ICF$_{\rm m}$(N$^{+}$/O$^{+}$) as a function of $\omega$\ for our BCRS and 
BCMS models. The dashed line represents the ICF suggested by KB94 and the solid line corresponds 
to the \icff\ of equation~(\ref{eq:n_1}) for $\log\upsilon$ = $-5$. 
(b) Values of $\Delta$/\icff(N$^{+}$/O$^{+}$) as a function of $\omega$ 
for our BCRS and BCMS models. The thick solid lines are $\varepsilon^{+}$ and $\varepsilon^{-}$. The dashed 
lines represent the uncertainties associated with N/O = N$^{+}$/O$^{+}$. The thin solid line represents 
\icfm(N$^{+}$/O$^{+}$) = \icff(N$^{+}$/O$^{+}$).
The color bar located on the side in both panels runs from low to high values of $\log\upsilon$ and the 
vertical line delimits the value of $\upsilon$ (0.95) below which our ICF is valid.\label{fig:n}}
\end{figure}

Nitrogen abundances are generally calculated assuming that N/O = N$^{+}$/O$^{+}$ (e.g., KB94). 
However, we see from Figure~\ref{fig:n}a that this procedure (corresponding to the dashed line) highly 
underestimates N/O in PNe with very low values of $\upsilon$ (i.e. low \efftemp). 
Indeed, while the ionization potentials of N$^+$ and O$^+$ are close, they are not identical (29.6 and 
35.1 eV respectively). This difference produces significant differences in the N$^+$/O$^+$ ratio in 
regimes where the photons just above such energies contribute significantly to the photoionization 
rates, which occurs at low values of \efftemp\footnote{Note that, in this case, the ICFs depend strongly 
on the model atmospheres.} (except in cases where the ionizing star is so cool that virtually 
no N$^{++}$ is produced).

The best ICF should take into account both $\upsilon$ and $\omega$ and we find that the best expression is given by:
\begin{equation}
\label{eq:n_1}
\log \mbox{ICF}_{\rm f}({\rm N}^{+}/{\rm O}^{+})  = -0.16\omega(1 + \log\upsilon). 
\end{equation}
We recommend to use this ICF only until $\omega = 0.95$. When \ion{He}{ii} lines are not detected $\log$ \icff = $0.64\omega$ 
should be used. This is equivalent to Equation~(\ref{eq:n_1}) when $\log\upsilon$ = -5 (represented with a solid line in 
Fig.~\ref{fig:n}a).

Figure~\ref{fig:n}b shows $\Delta$/\icff(N$^{+}$/O$^{+}$) values as a function of $\omega$. The uncertainties 
associated with our ICF, represented by the solid lines, are given by: 
\begin{equation}
\label{eq:n_5}
\varepsilon^{-}({\rm N}^{+}/{\rm O}^{+}) = 0.32\omega
\end{equation}
and 
\begin{equation}
\label{eq:n_6}
\varepsilon^{+}({\rm N}^{+}/{\rm O}^{+}) = 0.50\omega.
\end{equation}
The uncertainties related to N/O = N$^{+}$/O$^{+}$ are shown with dashed lines in the figure for comparison. 
The error bars for the models with $\log\upsilon\geq-2$ are similar for both correction schemes but for the rest 
of the models our error bars are significantly lower. 

The models with the highest discrepancies between ICF$_{\rm f}$(N$^{+}$/O$^{+}$) and 
ICF$_{\rm m}$(N$^{+}$/O$^{+}$) 
are those with low temperature central stars and a high percentage of the nebula trimmed away 
when constructing matter bounded models. This is a consequence of the ion N$^{+}$ being located 
in the outer regions of the nebula. In these PNe, nitrogen abundances are highly uncertain. However, 
this combination of low temperature central stars and matter bounded nebulae is very unlikely.

\subsection{Neon}

In PNe, Ne$^{++}$, Ne$^{+3}$, and Ne$^{+4}$ can be derived from [\ion{Ne}{iii}], 
[\ion{Ne}{iv}], and [\ion{Ne}{v}] optical collisionally excited lines. The [\ion{Ne}{iv}] 
$\lambda\lambda$4724,25 doublet is very weak and sensitive to the temperature. 
The [\ion{Ne}{v}] lines are emitted in the 3300--3450 range, often outside the observed 
wavelength range. Therefore total abundances of neon are often calculated using only 
Ne$^{++}$ and one ICF to account for the other ions. 

The usual correction scheme for neon, Ne/O = Ne$^{++}$/O$^{++}$, assumes that 
O$^{++}$ and Ne$^{++}$ are located in the same region of the nebula, based on their similar ionization potentials 
(54.93 and 63.45 eV respectively). In low ionization objects, a small fraction of Ne$^{+}$ could 
coexist with O$^{++}$ \citep[see, e.g.,][]{Peimbert_92}, making this ICF incorrect. In regions with a 
large amount of neutral hydrogen, charge exchange reactions become important, and O$^+$ may coexists 
with Ne$^{++}$ \citep{Pequignot_80}.

\begin{figure}
\subfigure{\includegraphics[width=\hsize,trim = 20 10 50 0,clip =yes]{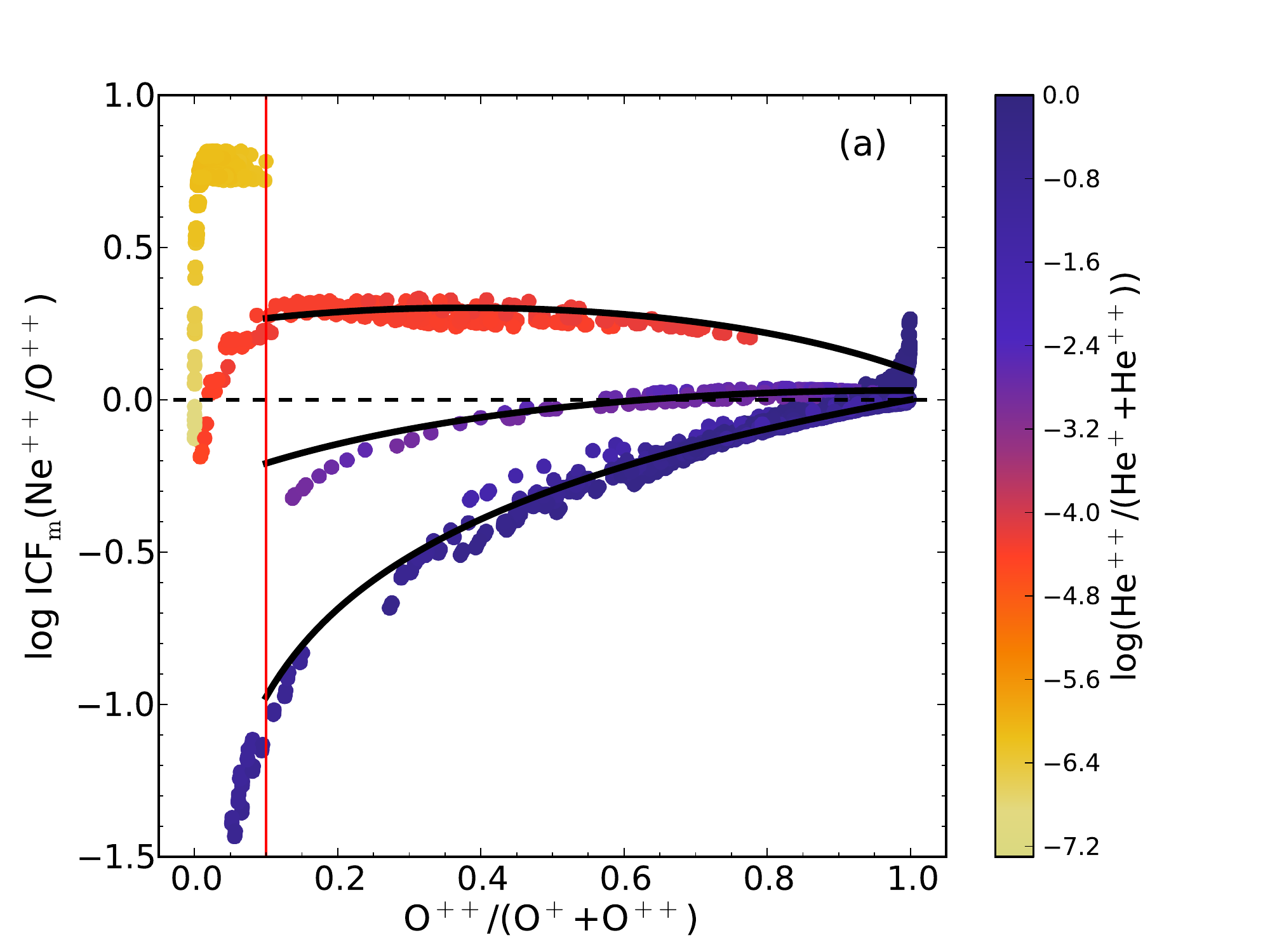}}
\subfigure{\includegraphics[width=\hsize,trim = 20 0 50 10,clip =yes]{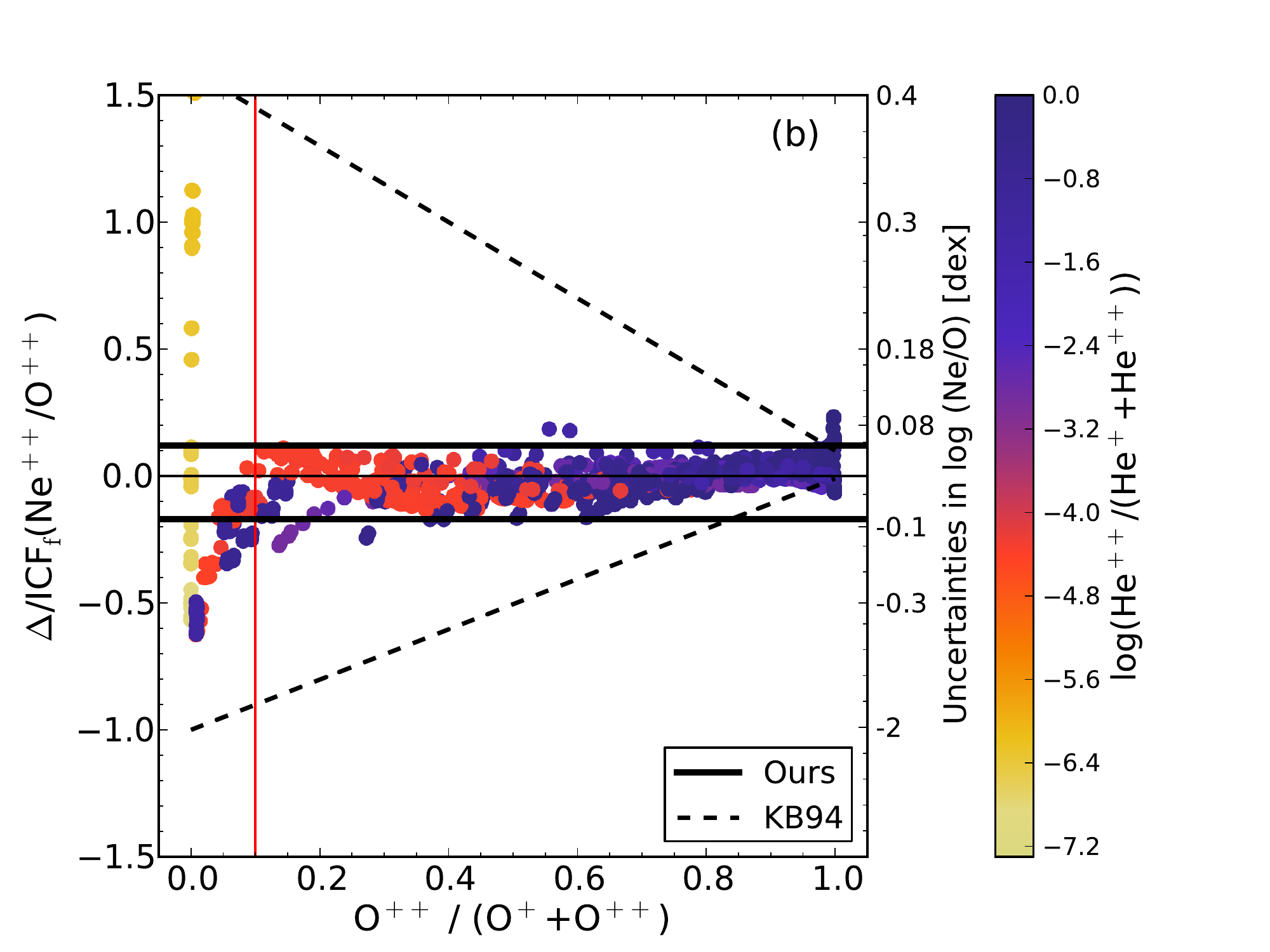}}
\caption{(a) Values of ICF$_{\rm m}$(Ne$^{++}$/O$^{++}$) as a function of $\omega$\ for our BCRS and 
BCMS models. The solid curves correspond, from top to bottom, to Equation~(\ref{eq:ne_1}) for 
$\upsilon'$ = 0.01, 0.015, and $\upsilon$. The dashed line represents Ne/O = Ne$^{++}$/O$^{++}$.  
(b) Values of $\Delta$/\icff(Ne$^{++}$/O$^{++}$) as a function of $\omega$ for our BCRS and BCMS models. 
The thick horizontal lines are $\varepsilon^{+}$ and $\varepsilon^{-}$. The dashed lines represent the uncertainties 
associated with the ICF by KB94. The thin horizontal line represents \icfm(Ne$^{+}$/O$^{++}$) = \icff(Ne$^{+}$/O$^{++}$). 
The color bar located on the side in both panels runs from low to high values of $\log\upsilon$ and the vertical line 
delimits the value of $\omega$ (0.1) above which our ICF is valid.\label{fig:ne_1}}
\end{figure}

Figure~\ref{fig:ne_1}a shows ICF$_{\rm m}$(Ne$^{++}$/O$^{++}$) values as a function of 
$\omega$. The ICF from KB94 is overplotted with a dashed line. We see from this figure that 
the ionic fractions of O$^{++}$ and Ne$^{++}$ are similar only for a small group of models with 
\efftemp\ $\sim$ 50000 K and $\omega$\ between 0.6 and 1. For the other models, the ICF 
from KB94 does not correct properly the contribution of unobserved ions, especially 
in those with central star temperatures below 50000 K.

As in the case of nitrogen, for neon an ICF based on $\upsilon$ and $\omega$ is needed. We suggest to use: 
\begin{equation}
\label{eq:ne_1}
\mbox{ICF}_{\rm f}(\mbox{Ne}^{++}/\mbox{O}^{++})  = \omega\ + \left(\frac{0.014}{\upsilon'} + 2\upsilon'^{2.7}\right)^3 \left(0.7 + 0.2\omega - 0.8\omega^2\right),
\end{equation}
where $\upsilon' = 0.01$ if no \ion{He}{ii} lines are observed, $\upsilon' = 0.015$ if \ion{He}{ii} lines 
are detected and $\upsilon$ is very small ($< 0.015$), and $\upsilon'$ = $\upsilon$ in all the other cases. 

The uncertainties associated with our ICF are:
\begin{equation}
\label{eq:ne_2}
\varepsilon^{-}({\rm Ne}^{++}/{\rm O}^{++}) = 0.17
\end{equation}
and 
\begin{equation}
\label{eq:ne_3}
\varepsilon^{+}({\rm Ne}^{++}/{\rm O}^{++}) = 0.12.
\end{equation}

Note that this expression is valid only when $\omega > 0.1$. Only some of our models with \efftemp\ = 25000 -- 35000 K 
reach values of $\omega$ below 0.1. The ICF for them can be very high (the contribution of Ne$^{+}$ to total neon 
abundances is more than 50\%) and we do not recommend to use objects with such low $\omega$ for mean abundance studies. 

As occurred for nitrogen, our ICF for neon implies a significant improvement in neon abundance determination, 
because Equation~(\ref{eq:ne_1}) is more adequate than the commonly adopted one Ne/O = Ne$^{++}$/O$^{++}$, and 
because the uncertainties associated with our ICF are much lower than the ones associated with the other 
correction scheme. 

One may wonder why our ICF for neon is so different from adopting Ne/O = Ne$^{++}$/O$^{++}$ as recommended by KB94 
(from their photoionization models), and as previously proposed by \citet{TorresPeimbert_77} (from similarities of ionization potentials of 
Ne$^+$ and O$^+$). As a matter of fact, the situation for neon is even more complicated than for nitrogen. First, the ionization 
potential of Ne$^+$ is 41.0 eV, sufficiently higher than the ionization potential of O$^+$ to decouple the frontiers between the 
once and twice ionized states of these ions when the star temperature is moderate. On the other hand, recombination of O$^{++}$ 
into O$^+$ is far more efficient than recombination of Ne$^{++}$ into Ne$^{+}$, and this has the reverse effect on the respective 
sizes of the O$^{+}$  and Ne$^{+}$ zones. When the stellar temperature is not high (25000K and 35000K in our grid of models), 
it is the first effect which dominates and the Ne$^{++}$ zone is smaller than the O$^{++}$ zone. When the stellar temperature is 
high and the ionization parameter is low, it is the second effect which dominates. In our model grid, the \efftemp = 50000 K case 
is intermediate, and leads to an ICF close to the canonical one. For the highest \efftemp, the canonical ICF applies only in cases 
where the ionization parameter is high (and therefore O$^{++}$/(O$^{+}$ + O$^{++}$) close to one. Why did KB 94 not find this? 
We suspect that this is because of their very restricted number of models, which did not allow them to find the subtle tendencies 
we see in our extended grid of models).

\begin{figure}
\subfigure{\includegraphics[width=\hsize,trim = 20 10 50 0,clip =yes]{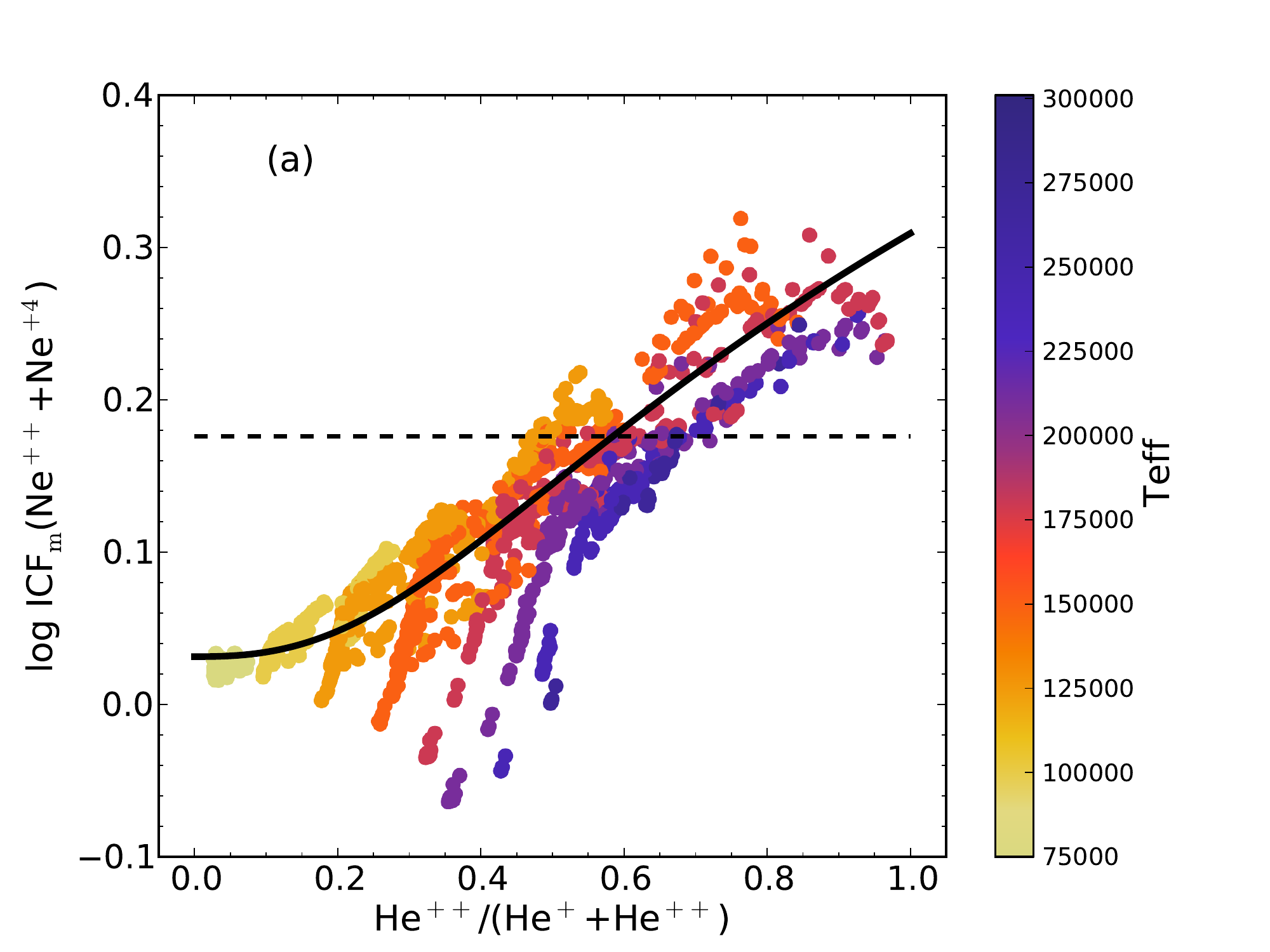}}
\subfigure{\includegraphics[width=\hsize,trim = 20 0 50 10,clip =yes]{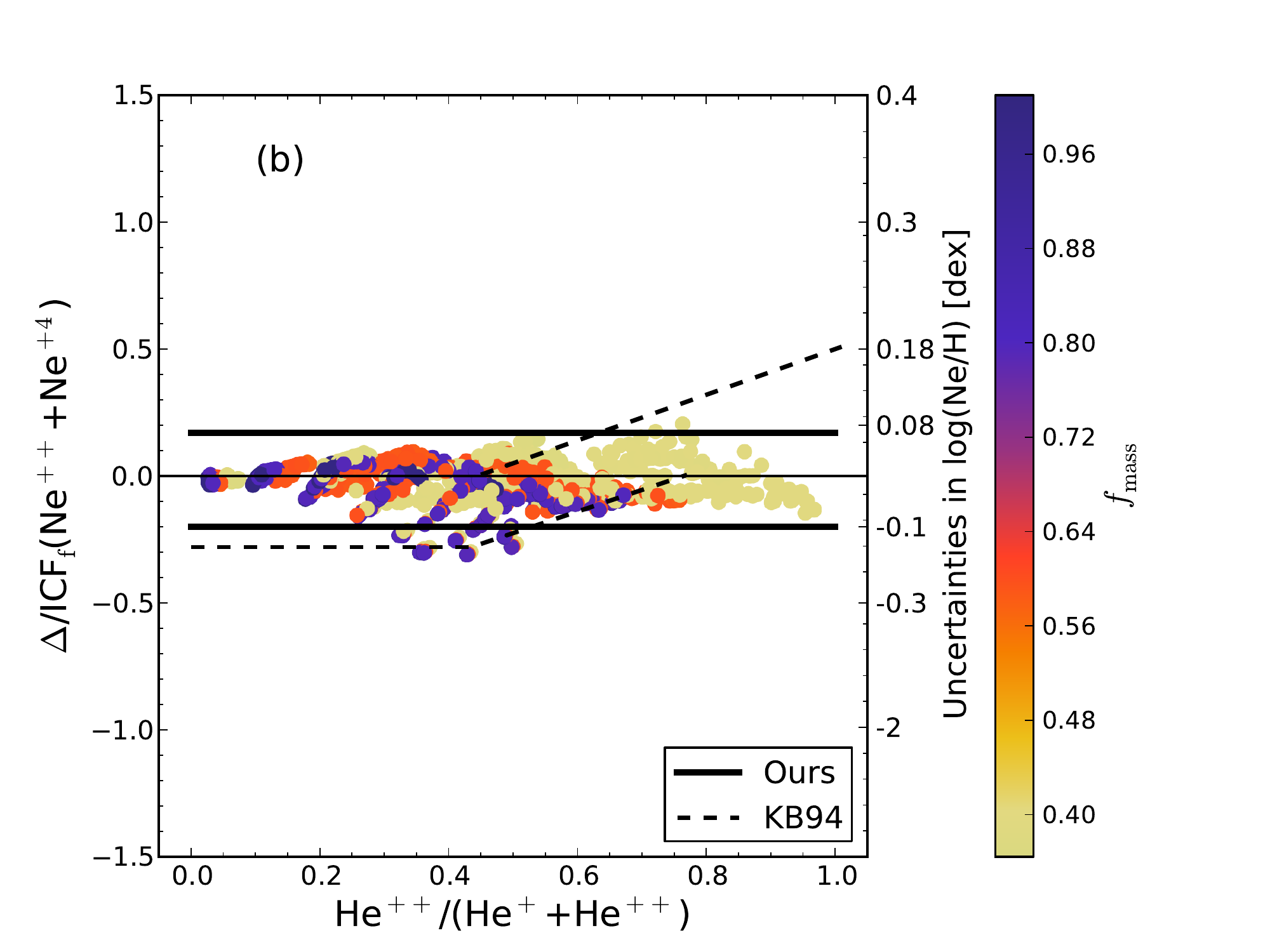}}
\caption{(a) Values of ICF$_{\rm m}$(Ne$^{++}$+Ne$^{+4}$) as a function of $\upsilon$\ for our BCRS and 
BCMS models with $\upsilon>0.02$ (see the text). 
The solid line is ICFs$_{\rm f}$(Ne$^{++}$+Ne$^{+4}$) while the dashed line corresponds to 
ICF = 1.5 (KB94). The color bar located on the side runs from low to high values of \efftemp. 
(b) Values of $\Delta$/\icff(Ne$^{++}$+Ne$^{+4}$) 
as a function of $\upsilon$ for our BCRS and BCMS models. The thick horizontal lines are $\varepsilon^{+}$ 
and $\varepsilon^{-}$. The dashed lines represent the uncertainties associated with the ICF by KB94. 
The thin horizontal line represents \icfm(Ne$^{++}$+Ne$^{+4}$) = \icff(Ne$^{++}$+Ne$^{+4}$). 
The color bar located on the side runs from low to high values of $f_{\rm mass}$.\label{fig:ne_2}}
\end{figure}

If [\ion{Ne}{iii}]  and [\ion{Ne}{v}] lines are observed, we propose to use:
\begin{equation}
\label{eq:ne_4}
\mbox{ICF}_{\rm f}(\mbox{Ne}^{++}+\mbox{Ne}^{+4})  = (1.31 + 12.68\upsilon^{2.57})^{0.27}. 
\end{equation}
This fit is shown with a solid line in Figure~\ref{fig:ne_2}a, together with the values of 
ICF$_{\rm m}$(Ne$^{++}$ + Ne$^{+4}$) as a function of $\upsilon$ for our BCRS and BCMS models. We 
exclude from the figure those models with $\upsilon < 0.02$ since [\ion{Ne}{v}] lines would be very weak 
(or absent) in their spectra. The figure also shows, with a dashed line, the constant ICF proposed by KB94. 
This ICF overestimates neon abundances when $\upsilon \lesssim 0.4$ and underestimates neon abundances 
when $\upsilon \gtrsim 0.6$. This is probably introducing a bias in the computed neon abundances.

Figure~\ref{fig:ne_2}b shows the uncertainties associated with our ICF (solid lines), that 
are: 
\begin{equation}
\varepsilon^- =  0.20
\end{equation}

\noindent and:

\begin{equation}
\varepsilon^+ = 0.17,
\end{equation}
and the ones related with the ICF from KB94 (dashed lines). 

Extreme caution should be taken when computing Ne$^{+4}$ ionic abundances since the adoption of an 
adequate \temp\ is crucial to obtain reliable values; if \temp[\ion{O}{iii}] is adopted, Ne$^{+4}$ abundances 
and thus, total neon abundances will be probably overestimated. 

We do not discuss the case when the auroral [\ion{Ne}{iv}] are observed since they very weak and sensitive to 
the electron temperature. In these cases, we strongly recommend to use the ICFs from equation~(\ref{eq:ne_1}) or (\ref{eq:ne_4}) 
to calculate neon abundances. 

\subsection{Sulfur}

The only sulfur ions that emit lines in the optical range are S$^{+}$ and S$^{++}$. The most frequently adopted 
ICF for sulfur is the one proposed by \citet{Stasinska_78}. We found from our models that this ICF can overestimate 
S/H by up to 0.08 dex in very low ionization PNe (with $\omega\lesssim0.2$) whereas in PNe with $\omega>0.2$, this ICF 
either overestimates S/H by up to 0.1 dex or underestimates it by up to 0.4 dex (or even more). Similar results were 
already found in \citet{Rodriguez_12}.

\begin{figure}
\subfigure{\includegraphics[width=\hsize,trim = 20 10 50 0,clip =yes]{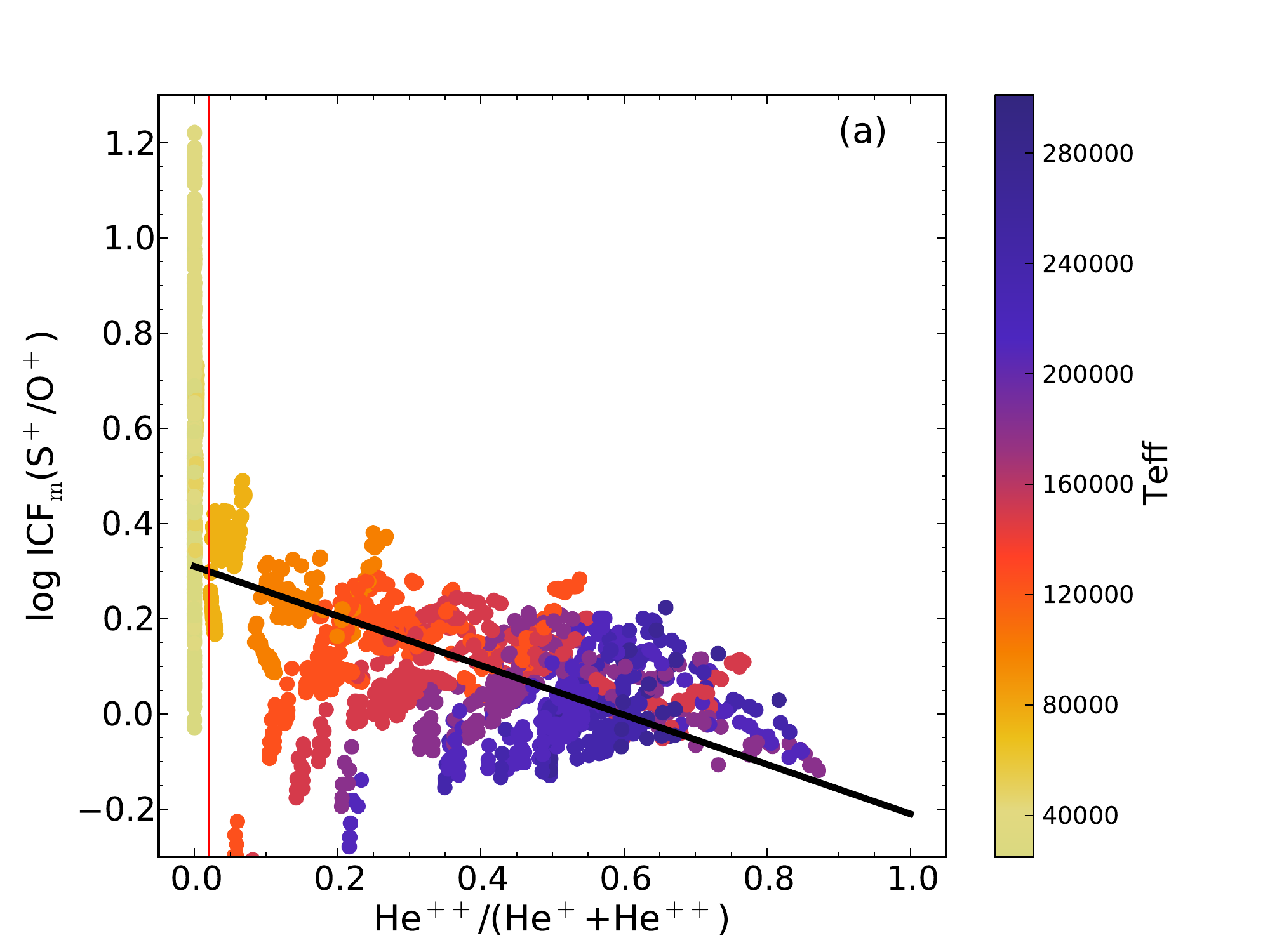}}
\subfigure{\includegraphics[width=\hsize,trim = 20 0 50 10,clip =yes]{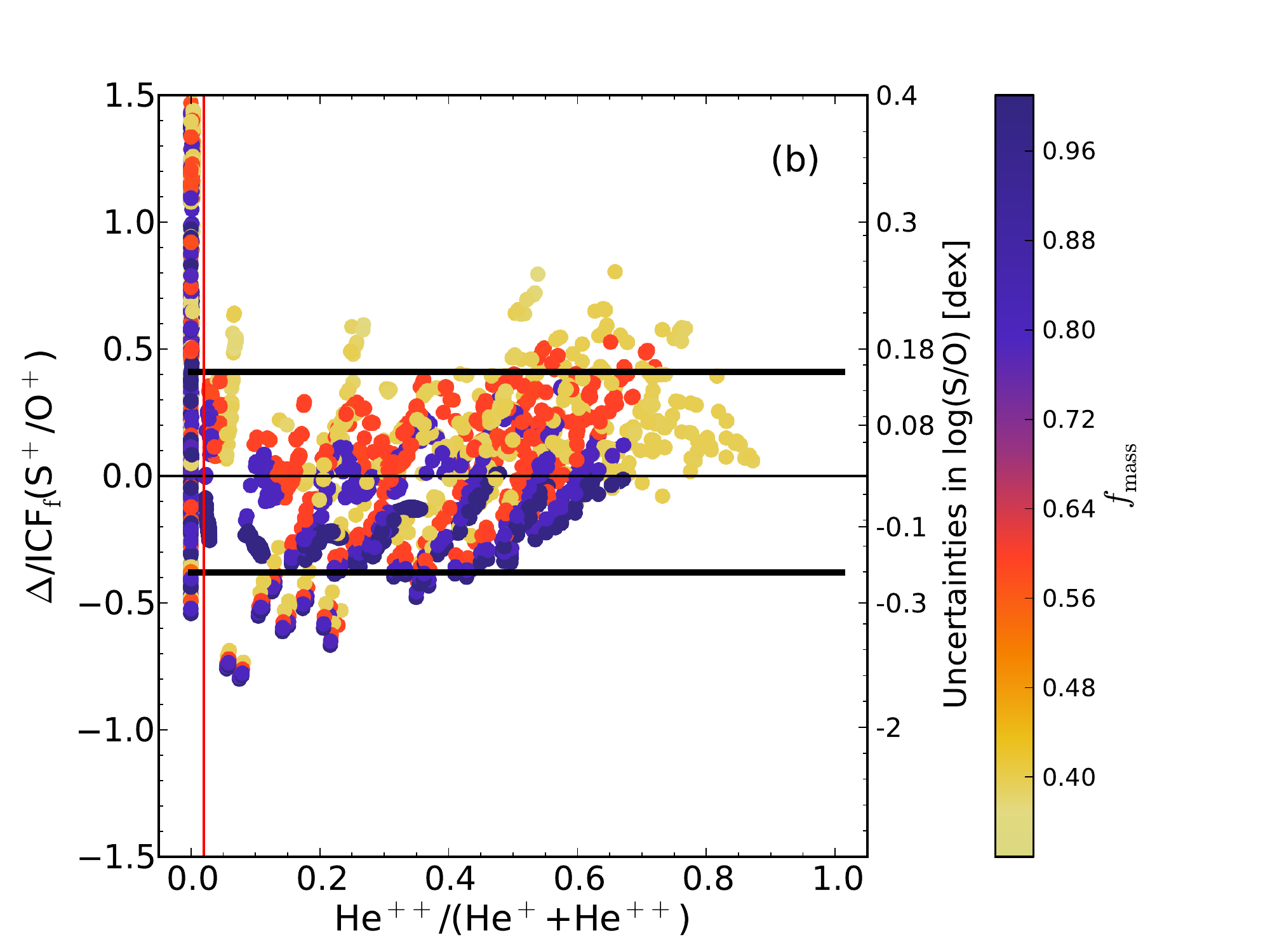}}
\caption{(a) Values of ICF$_{\rm m}$(S$^{+}$/O$^{+}$) as a function of $\upsilon$\ for our BCRS and 
BCMS models. The solid line is ICFs$_{\rm f}$(S$^{+}$/O$^{+}$). The color bar located on the side runs 
from low to high values of \efftemp. (b) Values of $\Delta$/\icff(S$^{+}$/O$^{+}$) as a function of $\upsilon$ 
for our BCRS and BCMS models. The thick horizontal lines are $\varepsilon^{+}$ and $\varepsilon^{-}$. The 
thin horizontal line represents \icfm(S$^{+}$/O$^{+}$) = \icff(S$^{+}$/O$^{+}$). The color bar located on the 
side runs from low to high values of $f_{\rm mass}$. The vertical line in both panels delimits the value 
of $\upsilon$ (0.02) above which our ICF is valid.\label{fig:s_1}}
\end{figure}

Figure~\ref{fig:s_1}a shows the values of ICF$_{\rm m}$(S$^{+}$/O$^{+}$) as a function of $\upsilon$. 
A fit to these values gives:
\begin{equation}
\label{eq:s_1}
\log \mbox{ICF}_{\rm f}({\rm S}^{+}/{\rm O}^{+})  = 0.31 - 0.52\upsilon,
\end{equation}
represented with a solid line. 
We do not recommend to use the above equation when $\upsilon$ $<$ 0.02 since the ICF for these objects 
is very uncertain. A tailored photoionization model should be constructed in these cases. 

Figure~\ref{fig:s_1}b shows the values of $\Delta$/\icff(S$^{+}$/O$^{+}$) as a function of 
$\upsilon$. The solid lines correspond to:
\begin{equation}
\label{eq:s_2}
\varepsilon^-({\rm S}^{+}/{\rm O}^{+})  =  0.38
\end{equation}
and
\begin{equation}
\label{eq:s_3}
\varepsilon^+({\rm S}^{+}/{\rm O}^{+})  = 0.41,
\end{equation}
which give an idea of the uncertainties associated with this ICF. Clearly, when the only observed ion of sulfur is 
S$^{+}$, the uncertainty in the S abundance is rather large.

\begin{figure}
\subfigure{\includegraphics[width=\hsize,trim = 20 10 50 0,clip =yes]{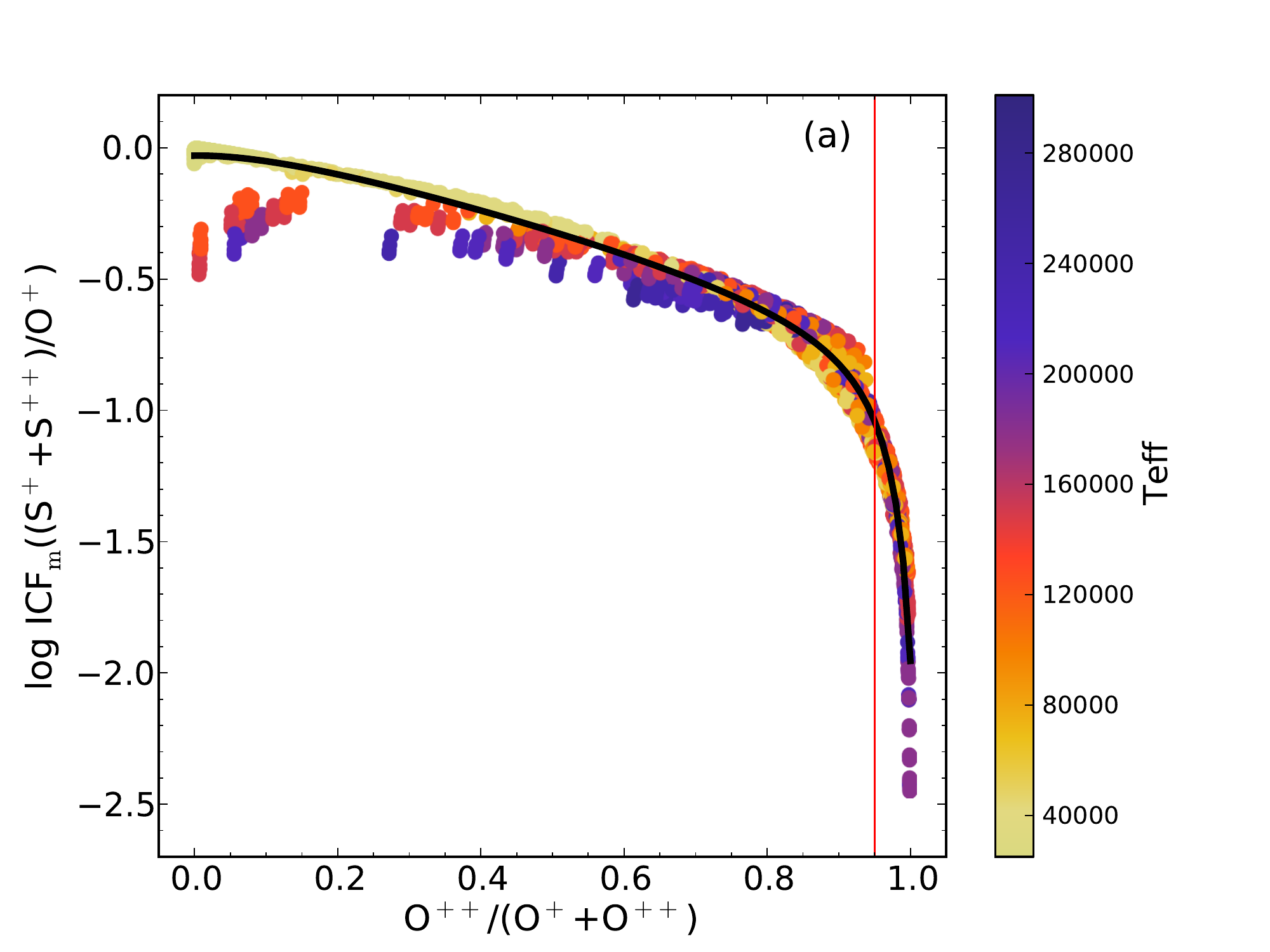}}
\subfigure{\includegraphics[width=\hsize,trim = 20 0 50 10,clip =yes]{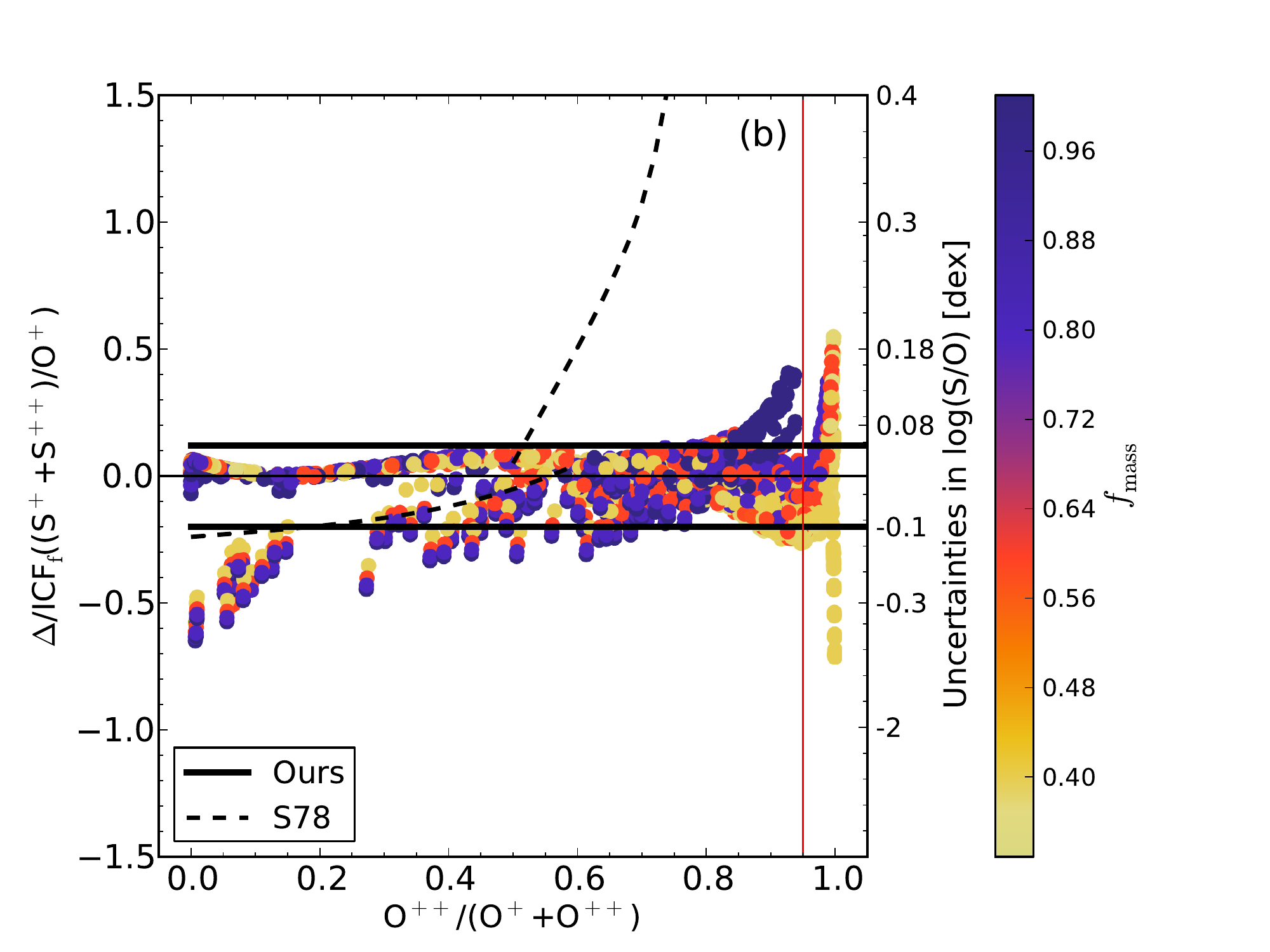}}
\caption{(a) Values of ICF$_{\rm m}$((S$^{+}$ + S$^{++}$)/O$^{+}$) as a function of $\omega$\ for our 
BCRS and BCMS models. The solid line is ICFs$_{\rm f}$((S$^{+}$ + S$^{++}$)/O$^{+}$). The color bar located 
on the side runs from low to high values of \efftemp. 
(b) Values of $\Delta$/\icff((S$^{+}$ + S$^{++}$)/O$^{+}$) as a function of $\omega$ for our BCRS and BCMS 
models. The thick horizontal lines are $\varepsilon^{+}$ and $\varepsilon^{-}$. The dashed lines represent 
the uncertainties associated with the ICF by \citet{Stasinska_78}. The thin horizontal line represents 
\icfm((S$^{+}$ + S$^{++}$)/O$^{+}$) = \icff((S$^{+}$ + S$^{++}$)/O$^{+}$). The color bar located on the side runs 
from low to high values of $f_{\rm mass}$. The vertical line in both panels delimits the value of $\omega$ (0.95)
below which our ICF is valid.\label{fig:s_2}}
\end{figure}

When [\ion{S}{iii}] lines are also observed, we suggest to use: 
\begin{equation}
\label{eq:s_4}
\log \mbox{ICF}_{\rm f}(({\rm S}^{+}+{\rm S}^{++})/{\rm O}^{+})  = \frac{-0.02 - 0.03\omega - 2.31\omega^2  + 2.19\omega^3}{0.69 + 2.09\omega - 2.69\omega^2},
\end{equation}
represented with a solid line in Figure~\ref{fig:s_2}a. We see from this figure that the contribution of S$^{\geq+3}$ ions 
is negligible in PNe with \inthe\ $< 0.02$, and thus, an ICF$_{\rm f}$((S$^{+}$ + S$^{++}$)/O$^{+}$) = 1 
can be adopted. The maximum uncertainties associated with our proposed ICF,  $\varepsilon^-(({\rm S}^{+}+{\rm S}^{++})/{\rm O}^{+})$ 
and $\varepsilon^+(({\rm S}^{+}+{\rm S}^{++})/{\rm O}^{+})$, are represented with solid lines in Figure~\ref{fig:s_2}b. They are given by:
\begin{equation}
\label{eq:s_2}
\varepsilon^-(({\rm S}^{+}+{\rm S}^{++})/{\rm O}^{+})  =  0.20
\end{equation}
and
\begin{equation}
\label{eq:s_3}
\varepsilon^+(({\rm S}^{+}+{\rm S}^{++})/{\rm O}^{+})  = 0.12.
\end{equation}
When lines from both S$^{+}$ and S$^{++}$ are observed, our models lead to a very small dispersion in 
\icfm((S$^{+}$ + S$^{++}$)/O$^{+}$) for a given value of $\omega$, so that the sulfur abundances should be quite accurate. 
This figure also shows the uncertainties related to the ICF proposed by \citet{Stasinska_78}, which, as we mentioned 
above introduces a bias in the computed sulfur abundances. 

\subsection{Chlorine}

In PNe, chlorine can be present in several ionization states. In the optical range, lines of [\ion{Cl}{ii}], 
[\ion{Cl}{iii}], and [\ion{Cl}{iv}] can be observed, depending on the degree of ionization of the object.
This element is not discussed in KB94 but other authors, such as \citet{Peimbert_77, Kwitter_01, Liu_00, Wang_07}, 
proposed correction schemes to estimate the contribution of unobserved ions. 

\begin{figure}
\subfigure{\includegraphics[width=\hsize,trim = 20 10 50 0,clip =yes]{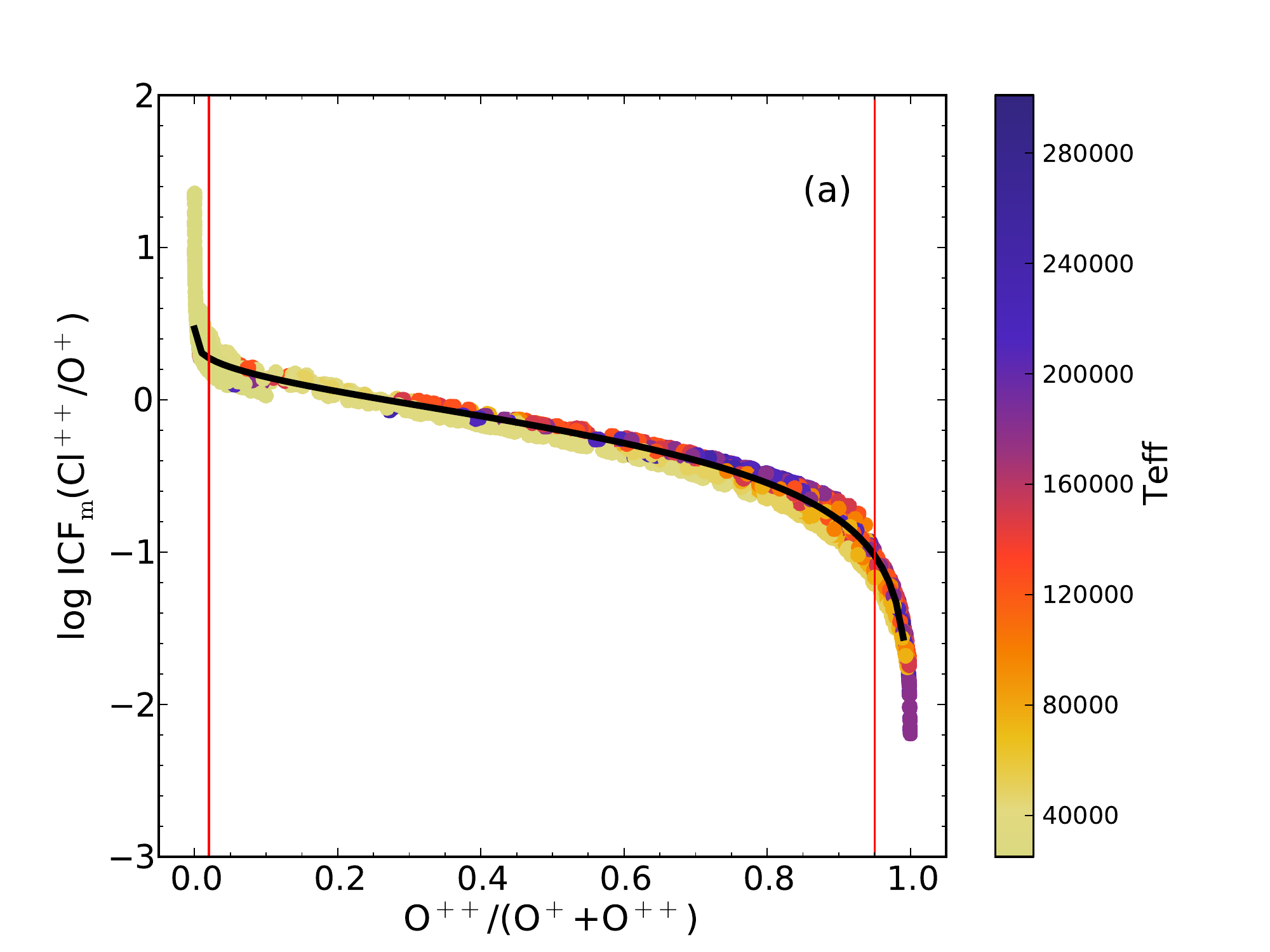}}
\subfigure{\includegraphics[width=\hsize,trim = 20 0 50 10,clip =yes]{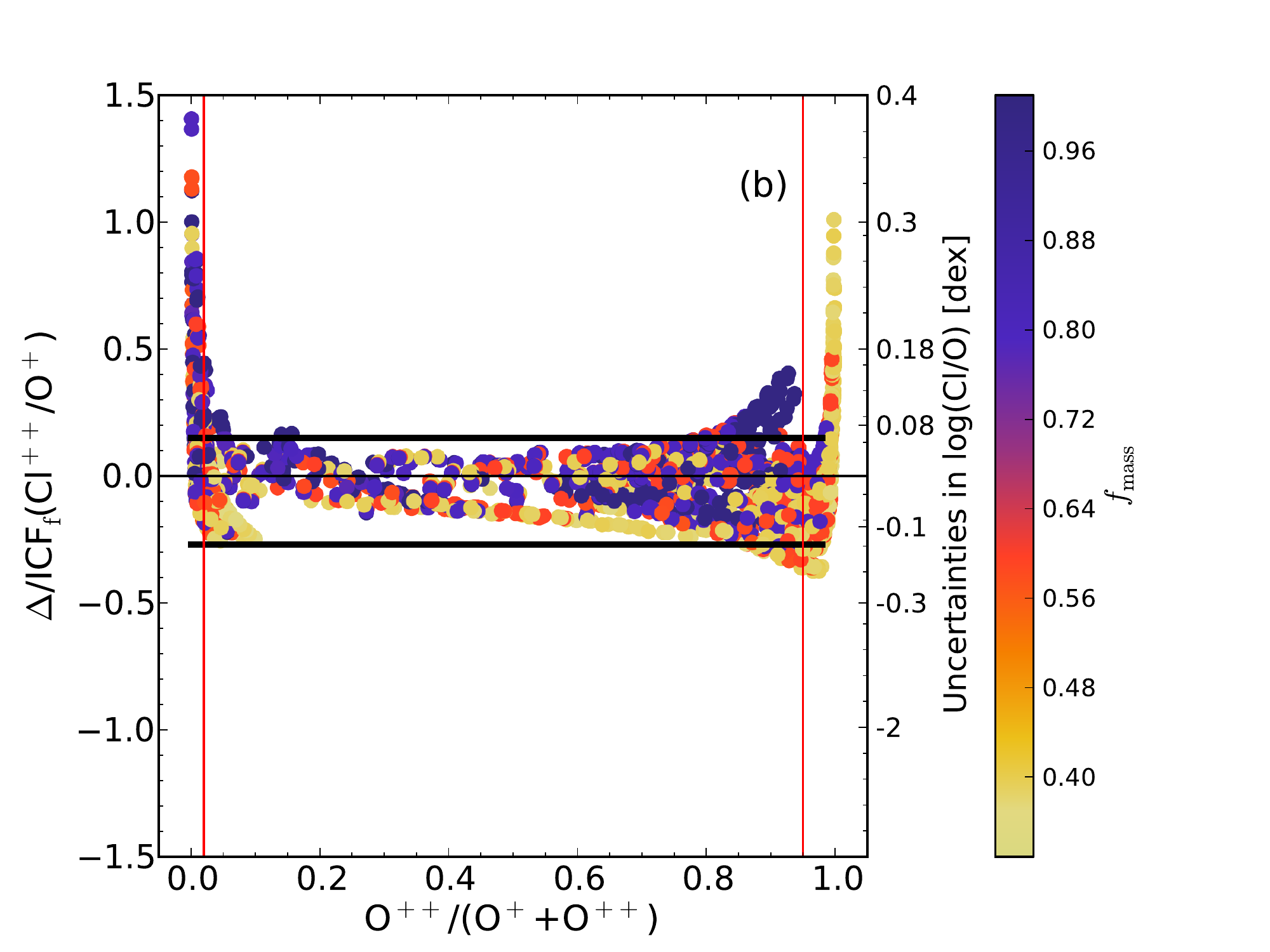}}
\caption{(a) Values of ICF$_{\rm m}$(Cl$^{++}$/O$^{+}$) as a function of $\omega$\ for our BCRS and BCMS 
models. The solid line is ICF$_{\rm f}$(Cl$^{++}$/O$^{+}$). The color bar located on the side runs from low to high 
values of \efftemp. (b) Values of $\Delta$/\icff(Cl$^{++}$/O$^{+}$) as a function of $\omega$ for our 
BCRS and BCMS models. The thick solid lines are $\varepsilon^+$ and $\varepsilon^-$. The thin horizontal line represents 
\icfm(Cl$^{++}$/O$^{+}$) = \icff(Cl$^{++}$/O$^{+}$). The color bar located on the side runs from low to high values 
of $f_{\rm mass}$. The vertical lines in both panels shows the range of validity of our ICF, $0.02 < \omega < 0.95$. 
\label{fig:cl_1}}
\end{figure}

Instead of proposing several ICFs to calculate chlorine abundances depending on the observed ions, 
we suggest to use only two ICFs. The first ICF can be used when only [\ion{Cl}{iii}] lines are observed:
\begin{equation}
\label{eq:cl_1}
\mbox{ICF}_{\rm f}({\rm Cl}^{++}/{\rm O}^{+})  = (4.1620 - 4.1622\omega^{0.21})^{0.75}.
\end{equation}
This ICF is showed with a solid line in Figure~\ref{fig:cl_1}a and is valid when $0.02 < \omega < 0.95$. 
The uncertainties related to the ICF are:
\begin{equation}
\varepsilon^- = 0.27
\end{equation}
and
\begin{equation}
\varepsilon^+= 0.15. 
\end{equation}
In PNe with $\omega \lesssim 0.02$, where the Equation~\ref{eq:cl_1} is not valid, the Cl/O values can be computed 
simply as (Cl$^{+}$ + Cl$^{++}$)/O$^{+}$ if [\ion{Cl}{ii}] and [\ion{Cl}{iii}] lines are observed. The related uncertainties are 
$\varepsilon^+=0$ and $\varepsilon^-=0.13$. In PNe with $\omega > 0.95$, the Equation~\ref{eq:cl_1} is neither valid, 
and we recommend to construct a tailored photoionization model to derive chlorine abundances (or use the 
Equation~\ref{eq:cl_2}, described below, if the three ions are observed).

The second ICF is valid when [\ion{Cl}{ii}], [\ion{Cl}{iii}], and [\ion{Cl}{iv}] lines are observed:
\begin{equation}
\label{eq:cl_2}
\mbox{ICF}_{\rm f}({\rm Cl}^{+}+{\rm Cl}^{++}+{\rm Cl}^{+3})  = 0.98 + (0.56 - 0.57\upsilon)^{7.64}.
\end{equation}
The uncertainties associated with this fit are:
\begin{equation}
\varepsilon^- = 0.30
\end{equation}
and
\begin{equation}
\varepsilon^+= 0.33. 
\end{equation}

\begin{figure}
\subfigure{\includegraphics[width=\hsize,trim = 20 10 50 0,clip =yes]{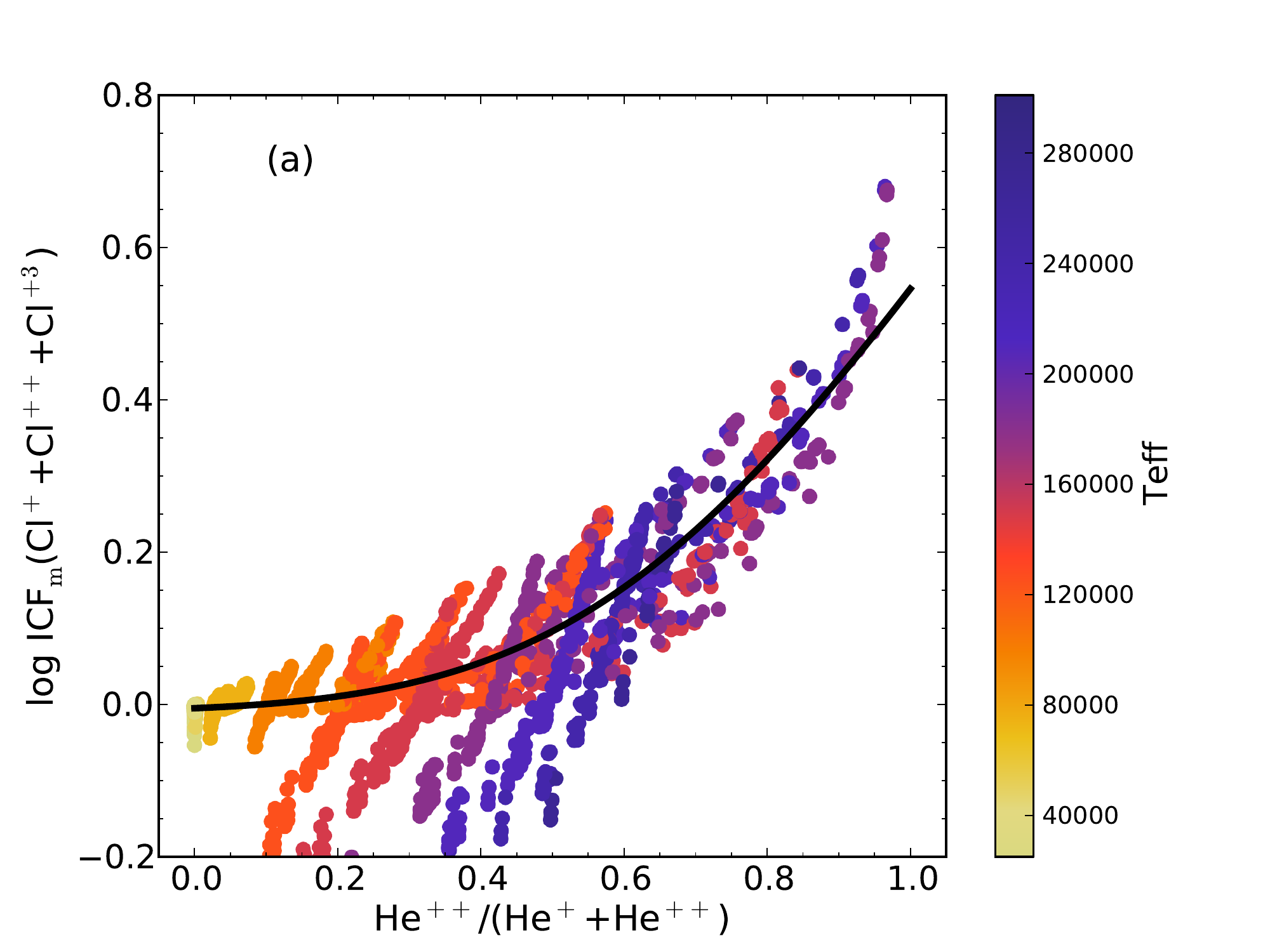}}
\subfigure{\includegraphics[width=\hsize,trim = 20 0 50 10,clip =yes]{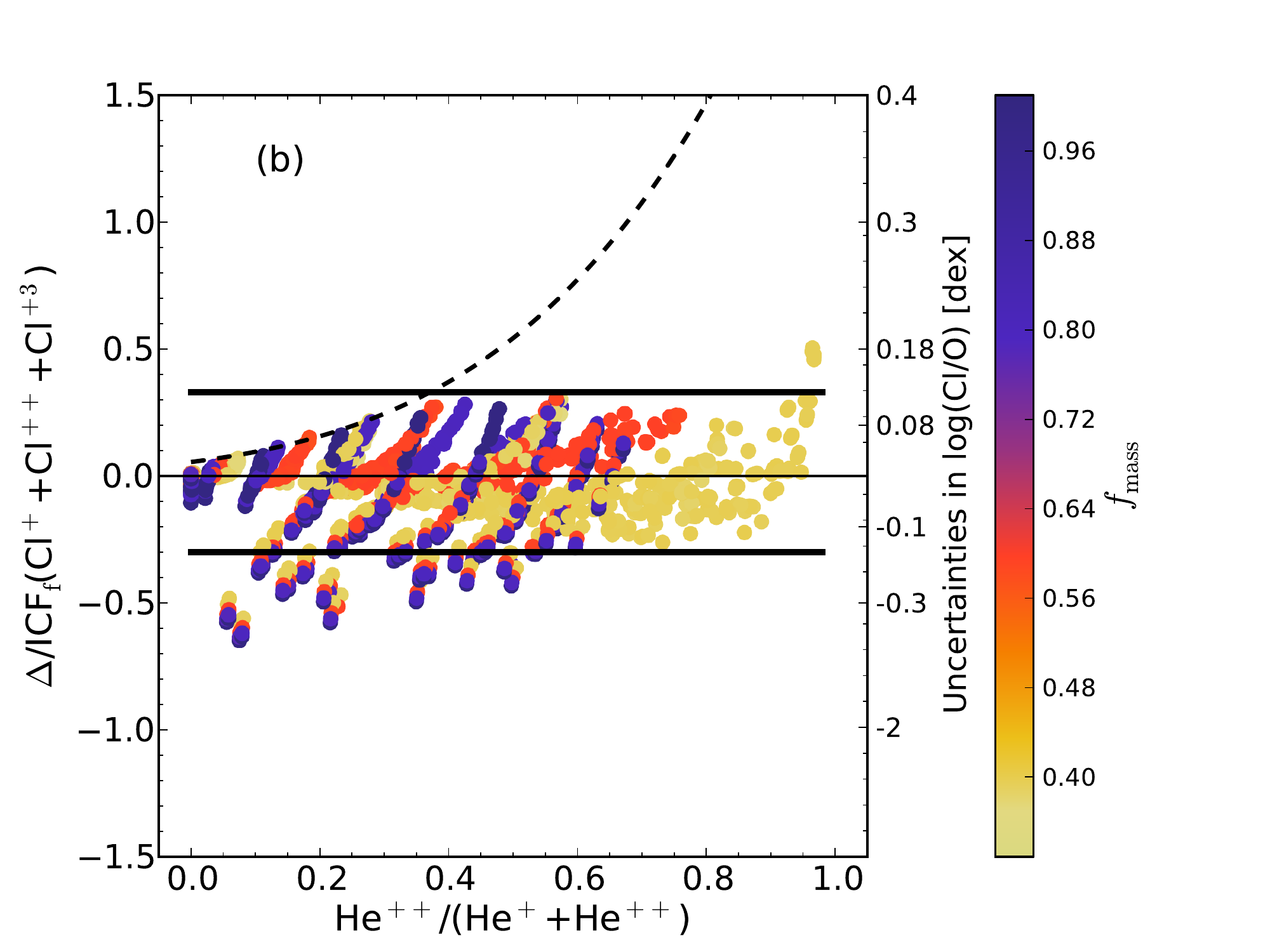}}
\caption{(a) Values of ICF$_{\rm m}$(Cl$^{+}$ + Cl$^{++}$ + Cl$^{+3}$) as a function of $\upsilon$\ for our BCRS and BCMS 
models. The solid line is ICF$_{\rm f}$(Cl$^{+}$ + Cl$^{++}$ + Cl$^{+3}$). The color bar located on the side runs from low to high 
values of \efftemp. (b) Values of $\Delta$/\icff(Cl$^{+}$ + Cl$^{++}$ + Cl$^{+3}$) as a function of $\omega$ for our 
BCRS and BCMS models. The thick solid lines are $\varepsilon^+$ and $\varepsilon^-$. The dashed lines represent the 
uncertainties associated with ICF = 1. The thin horizontal line represents 
\icfm(Cl$^{+}$ + Cl$^{++}$ + Cl$^{+3}$) = \icff(Cl$^{+}$ + Cl$^{++}$ + Cl$^{+3}$). The color bar located on the side runs from 
low to high values of $f_{\rm mass}$. \label{fig:cl_2}}
\end{figure}

Figure~\ref{fig:cl_2}b we can see that if we calculate Cl/H simply by adding up the ionic abundances 
Cl$^{+}$, Cl$^{++}$, and Cl$^{+3}$, the total chlorine abundances are probably underestimated by more than 0.2 dex in PNe 
with $\upsilon \gtrsim 0.5$.

\subsection{Argon}

Argon lines of Ar$^{++}$, Ar$^{+3}$, and Ar$^{+4}$ can be observed in the optical range. In low ionization 
PNe, Ar$^{+}$ may be an important contributor to the total abundance of argon whereas in very high ionization 
PNe the contribution of Ar$^{\geq+5}$ ions may be significant. From our models we find 
that the ionic fraction of Ar$^{+}$ can be more than 60\% when $\omega \lesssim 0.1$, while the ionic fraction of 
Ar$^{+5}$ can be more than 30$\%$ when $\omega \gtrsim 0.9$. Therefore, we need to account for these ions 
in order to calculate total abundances of argon. 

\begin{figure}
\subfigure{\includegraphics[width=\hsize,trim = 20 10 50 0,clip =yes]{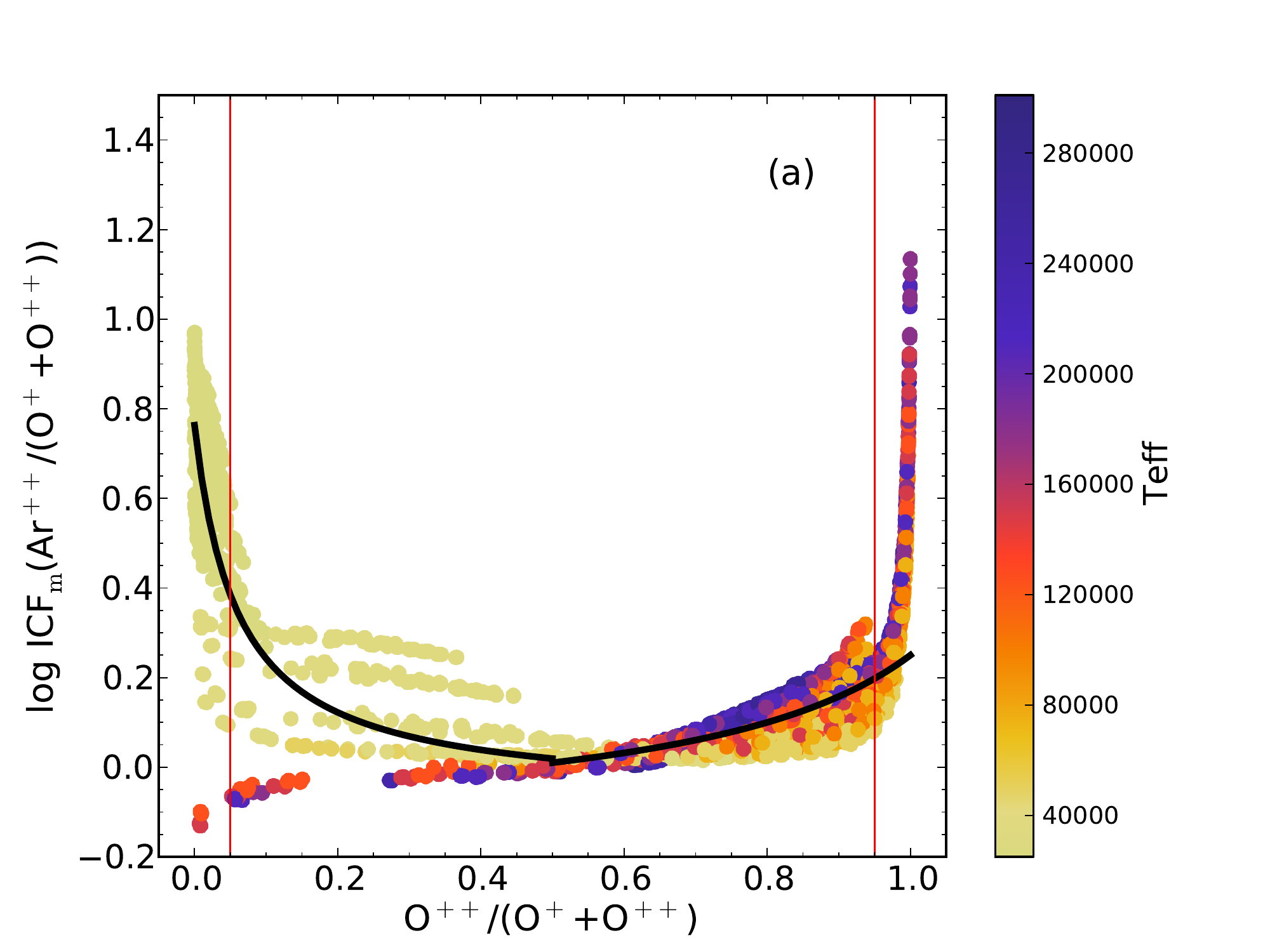}}
\subfigure{\includegraphics[width=\hsize,trim = 20 0 50 10,clip =yes]{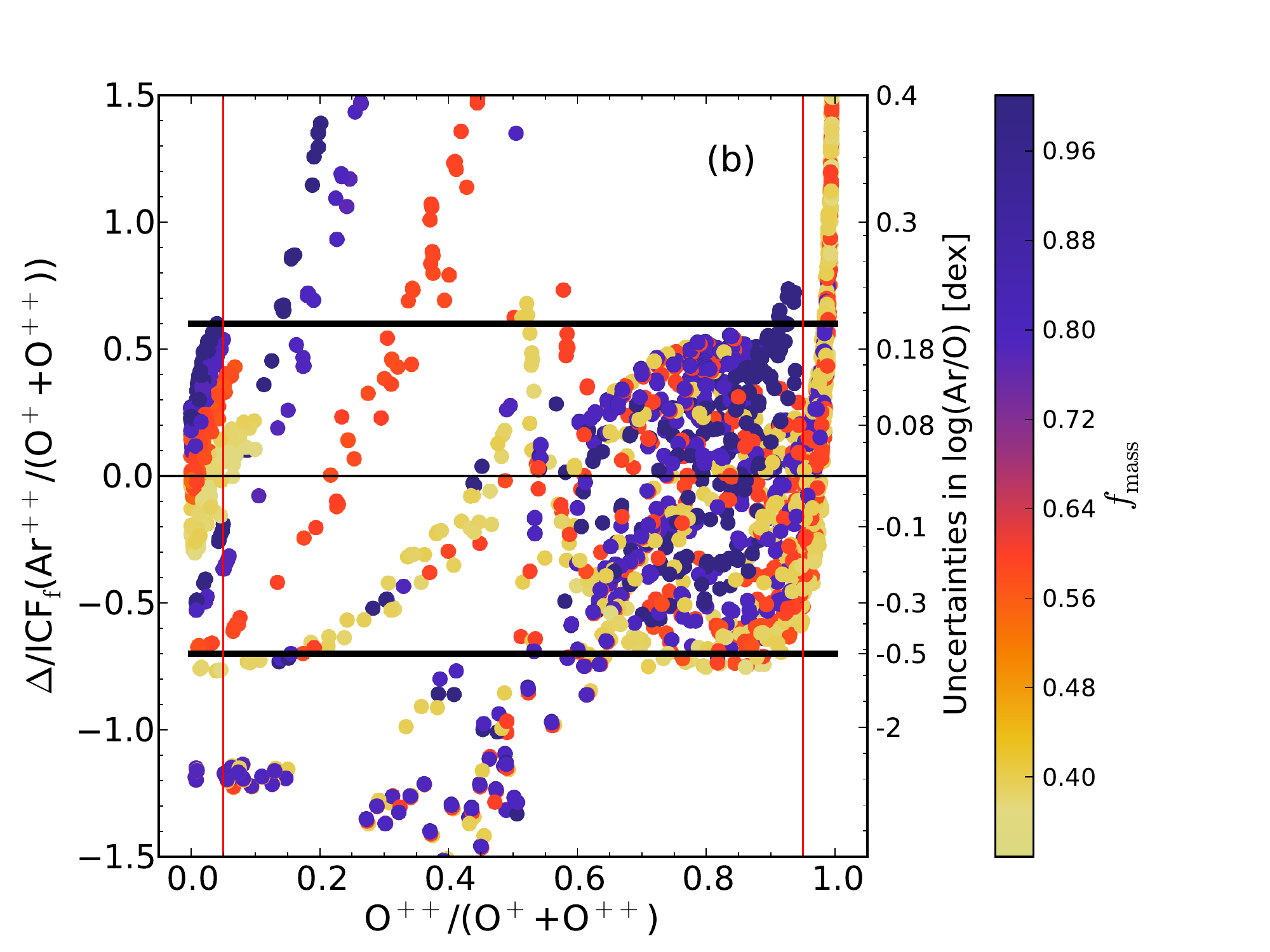}}
\caption{(a) Values of ICF$_{\rm m}$(Ar$^{++}$/(O$^{+}$ + O$^{++}$)) as a function of 
$\omega$\ for our BCRS and BCMS models. The solid lines corresponds to 
ICF$_{\rm f}$(Ar$^{++}$/(O$^{+}$ + O$^{++}$)). The color bar located on the side runs from low to high values of \efftemp. 
(b) Values of $\Delta$/\icff(Ar$^{++}$/(O$^{+}$ + O$^{++}$)) as a function of $\omega$ for our 
BCRS and BCMS models.The solid lines are $\varepsilon^+$ and $\varepsilon^-$. The thin horizontal line represents 
\icfm(Ar$^{++}$/(O$^{+}$ + O$^{++}$)) = \icff(Ar$^{++}$/(O$^{+}$ + O$^{++}$)). The color bar located on the side in both 
panels runs from low to high values of  $f_{\rm mass}$. The vertical line in both 
panels shows the range of validity of our ICF, $0.05 < \omega < 0.95$. \label{fig:ar_1}}
\end{figure}

The best correlation we found between argon ions and $\upsilon$ or $\omega$ is showed in Figure~\ref{fig:ar_1}a. 
We performed two fits to these values. The first one is valid when $\omega\leq0.5$:
\begin{equation}
\log\mbox{ICF}_{\rm f}({\rm Ar}^{++}/({\rm O}^{+}+{\rm O}^{++})) =  \frac{0.05}{0.06+\omega}-0.07.
\end{equation}
And the second one is valid when $\omega>0.5$:
\begin{equation}
\log\mbox{ICF}_{\rm f}({\rm Ar}^{++}/({\rm O}^{+}+{\rm O}^{++})) =  \frac{0.03\omega}{0.4-0.3\omega}-0.05.
\end{equation}

Figure~\ref{fig:ar_1}b shows the uncertainties associated with our ICFs, which are much higher 
than the ones obtained for the other elements. They can be estimated in general as:
\begin{equation}
\varepsilon^- = 0.70
\end{equation}
and
\begin{equation}
\varepsilon^+ = 0.60.
\end{equation}

This correction scheme has the advantage that only [\ion{Ar}{iii}] lines need to be observed. However, 
it must be noted that it would provide extremely uncertain argon abundances if it were uses for extreme values of 
$\omega$ (say $\omega < 0.05$ or $\omega > 0.95$).

\subsection{Carbon}

Collisionally excited lines from carbon are emitted only in the ultraviolet range but several recombination 
lines of \ion{C}{ii}, \ion{C}{iii}, and \ion{C}{iv} can be observed in optical spectra. The \ion{C}{ii} line at 
4267 $\AA$ is the brightest and often the only one detected. \ion{C}{iii} lines may be observed at 4187 $\AA$ and 4650 $\AA$ 
and \ion{C}{iv} at 4658 $\AA$. The \ion{C}{iii} $\lambda$4650 line can be blended with nearby
\ion{O}{ii} lines, and the \ion{C}{iv} $\lambda$4658 line can be contaminated with the [\ion{Fe}{iii}] line at 
the same wavelength. Therefore, carbon abundances are usually computed using only C$^{++}$ abundances 
and one ICF to account for the other ions. 

\begin{figure}
\subfigure{\includegraphics[width=\hsize,trim = 20 10 50 0,clip =yes]{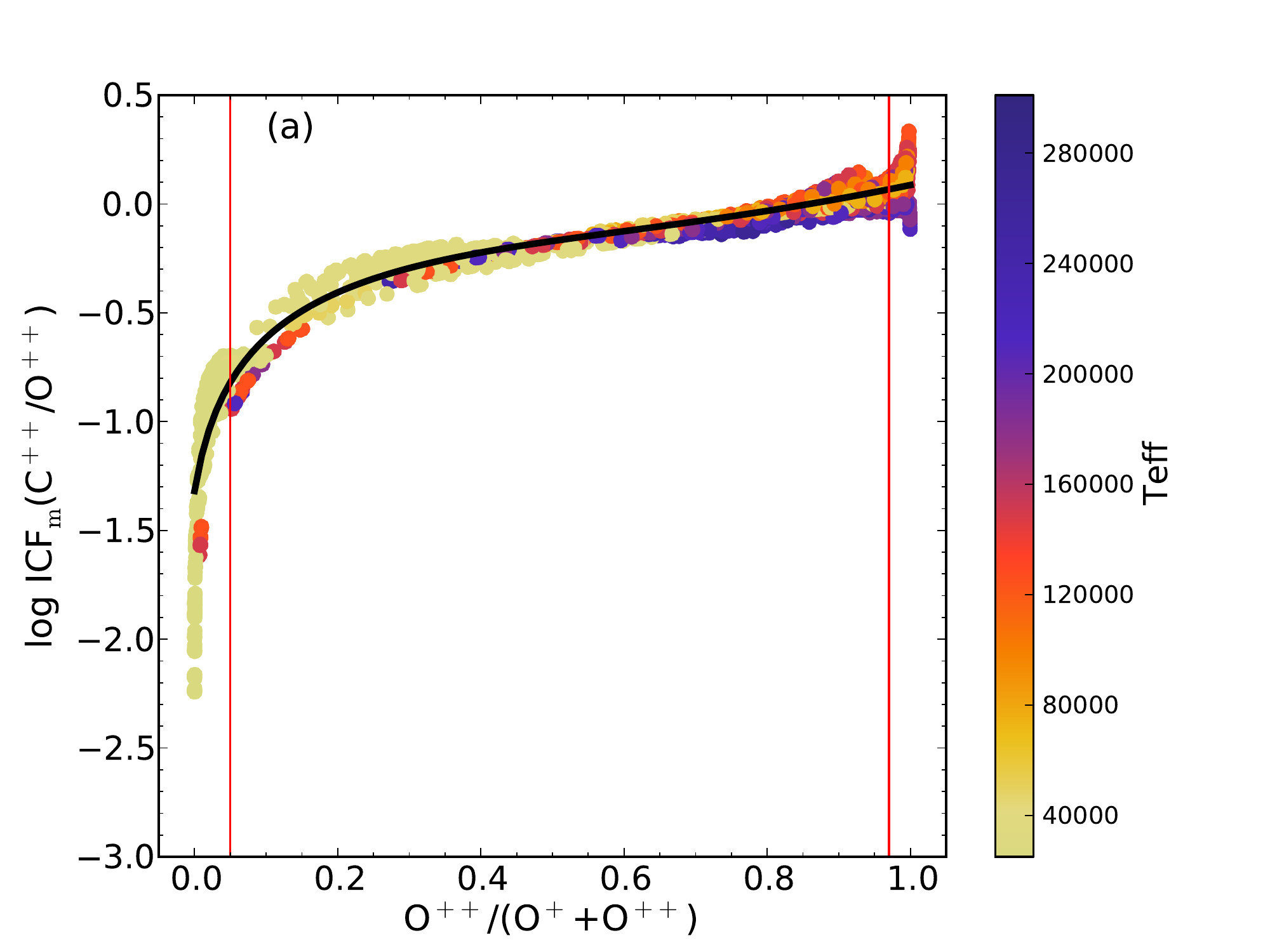}}
\subfigure{\includegraphics[width=\hsize,trim = 20 0 50 10,clip =yes]{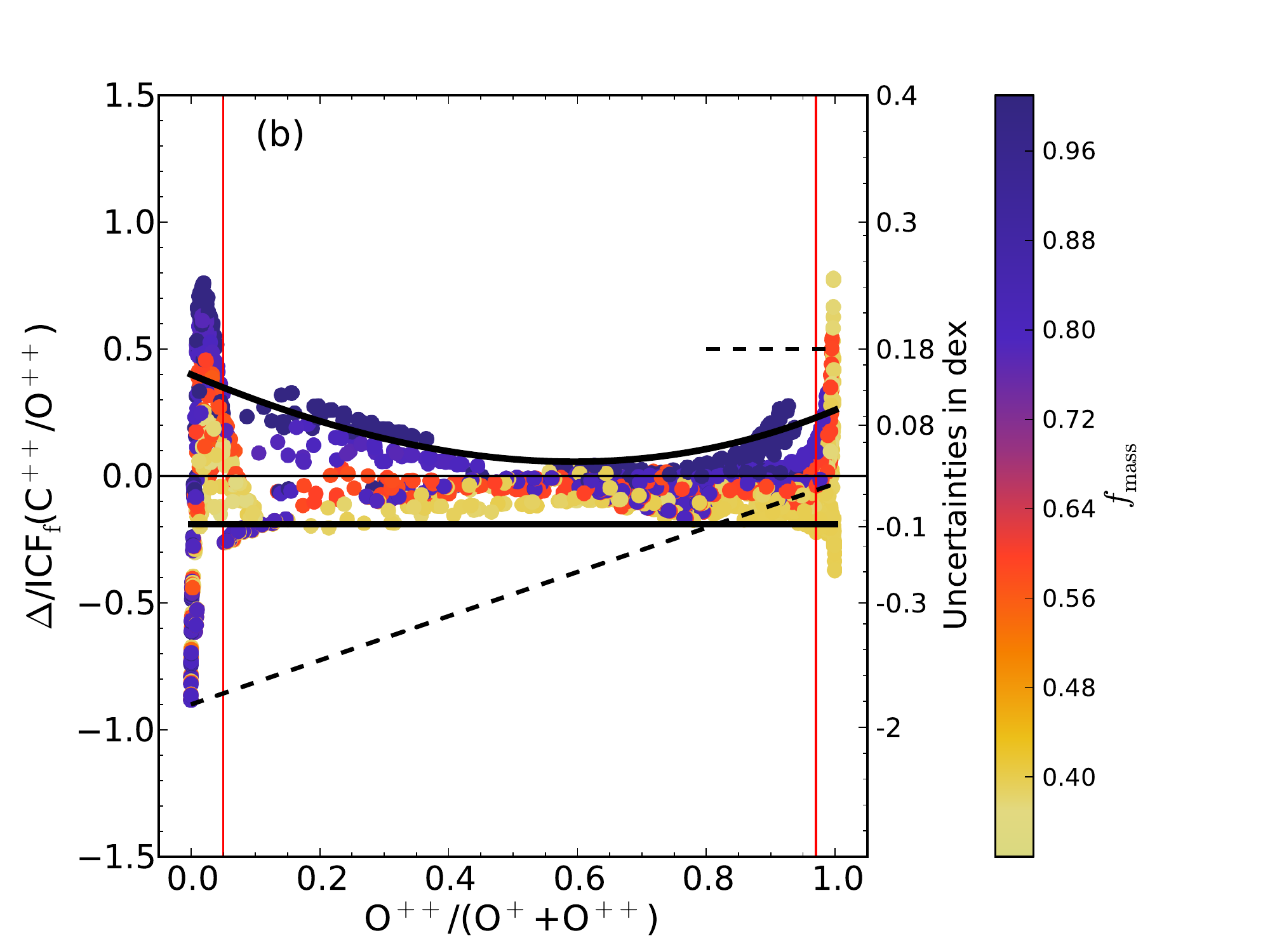}}
\caption{(a) Values of ICF$_{\rm m}$(C$^{++}$/O$^{++}$) as a function of 
$\omega$\ for our BCRS and BCMS models. The solid line corresponds to 
ICF$_{\rm f}$(C$^{++}$/O$^{++}$). The color bar located on the side runs from low to high values of \efftemp. 
(b) Values of $\Delta$/\icff(C$^{++}$/O$^{++}$) as a function of $\omega$ for our 
BCRS and BCMS models.The solid lines are $\varepsilon^+$ and $\varepsilon^-$. 
The dashed lines represent the uncertainties associated with C/O = C$^{++}$/O$^{++}$. 
The thin horizontal line represents \icfm(C$^{++}$/O$^{++}$) = \icff(C$^{++}$/O$^{++}$). 
The color bar located on the side in both panels runs from low to high values of  $f_{\rm mass}$. 
The vertical line in both panels shows the range of validity of our ICF, $0.05 < \omega < 0.97$.\label{fig:c_1}}
\end{figure}

In Figure~\ref{fig:c_1}a we show the values of ICF$_{\rm m}$(C$^{++}$/O$^{++}$) as a function of $\omega$ for our 
models. It is generally assumed that C/O = C$^{++}$/O$^{++}$ but, as seen in the figure, $x$(O$^{++}$)/$x$(C$^{++}$) $\sim$ 1 only when $\omega \sim 0.9$.
We suggest to use:
\begin{equation}\label{eq:c_1}
\mbox{ICF}_{\rm f}({\rm C}^{++}/{\rm O}^{++})  = 0.05 + 2.21\omega - 2.77\omega^2 + 1.74\omega^3.
\end{equation}
The error bars associated with the ICF of equation~(\ref{eq:c_1}) are estimated as:
\begin{equation}
\varepsilon^- = 0.19
\end{equation}
and
\begin{equation}
\varepsilon^+ = 0.4 -1.06\omega + 0.65\omega^2 + 0.27\omega^3. 
\end{equation}
Figure~\ref{fig:c_1}b shows that the C/O values calculated as C$^{++}$/O$^{++}$ (dashed lines) can be 
overestimated by up to one dex or underestimated by up 0.2 dex depending on the degree of ionization of 
the PN. 

As for the case of argon, we do not recommend to compute the carbon abundance for objects with $\omega < 0.05$ or objects with $\omega > 0.95$.
Note that the commonly used recipe that  C/O = C$^{++}$/O$^{++}$  significantly overestimates the carbon abundance in the vast majority of cases. 

\renewcommand{\arraystretch}{1.2}
\begin{table*}
\centering
\begin{minipage}{200mm}
\caption{Our ICFs and their uncertainties\label{tab:3}.}
\begin{tabular}{clccc}
\hline \hline \\[-2ex]
\multicolumn{1}{c}{Abundance} & \multicolumn{1}{c}{\icff$^1$} & \multicolumn{2}{c}{Uncertainties} & \multicolumn{1}{c}{Ranges of}\\
\multicolumn{1}{c}{ratios} & &  \multicolumn{1}{c}{$\varepsilon^-$} & \multicolumn{1}{c}{$\varepsilon^+$} &  \multicolumn{1}{c}{validity} \\
[0.2ex] \hline \\
[-1.8ex]
He/H   & ICF(He$^{+}$ + He$^{++}$) = 1    &  0 & \TabV{$\frac{\displaystyle0.1}{\displaystyle0.03 + \omega^{1.1}}$} & $0.05 < \omega <  0.5$  \\
           & ICF(He$^{+}$ + He$^{++}$) = 1    &  $<<1$ & $<<1$ &  $\omega > 0.5$  \\
[0.2ex] \hline \\
[-1.8ex]
O/H  & $\log$ ICF(O$^{+}$ + O$^{++}$) =  \TabV{$\frac{\displaystyle(0.08\upsilon + 0.006\upsilon^2)}{\displaystyle(0.34 - 0.27\upsilon)}$} & $0.03 + 0.5\upsilon - 0.2\upsilon^2$ & $0.03 + 0.5\upsilon - 0.2\upsilon^2$ & $\upsilon<0.95$\\
[0.2cm] \hline \\
[-1.8ex]
N/O  &  $\log$ ICF(N$^{+}$/O$^{+}$) = \TabV{$-$0.16$\omega$(1 + $\log\upsilon$)} &  0.32$\omega$ & $0.50\omega$ & $\omega<0.95$\\
        &  $\log$ ICF(N$^{+}$/O$^{+}$) = \TabV{0.64$\omega$} &  0.32$\omega$ & $0.50\omega$ & No \ion{He}{ii} lines\\
[0.2ex] \hline \\
[-1.8ex]
Ne/O   &  ICF(Ne$^{++}$/O$^{++}$) =  \TabV{$\omega$+$\left(\frac{\displaystyle0.014}{\displaystyle\upsilon'} + 2\upsilon'^{2.7} \right)^3(0.7 + 0.2\omega - 0.8\omega^{2})$} & 0.17 & 0.12 & $\omega>0.1$  \\
Ne/H   &  ICF(Ne$^{++}$+Ne$^{+4}$) =  \TabV{$(1.31 + 12.68\upsilon^{2.57})^{0.27}$} & 0.20 & 0.17 & $\upsilon > 0.02$\\
[0.2ex] \hline \\
[-1.8ex]
S/O    &  $\log$ ICF(S$^{+}$/O$^{+}$) = \TabV{$0.31 - 0.52\upsilon$ }& 0.38 & 0.41 & $\upsilon > 0.02$\\
S/O    &  $\log$ ICF((S$^{+}$ + S$^{++}$)/O$^{+}$) = \TabV{$\frac{\displaystyle-0.02 - 0.03\omega - 2.31\omega^2 + 2.19\omega^3}{\displaystyle0.69 + 2.09\omega - 2.69\omega^2}$} & 0.20 & 0.12 & $\omega < 0.95$ \\ 
[0.2cm] \hline \\
[-1.8ex]
Cl/O & ICF(Cl$^{++}$/O$^{+}$) = \TabV{($4.1620 - 4.1622\upsilon^{0.21})^{0.75}$} & 0.27 & 0.15 & $0.02 < \upsilon < 0.95$ \\
Cl/O & ICF((Cl$^{+}$+Cl$^{++}$)/O$^{+}$) = 1 & 0.13 & 0 & $\upsilon \leq 0.02$ \\
Cl/O & ICF(Cl$^{+}$+Cl$^{++}$+Cl$^{+3}$) = \TabV{$0.98 + ({0.56} - {0.57}\upsilon)^{7.64}$} & 0.30 & 0.33 & $0 \leq \upsilon \leq 1 $\\
[0.2ex] \hline \\
[-1.8ex]
Ar/O &  $\log$ ICF(Ar$^{++}$/(O$^{+}$+O$^{++}$)) =  \TabV{$\frac{\displaystyle0.05}{\displaystyle0.06+\omega}-0.07$} & 0.70 & 0.60 & $0.05<\omega\leq0.5$\\
&  $\log$ ICF(Ar$^{++}$/(O$^{+}$+O$^{++}$)) =  \TabV{$\frac{\displaystyle0.03\omega}{\displaystyle0.4-0.3\omega}-0.05$} & 0.70 & 0.60 & $0.5<\omega<0.95$\\
[0.2cm] \hline \\
[-1.8ex]
C/O &  ICF(C$^{++}$/O$^{++}$) = \TabV{0.05 + 2.21$\omega$ - 2.77$\omega^2$ + 1.74$\omega^3$} & 0.19 & $0.4 - 1.06\omega + 0.65\omega^2 + 0.27\omega^3$ & $0.05 < \omega < 0.97$ \\
\hline \hline \\[-2ex]
\end{tabular}
\hspace{2.3cm}${^1}$ The ICFs are expressed as a function of $\upsilon$ = \gihe\ or $\omega$ = \gio, the values of $\upsilon^\prime$ are defined in the text.
\end{minipage}
\end{table*}

%%%%%%%%%%%%%%%%%%%%%%%%%%%%%%%%%%%%%%%%%%%%%%%%%%%%%%%%%%%%%%%%%%%%%%%%%%
\section{Testing the robustness of our ICFs}
\label{sec:effect}

\begin{figure*}
\centering
\includegraphics[width=\hsize,trim = 60 80 40 0,clip =yes]{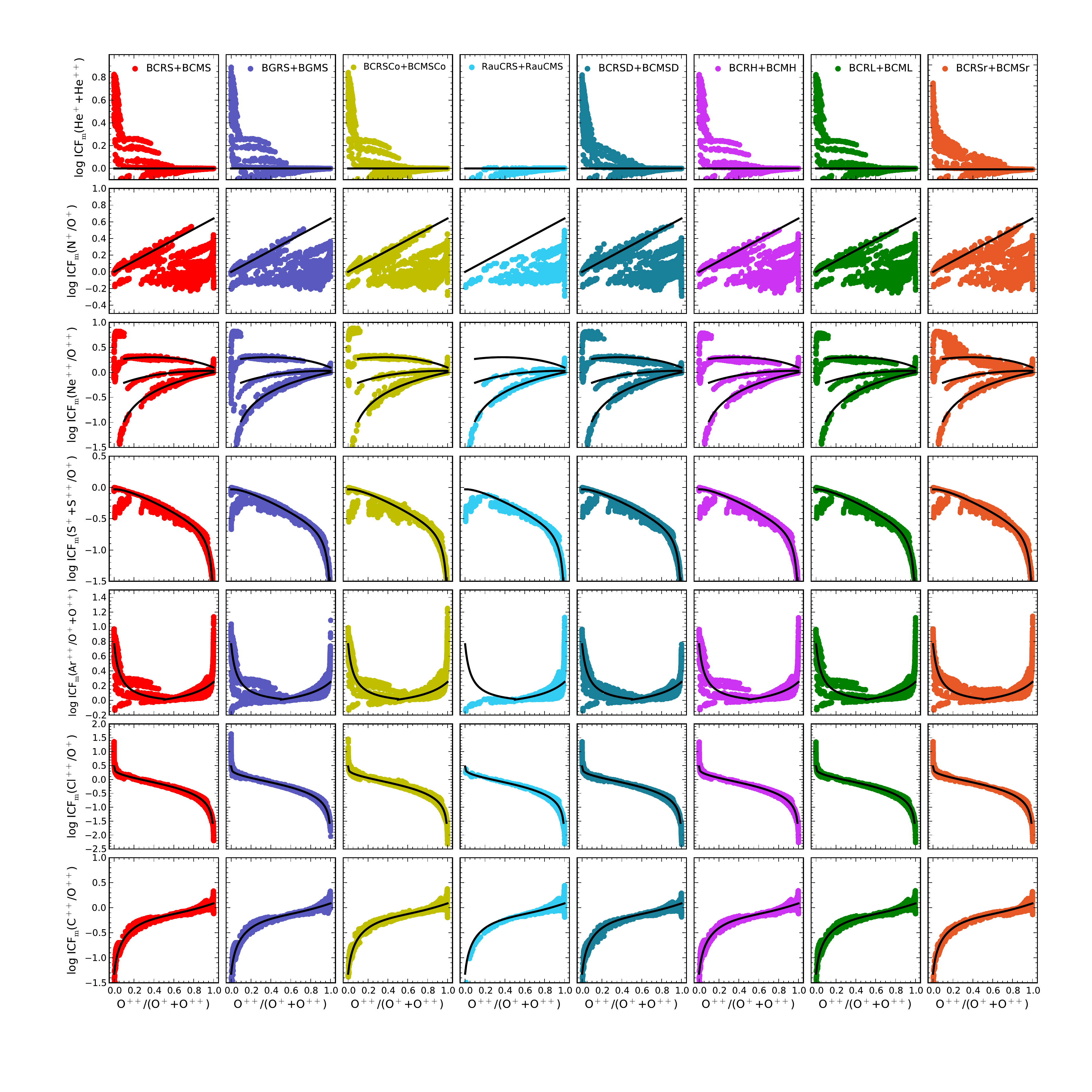}
\caption{Values of \icfm as a function of $\omega$ for the different families of models. 
The solid curves are our \icff. \label{fig:effect_1}}
\end{figure*}

\begin{figure*}
\centering
\includegraphics[width=\hsize,trim = 60 20 40 0,clip =yes]{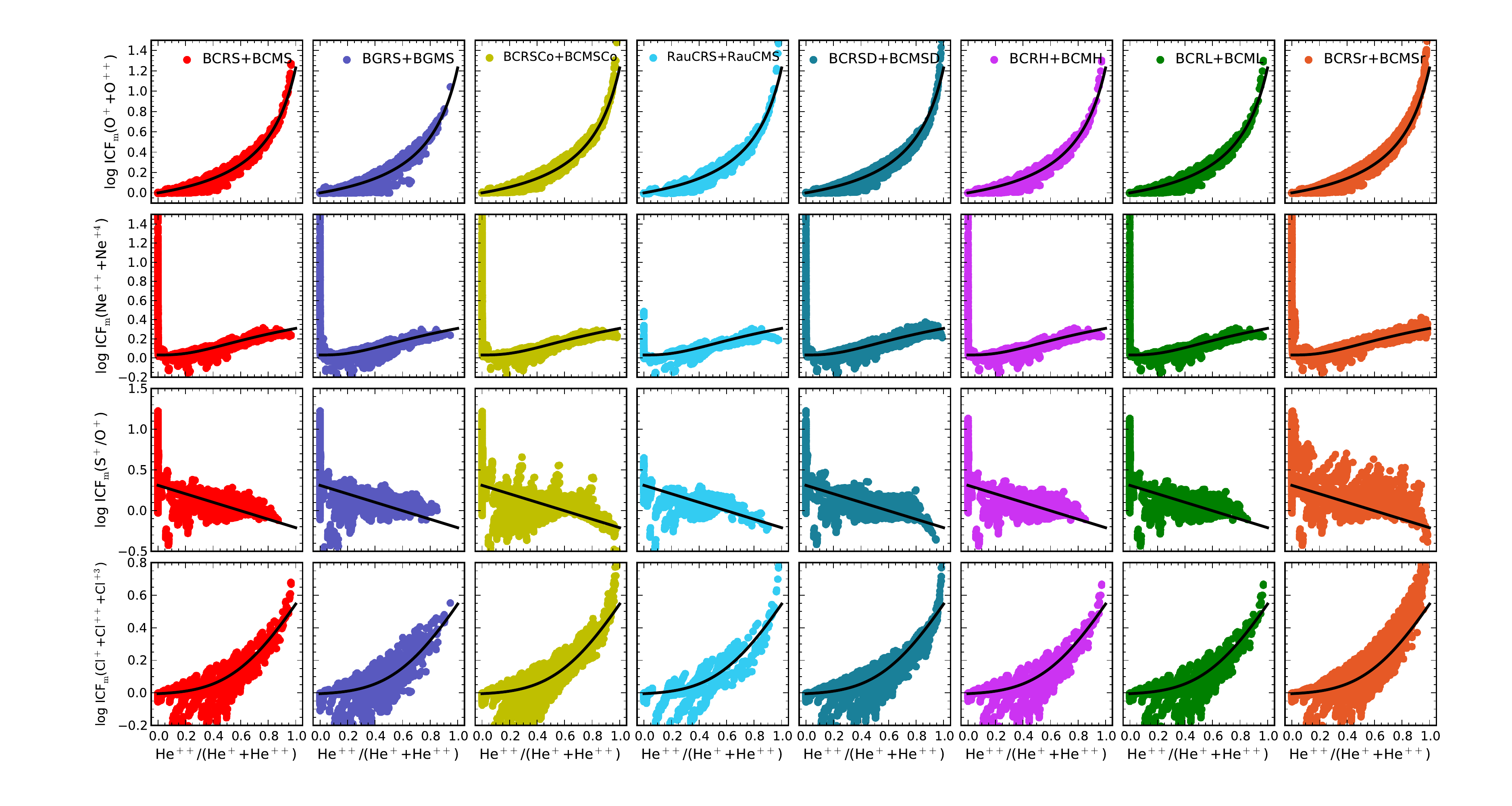}
\caption{Values of \icfm as a function of $\upsilon$ for the different families of models. 
The solid curves are our \icff. \label{fig:effect_2}}
\end{figure*}

To explore the robustness of our ICFs, we consider the additional families of models presented in Section~\ref{sec:mod} and 
Table~\ref{tab:2}. Figures~\ref{fig:effect_1} and \ref{fig:effect_2} show the values of \icfm\ as a function of $\omega$\ or $\upsilon$ 
for all the models. 
Each row concerns a different ICF and each column a different family of models. From left to right:

\renewcommand{\theenumi}{\arabic{enumi}}
\begin{enumerate}
\item[(1)] blackbody models that were used to derive \icff\ for each element (BCRS and BCMS), 
\item[(2)] non constant density models (BGRS and BGMS), 
\item[(3)] models combined as explained in Section~\ref{sec:mod} (BCRSCo and BCMSCo), 
\item[(4)] models with Rauch atmospheres (RCRS and RCMS), 
\item[(5)] models with ISM dust grains (BCRSD and BCMSD), 6) models with Z = 2$\times$Z$_\odot$ (BCRH and BCMH), and 
\item[(6)] models with Z = Z$_\odot$/2 (BCRL and BCML). 
\end{enumerate}

The last column concerns observations of BCRS and BCMS models integrated on the line of sight passing 
through the center instead of integrated over the entire nebular volume. In other words, the integrals of Equation~\ref{eq:x} 
are performed over the radius. This is important for extended PNe. As a matter of fact for such 
cases observed spectra are intermediate between this pencil beam case (last column) and the volume integrated 
case shown in the first column.  

Figures~\ref{fig:effect_1} and \ref{fig:effect_2} clearly show that our ICFs are valid for a wide range of model 
assumptions. Even our error bars, which are rather conservative, roughly apply to all cases. Note that in this 
paper we do not consider truly aspherical PNe as well as pencil beam observations across the face of large 
nebulae. These two cases deserve a dedicated study using three dimensional maps. But so far, we can say that 
our ICFs are robust and can be safely used in a large variety of PN abundance studies.

%%%%%%%%%%%%%%%%%%%%%%%%%%%%%%%%%%%%%%%%%%%%%%%%%%%%%%%%%%%%%%%%%%%%%%%%%%
\section{How do our ICFs change previously computed abundance ratios}

In this section we use our ICFs to calculate total abundances and the uncertainties associated with our ICFs 
in the sample of PNe described in Section~\ref{sec:sample} and compare them to abundances obtained using previous ICFs

\subsection{Determination of the ionic abundances}
\label{sec:ion_ab}

We use the python tool PyNeb\footnote{http://www.iac.es/proyecto/PyNeb/} \citep{Luridiana_12} to derive 
the physical conditions (electron temperatures and densities) and the ionic abundances of the PNe. 
We adopt the atomic data listed in Table~\ref{tab:4}, which should be the best ones presently 
available\footnote{Note that some of the transition probabilities and collision strengths used here are different from 
the ones used in the version of cloudy we employed to compute our models. This is not a problem, since the ICFs depend 
basically on atomic data that influence the ionization structure.}. 

The spectral resolution, the number of observed lines, and their signal-to-noise ratio vary greatly from one sample to another. 
In order to make a reliable comparison between the computed abundances, we follow a procedure as homogeneous as 
possible for all the objects, using the same lines to compute all the physical quantities. 
We derive one electron density and one electron temperature from the [\ion{S}{ii}] $\lambda$6716/$\lambda$6731 and 
[\ion{O}{iii}] $\lambda$4363/($\lambda$4959+$\lambda$5007) ratios respectively, and we 
use them to calculate all the ionic abundances. When the [\ion{S}{ii}] lines are not available, we use the [\ion{O}{ii}] $\lambda$3726/$\lambda$3729, 
[\ion{Cl}{iii}] $\lambda$5518/$\lambda$5538 or [\ion{Ar}{iv}] $\lambda$4711/$\lambda$4740 ratios to estimate the electron density. 

\begin{table}
\caption{Atomic Data}             
\label{tab:4}      
\centering          
\begin{tabular}{l l l }
\hline\hline       
Ion & Transition Probabilities & Collisional Strengths\\ 
\hline                    
\aboii   & \citet{Zeippen_82} & \citet{Kisielius_09}\\
\aboiii  & \citet{Storey_2000} & \citet{Aggarwal_99}\\
            & \citet{Wiese_96} & \\
\abnii   & \citet{Galavis_97} & \citet{Tayal_11}\\
\absii   & \citet{Tayal_10} & \citet{Tayal_10}\\
\absiii  & \citet{Podovedova_09} & \citet{Tayal_99}\\
\abneiii & \citet{Galavis_97} & \citet{Mc_00}\\
\abnev & \citet{Galavis_97} & \citet{Griffin_00}\\
\abariii & \citet{Mendoza_83} & \citet{Galavis_95}\\
\abclii & \citet{Mendoza_83} & \citet{Krueger_70}\\
\abcliii & \citet{Mendoza_82a} & \citet{Krueger_70}\\
\abcliv & \citet{Mendoza_82b} & \citet{Krueger_70}\\
\hline                  
\end{tabular}
\end{table}

As shown in Figure~\ref{fig:obs_1}, where  \gihe\ is plotted as a function of \gio, the models cover almost the same region 
as the observational sample of PNe. We can also see in this figure that most of the observed PNe have relatively high degrees of 
ionization, $\omega > 0.5$.

\subsection{Total abundances}

The total abundances and their associated uncertainties are computed using the analytical expressions compiled in Table~\ref{tab:3}. 
For each element, we only consider those objects with $\upsilon$ or $\omega$ within the range of validity. The error bars in the plots 
arise only from the uncertainties associated with our ICFs, not from the uncertainties in the line fluxes.

\begin{figure}
\centering
\includegraphics[width=\hsize,trim = 30 0 30 0,clip =yes]{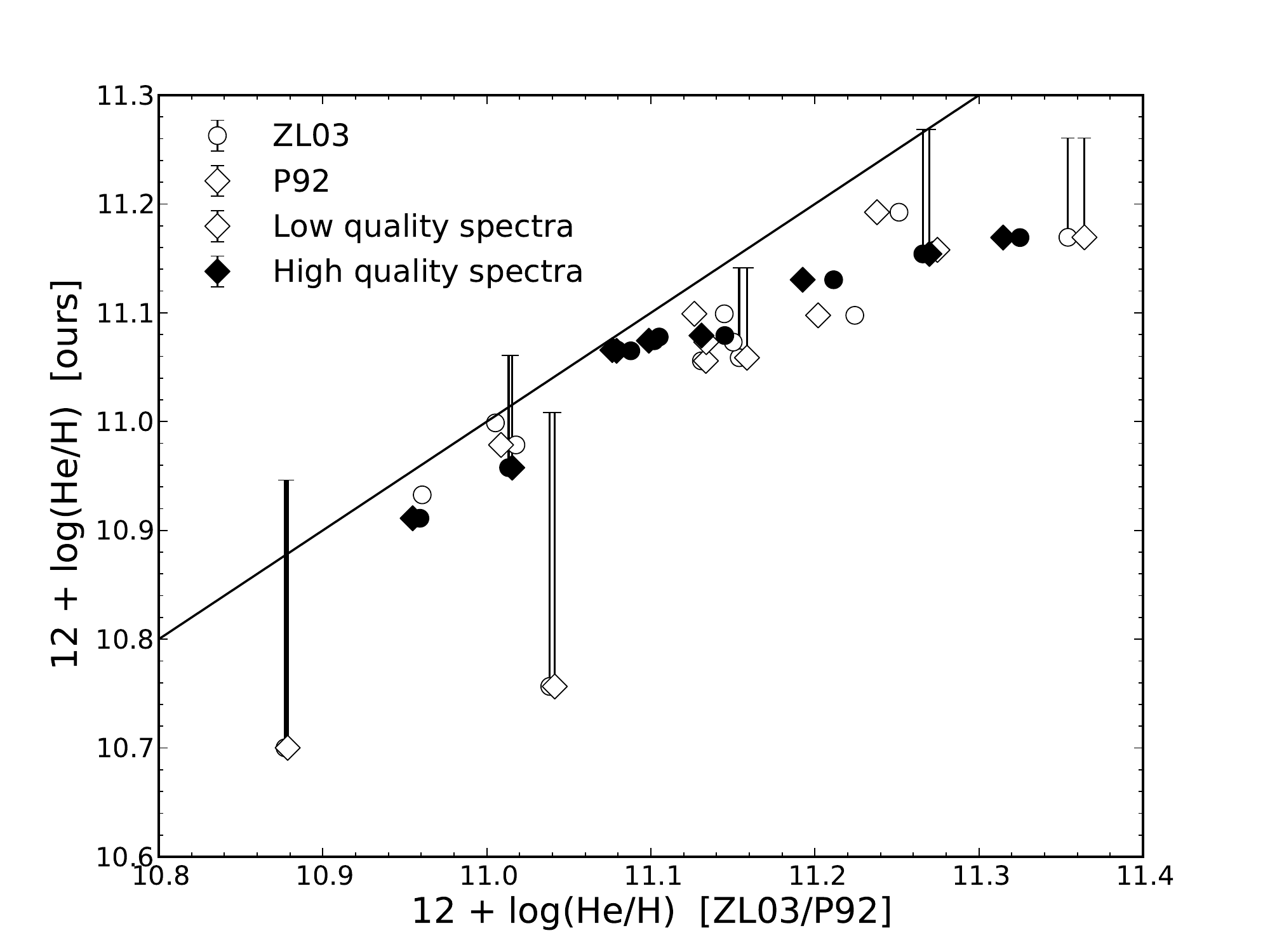}
\caption{Comparison between He/H values computed with our ICF and with the ICFs 
from \citet{Zhang_03} (ZL03) and \citet{Peimbert_92} (P92). The error bars only include 
the uncertainties associated with our ICF. \label{fig:obstot_1}}
\end{figure}

Figure~\ref{fig:obstot_1} compares the helium abundances obtained through Equation~(\ref{eq:he_1}) and (\ref{eq:he_2}) and 
the ones computed with the ICFs by \citet{Peimbert_92} and \citet{Zhang_03}. The error bars in the plot arise from the 
uncertainties in our ICF. We are not considering here the uncertainties related to line fluxes, which are in general around 
$\pm0.02$ dex for helium \citep[e.g.,][]{DI_14}. Only the PNe without \ion{He}{ii} lines in the spectra are included in the plot since the 
abundance of He$^0$ is probably negligible in the others. 

Our values of He/H, including the error bars, are generally lower than those computed with the ICFs by \citet{Peimbert_92} and 
\citet{Zhang_03}. These values of He/H are probably overestimated, at least in the PNe 
with $\omega > 0.5$ (represented in the figure by symbols without upper error bars) where we do not expect a significant 
contribution of neutral helium (as seen in Fig.~\ref{fig:he}b).

\begin{figure}
\centering
\subfigure{\includegraphics[width = \hsize,trim = 0 20 0 0,clip =yes]{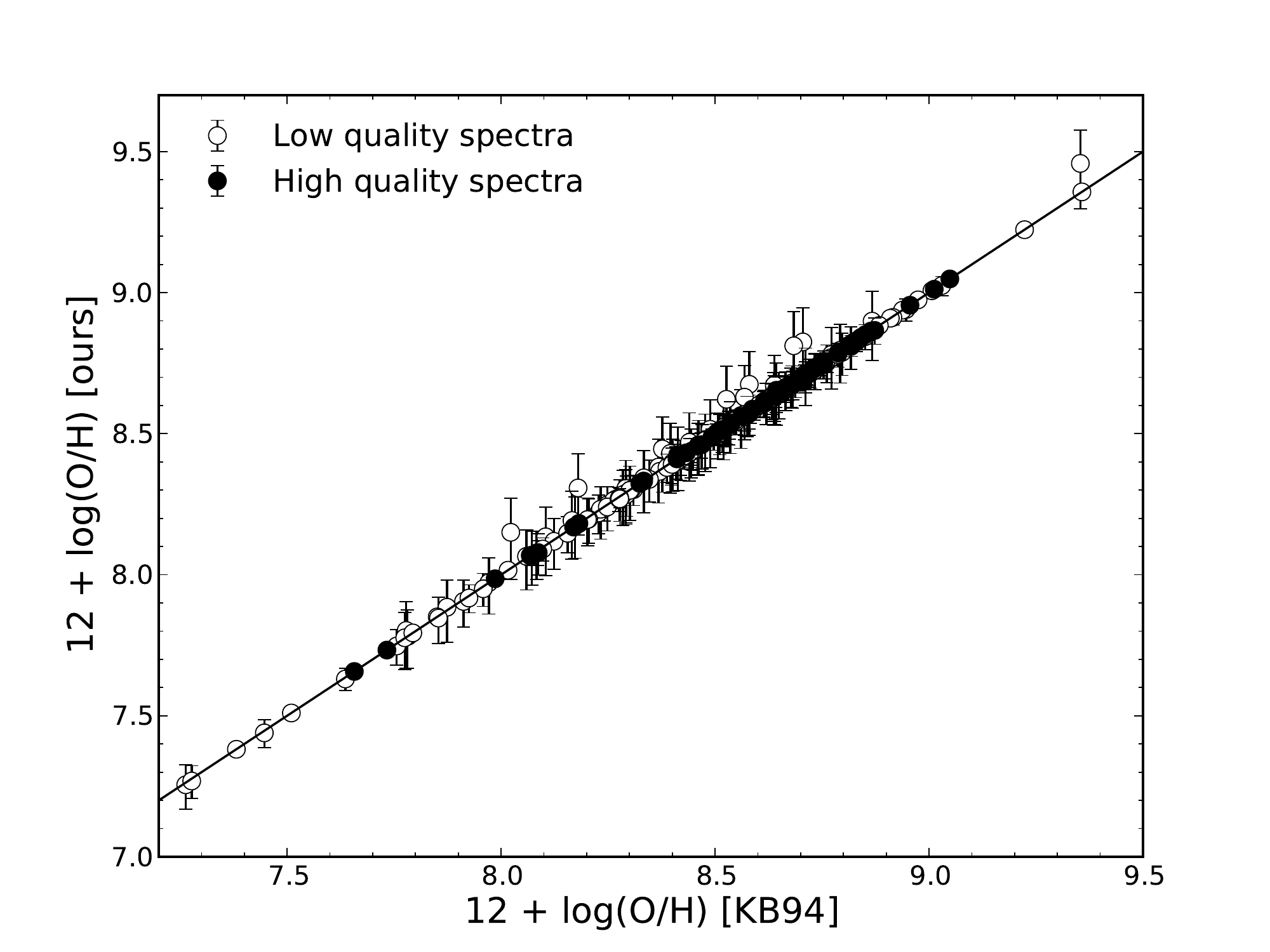}}\\
\subfigure{\includegraphics[width= \hsize,trim = 0 20 0 10,clip =yes]{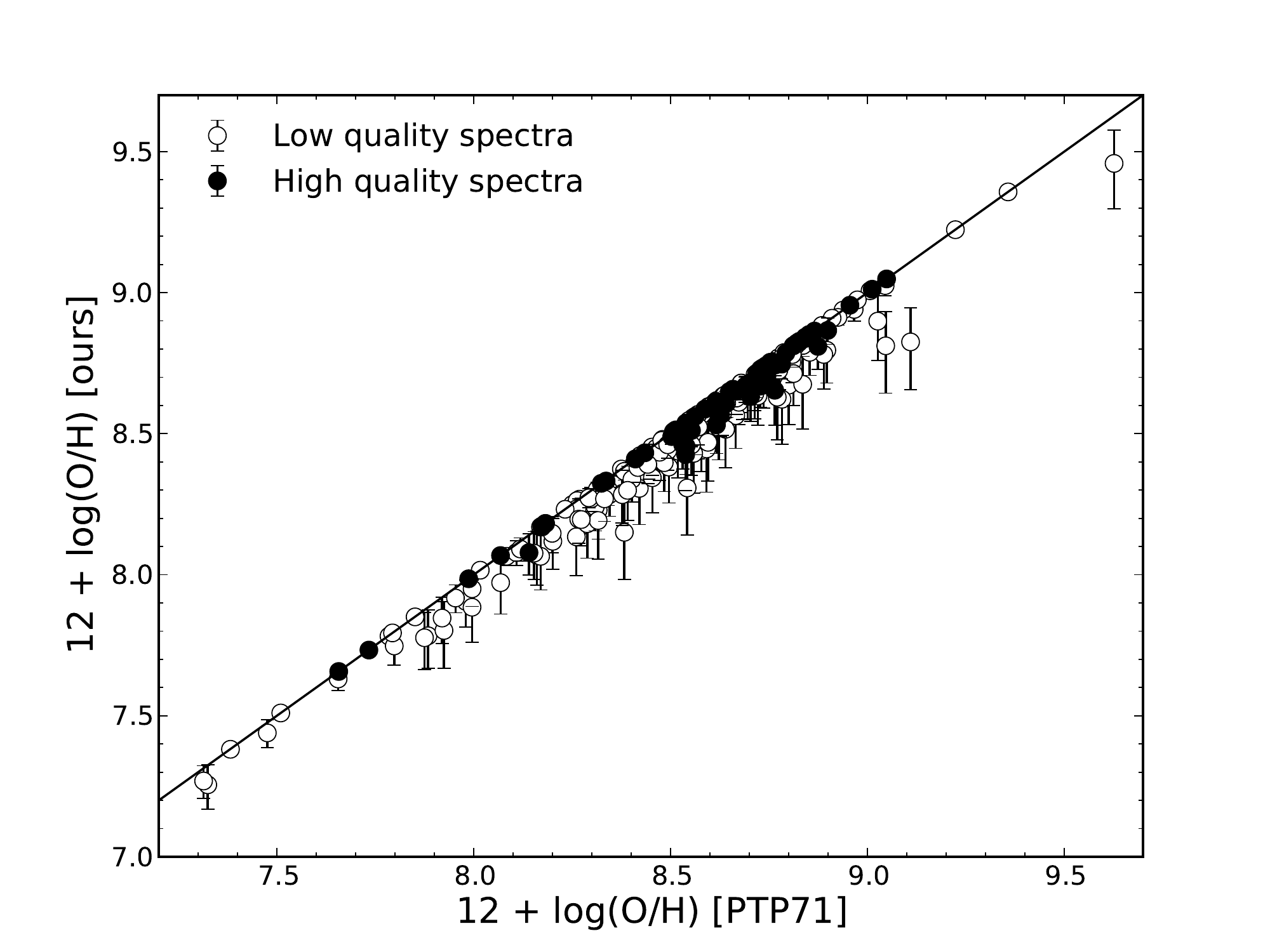}}
\caption{Comparison between O/H values computed with our ICF and with the ICF 
from KB94 (upper panel) and \citet{Peimbert_71} (PTP71, lower panel). The error bars only include
the uncertainties associated with our ICF. \label{fig:obstot_2}}
\end{figure}

The oxygen abundances computed with three different ICFs are compared in Figure~\ref{fig:obstot_2}. 
Our ICF leads to O/H values similar to the ones obtained 
from the ICF by KB94 whereas O/H values computed with the ICF by \citet{Peimbert_71} are generally higher than ours. 
As we have shown in Figure~\ref{fig:o}b, the latter ICF overestimates the oxygen abundances in cases where He$^{++}$ 
is present, since it assumes that no O$^{++}$ can be present in the He$^{++}$ zone, which is not correct. The error bars 
in these plots correspond to the uncertainties associated with our ICF, estimated with Equation~(\ref{eq:o_2}). Note that in 
high ionization objects, the error bars can be greater than the usual error bars which result only from the propagation of 
the line intensity errors into the physical conditions and ionic abundances \citep[around $\sim$0.05 dex, see, e.g., ][]{DI_14}.

\begin{figure*}
\centering
\subfigure{\includegraphics[width=0.45\hsize,trim = 0 20 0 0,clip =yes]{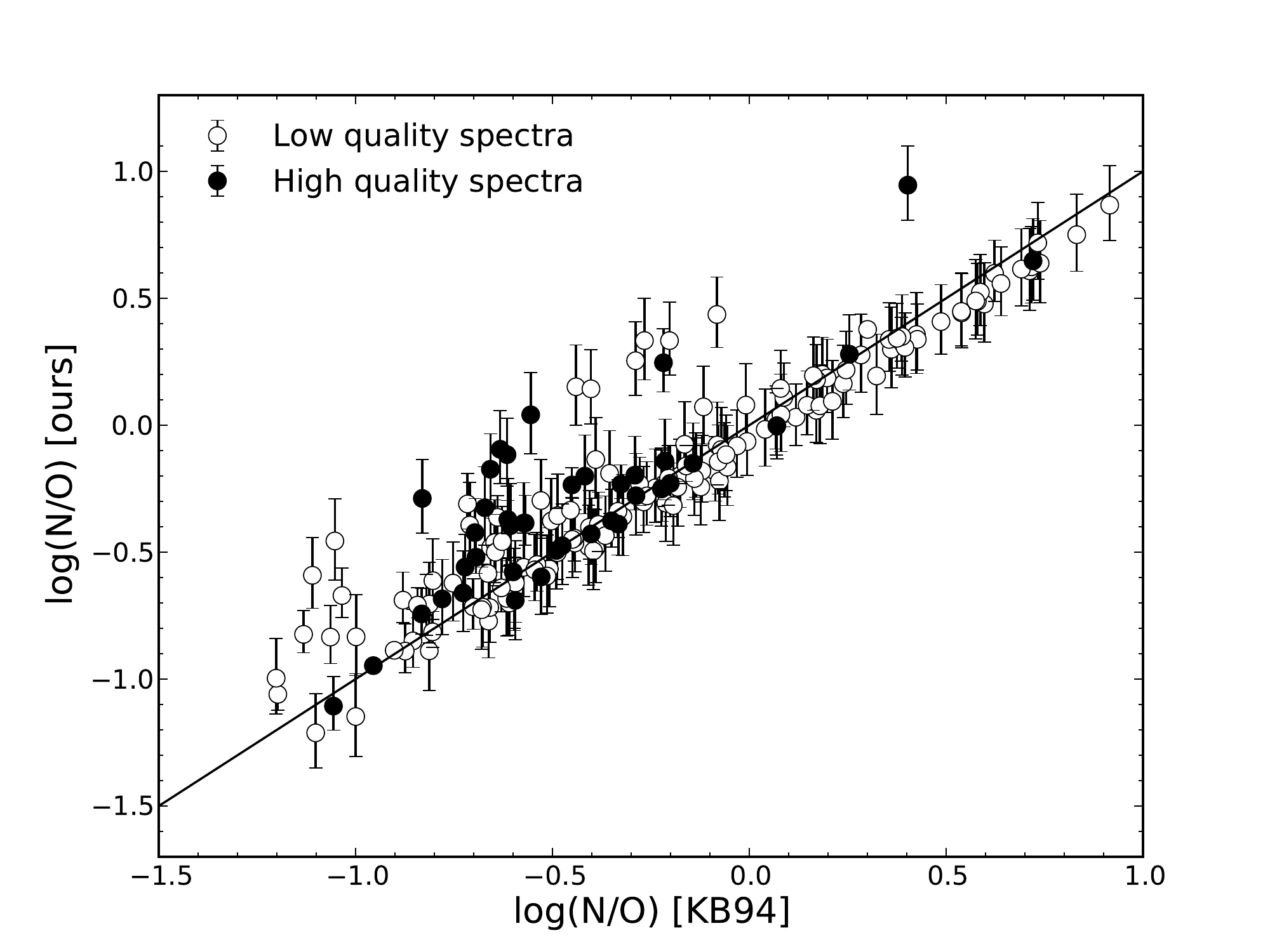}}
\subfigure{\includegraphics[width=0.45\hsize,trim = 0 20 0 0,clip =yes]{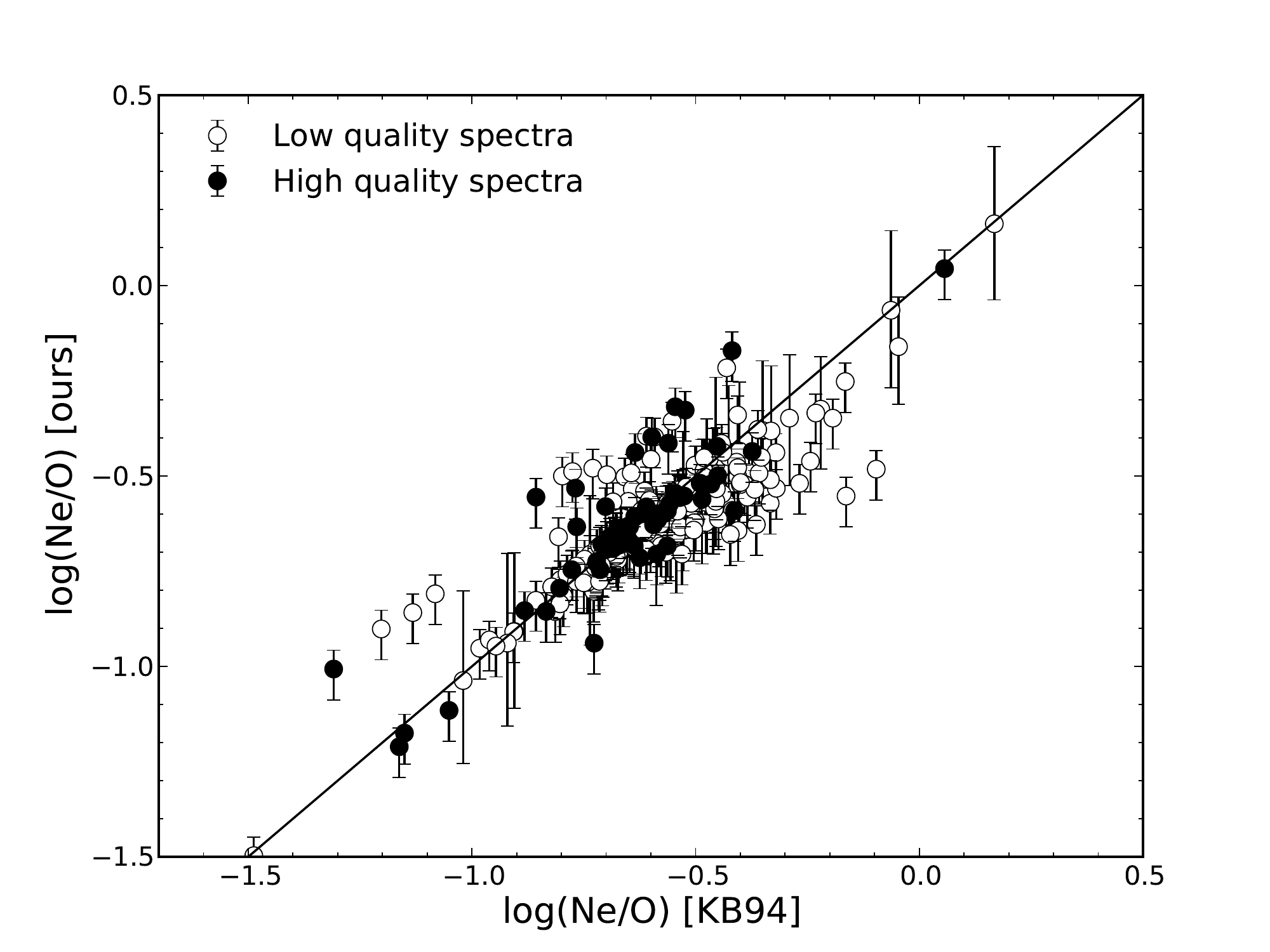}}
\subfigure{\includegraphics[width=0.45\hsize,trim = 0 20 0 0,clip =yes]{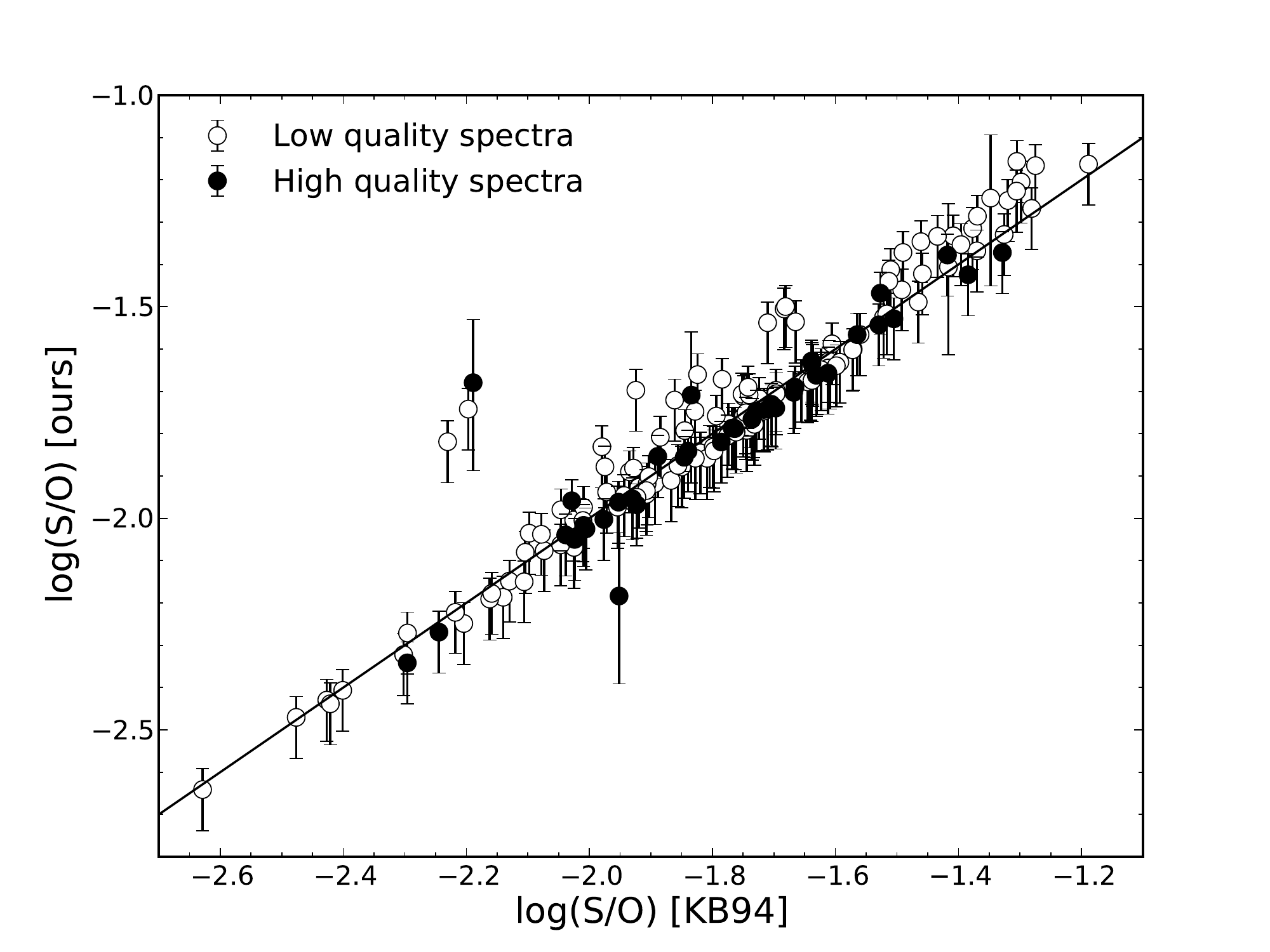}}
\subfigure{\includegraphics[width=0.45\hsize,trim = 0 20 0 0,clip =yes]{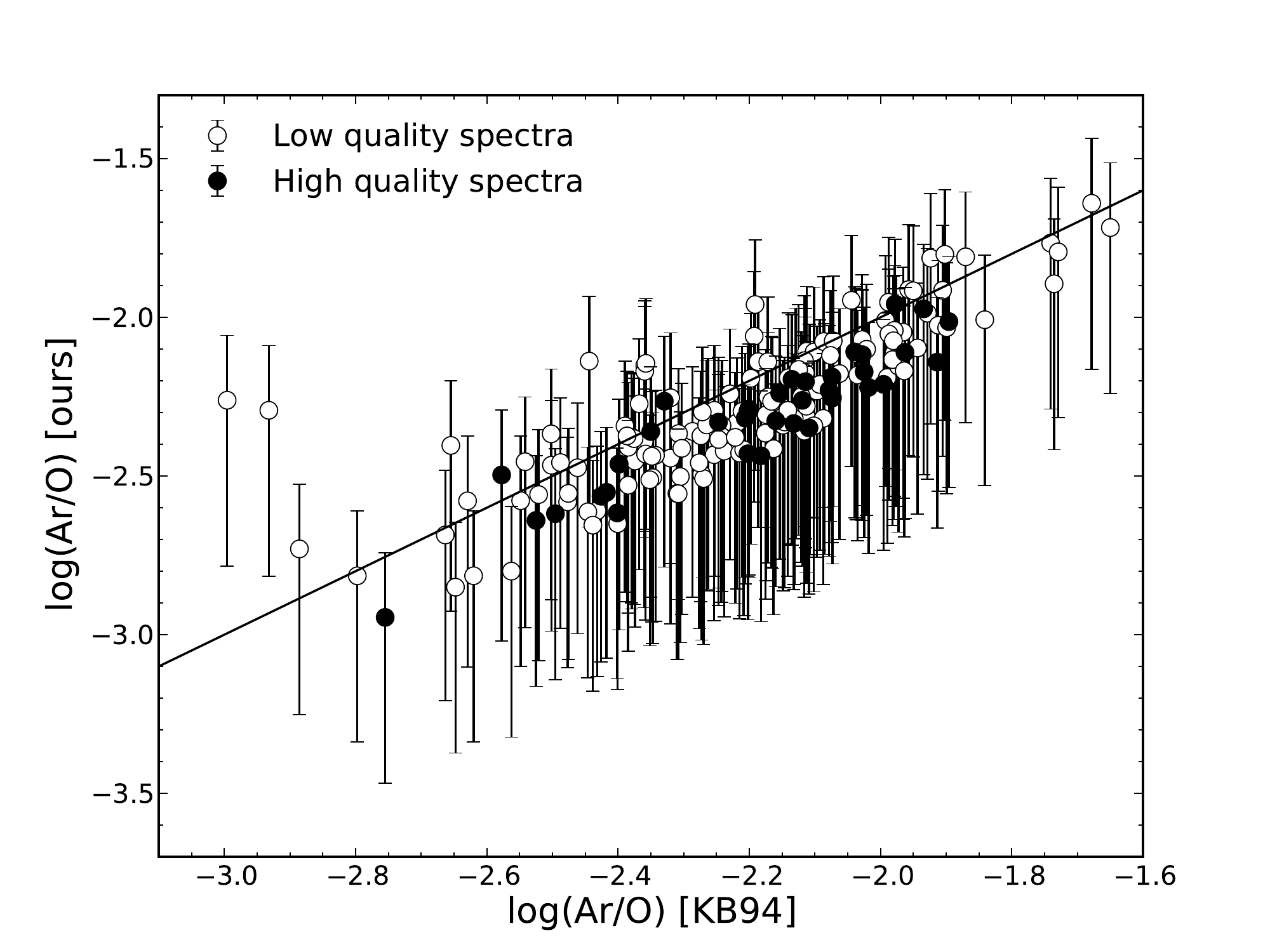}}
\subfigure{\includegraphics[width=0.45\hsize,trim = 0 25 0 0,clip =yes]{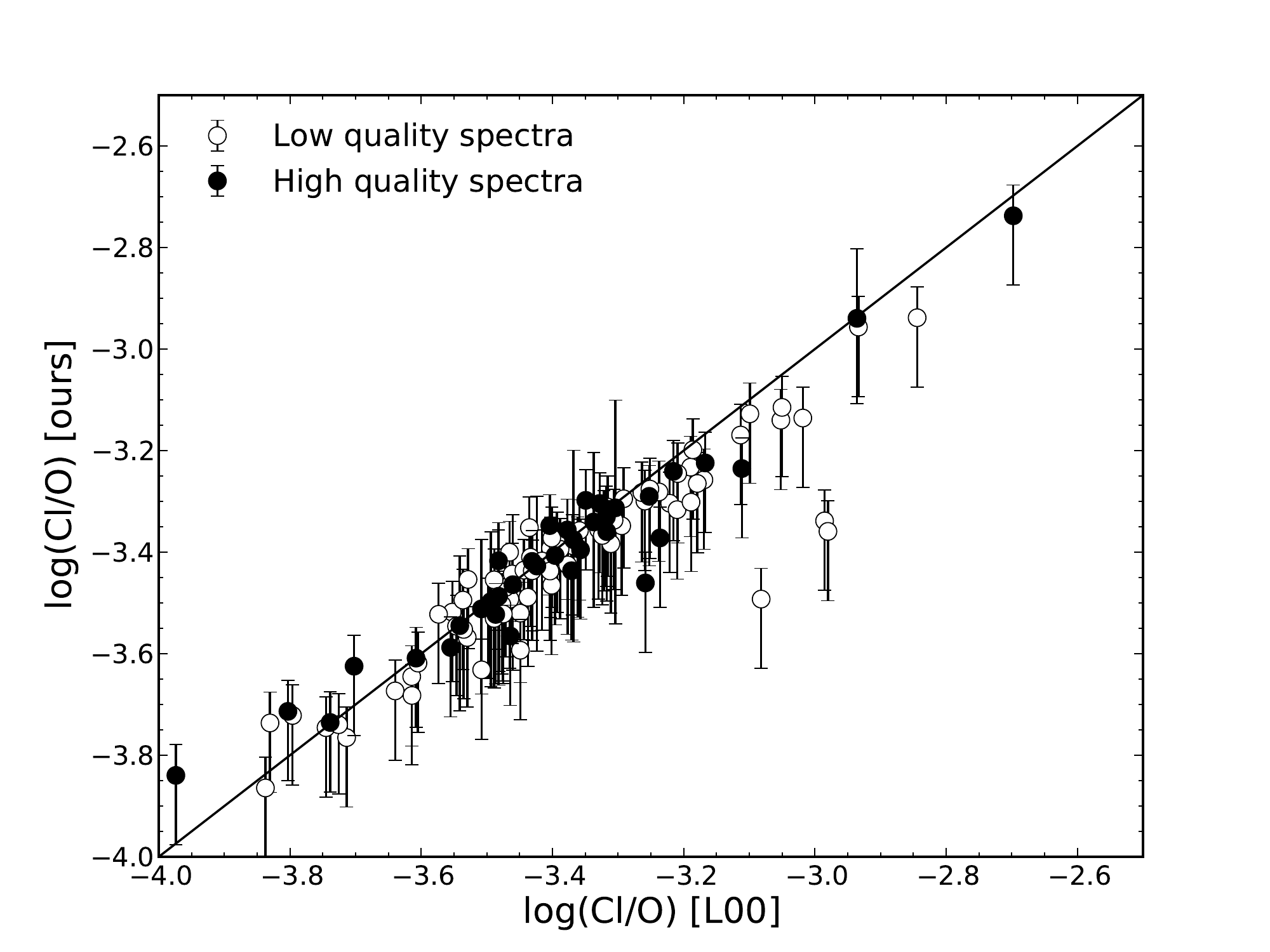}}
\caption{Comparison between N/O, Ne/O, S/O, Ar/O, and Cl/O values computed with our ICFs and with 
the ICFs by KB94 or \citet{Liu_00} (L00). The error bars only include the uncertainties associated with 
our ICF.\label{fig:obstot_3}}
\end{figure*}

Figure~\ref{fig:obstot_3} compares our values of N/O, Ne/O, S/O, Ar/O, and Cl/O with the ones computed 
using ICFs from other authors. For chlorine we use the ICF proposed by \citet{Liu_00} because this element is not 
considered in KB94. For the other four elements we use the ICFs by KB94. The differences in the computed 
values of N/O, Ne/O, and Ar/O are significant. 

\begin{figure}
\centering
\includegraphics[width=\hsize,trim = 0 0 0 0,clip =yes]{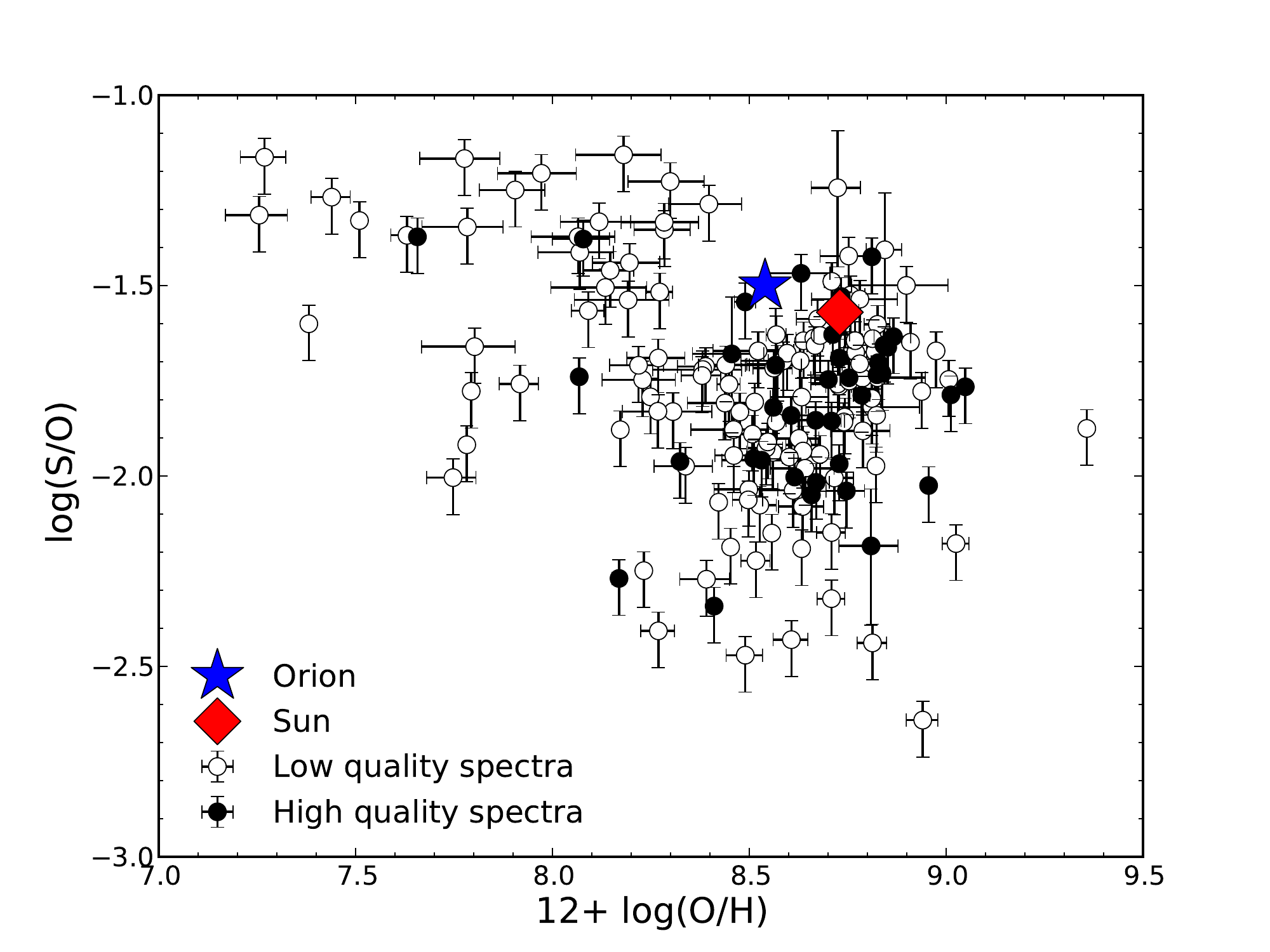}
\caption{Values of S/O as a function of O/H for the sample of Galactic PNe. The error bars only include 
the uncertainties associated with our ICF. \label{fig:obstot_4}}
\end{figure}

In general, the S/O values derived from our ICF and from the ICF by KB94 (originally proposed by 
\citealt{Stasinska_78}) are similar within errors. There are some exceptions showing higher S/O values 
when using our ICF. This result is relevant in regard to the so-called 
sulfur anomaly \citep{Henry_04, Henry_10}, the fact that many PNe show low sulfur abundances compared 
to H~II regions of the same metallicity (measured by O/H). 
Figure~\ref{fig:obstot_4} shows the values of S/O as a function of O/H. The values for the Sun and for Orion 
\citep{Lodders_10,Esteban_04} are also plotted for comparison. It can be seen that the S/O values in most of 
our sample PNe are lower than those in Orion and in the Sun. Several hypotheses have been proposed 
to explained this result, the most accepted one being the incorrect correction for unobserved ions of sulfur. 
Our results suggest that the ICF is, at least, not the main cause of this anomaly. 

\section{Testing our ICFs with observational data}

Here, we want to test the validity of our ICFs by checking that some of the computed abundances ratios do not show 
any specific trend with the ionization degree. It is important to consider PNe covering a wide range of physical conditions, 
therefore we use the whole sample of PNe. 

We consider only the results for Ne/O, S/O, Ar/O, Cl/O, and Ar/S. These five elements are mainly produced 
in massive stars and they are expected to evolve in lockstep. Therefore, their relative ratios are not expected to depend on the 
degree of ionization and they can be used to see if our ICFs show any trend with the degree of ionization. 
However, several nucleosynthesis models do predict some oxygen production/destruction in low metallicity environments 
\citep[see, e.g.][]{Karakas_07} which could perhaps lead to a weak tendency of the above abundance rations with the nebular 
excitation. Moreover, given that several PNe in the solar neighborhood have higher oxygen abundances than nearby H~II 
regions \citep{Rodriguez_11}, even at solar metallicities oxygen production in PNe could be occurring. 

The results for helium, carbon, and nitrogen are not discussed here since these elements can be produced in the interiors of 
asymptotic giant branch stars, depending on the progenitor mass, and the ionization of PNe depends on the mass of the central star \citep{Stasinska_98}. 
A trend between N/O, C/O or He/H and the degree of ionization would not be surprising and thus, these abundance ratios are not 
useful to test our ICFs.

\begin{figure*}
\centering
\subfigure{\includegraphics[width=\hsize,trim = 0 20 0 0,clip =yes]{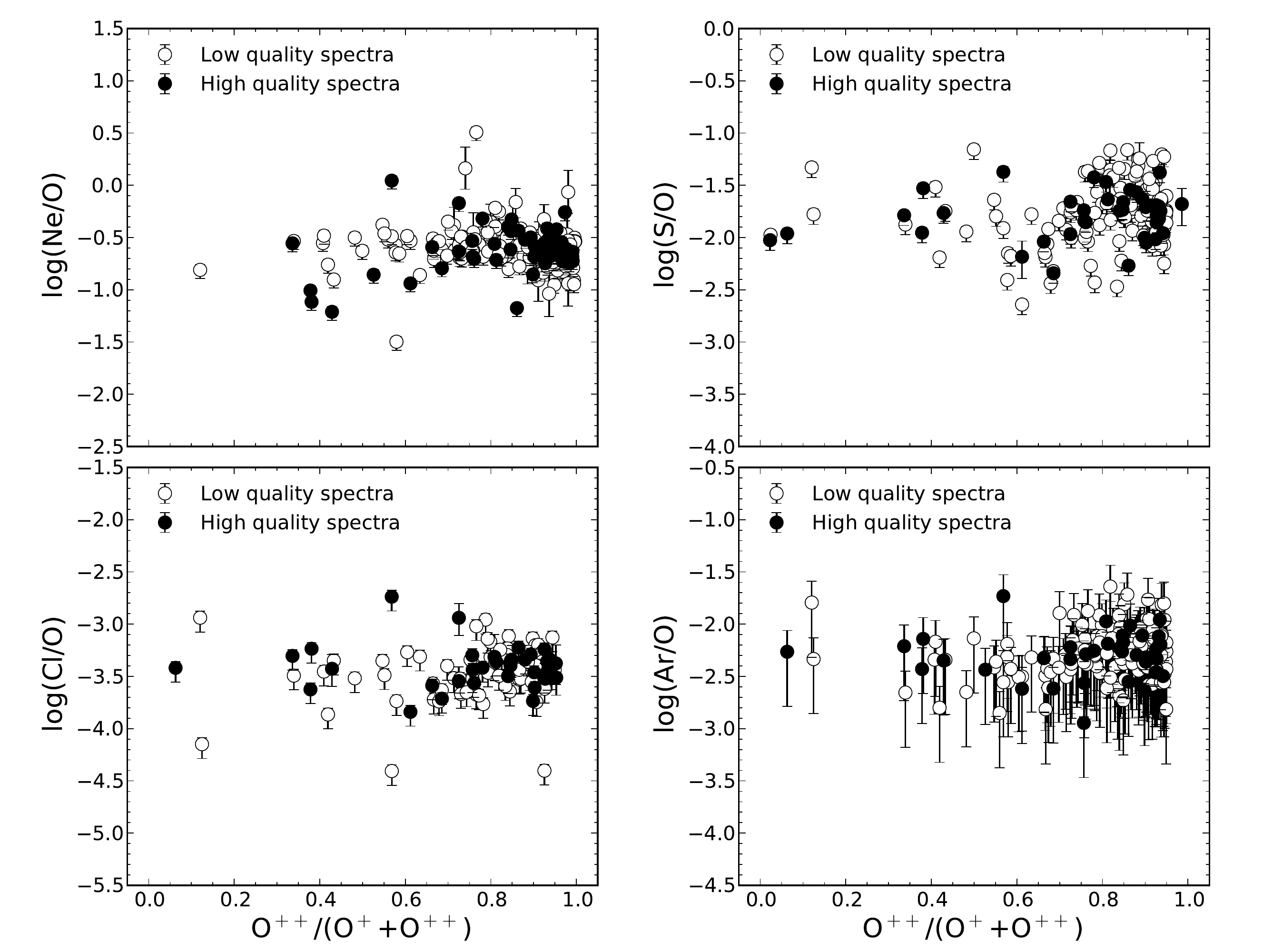}}
\caption{Values of Ne/O, S/O, Cl/O, and Ar/O derived with our ICFs as a function of $\omega$. The error bars only include 
the uncertainties associated with our ICF. \label{fig:obstot_5}}
\end{figure*}

In Figure~\ref{fig:obstot_5} we display the values of Ne/O, S/O, Ar/O, and Cl/O as a function of the degree of ionization. 
The absence of a trend in this plot argues in favor of our ICF recipes. This figure shows that the error bars associated with 
the ICFs can be higher than the usual uncertainties in the abundance ratios which arise only from the uncertainties in the line 
fluxes and range from $\pm0.05$ dex to $\pm$0.2 dex depending on the element (and the quality of the observations). 

\begin{figure}
\centering
\includegraphics[width=\hsize,trim = 0 0 0 0,clip =yes]{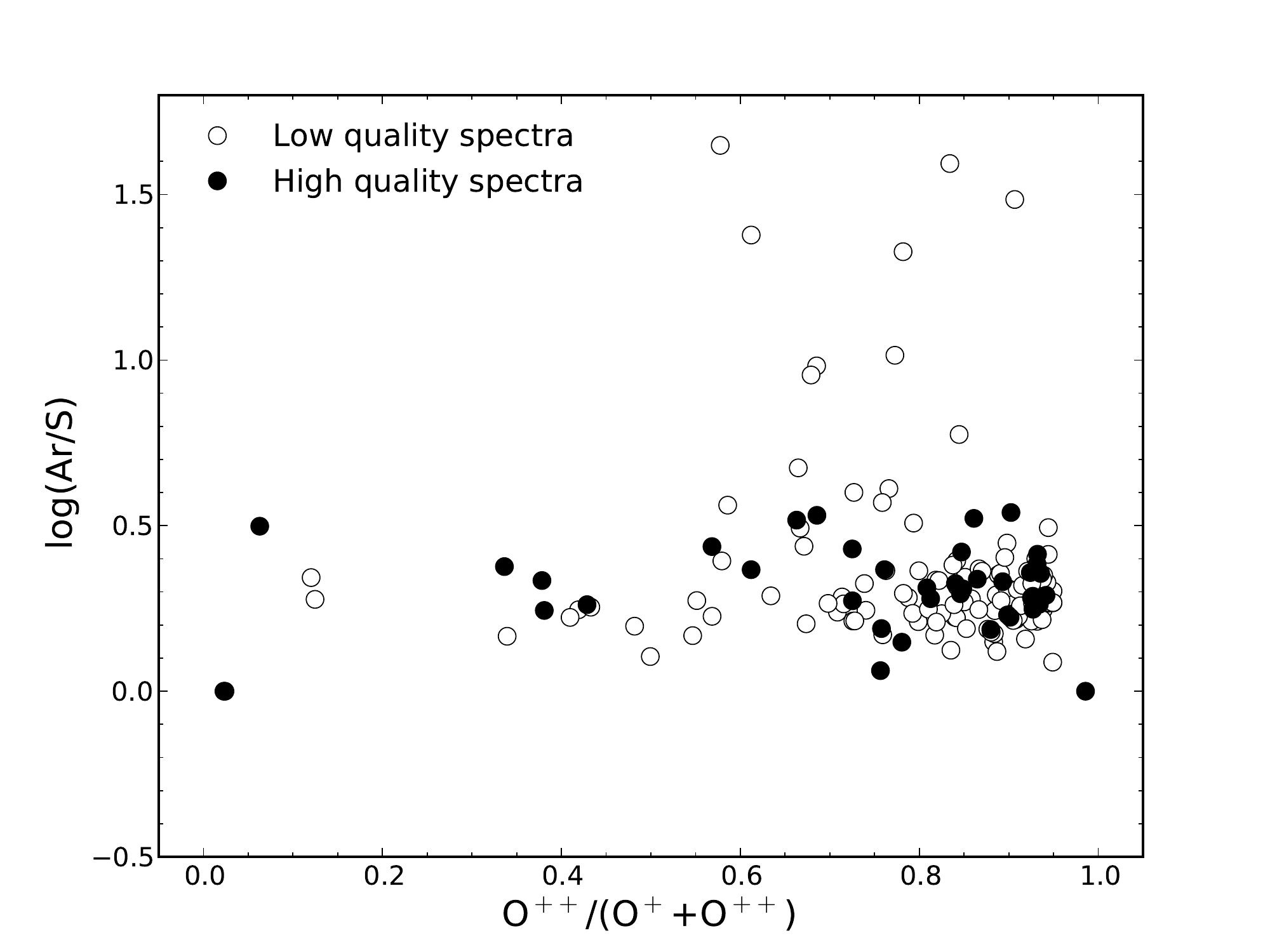}
\caption{Values of Ar/S as a function of the degree of ionization for the sample of Galactic PNe.\label{fig:obstot_6}}
\end{figure}

The abundance ratios Ar/S are plotted in Figure~\ref{fig:obstot_6} as a function of the degree of ionization. This abundance ratio is 
even less suspect of being affected by nucleosynthesis processes in the progenitor star than the abundance ratios involving oxygen. 
A certain amount of sulfur atoms in some PNe can be trapped in dust grains \citep[such as MgS][]{DI_14} or molecules \citep{Henry_12}. There is no obvious trend between the Ar/S values and the degree of ionization, which tends to support our ICFs.
The high Ar/S values obtained in eight PNe could be a consequence of sulfur depletion into dust grains 
(such as MgS, see \citealt{DI_14}) or molecules \citep{Henry_12}.  

%%%%%%%%%%%%%%%%%%%%%%%%%%%%%%%%%%%%%%%%%%%%%%%%%%%%%%%%%%%%%%%%%%%%%%%%%%
\section{Summary}
\label{sec:concl}

We have computed a large grid of photoionization models covering a wide range of physical parameters representative of most 
of the observed PNe. Using this grid of models, we derived new recipes for the ICFs of He, O, N, Ne, 
S, Ar, Cl, and C that should be used for optical observations. We also provide analytical expressions for the associated 
uncertainties. This should be useful for empirical abundance studies since errors in ICFs are usually not considered when 
estimating errors in element abundances. 

The ICFs are valid for all the computed families of models, which include models with different input 
metallicities, different gas density distributions, presence or absence of dust, or different spectral energy distribution of the ionizing 
source. Besides, our ICFs are valid for volume integrated observations as well as for pencil beam observations in the vicinity of 
the central star. Therefore they can be safely used in a large variety of PNe observations. 
  
For each correction scheme, we define the region of validity, outside this zone the error bars are so large that no meaningful 
abundance can be computed. Objects presenting such \gihe\ and \gio\ should be removed from abundance computations 
using our ICFs for the corresponding element. A tailored photoionization model is recommended for such cases.

In the case of helium we do not suggest the derivation of an ICF based on other ions since the relative populations of 
helium ions depend essentially on the \efftemp\ of the central star while the ones of metal ions also depend on the ionization 
parameter. We recommend to use ICF(He$^{+}$ + He$^{++}$) = 1 and, for those objects with $\omega \lesssim 0.45$, 
incorporate the correction for unobserved ions in the error bars.

For nitrogen and neon, we propose ICFs based both on \gio\ and \gihe. The uncertainties associated with our computed 
N/O and Ne/O values are significantly lower (especially for PNe with low \efftemp\ central stars) than those derived using 
the ICFs suggested by KB94. 

As for the other elements we suggest ICFs based on \gio\ or \gihe. In general, the uncertainties related to our ICFs are 
significantly lower than the ones related to other ICFs in the literature.

We have shown that the error bars associated with ICFs should be taken into account in the total error budget on abundances 
since they represent a significant budget of total errors abundances.

Although our sulfur abundances are somewhat higher than those derived from the ICF by KB94, the S/O values obtained in most 
of the studied PNe are still lower than those in Orion or in the Sun. This suggests that the ICF is not the principal cause of the 
sulfur anomaly. 

The abundance ratios Ne/O, S/O, Cl/O, and Ar/O computed with our ICFs in a wide sample of observed PNe (covering a wide 
range of ionization, central star temperature, and metallicity) do not show any noticeable trend with the degree of ionization. 
This validates our ICFs for abundance studies. 

The ICFs defined in this paper will be included in the next release of the Pyneb package.

\section*{Acknowledgments}
The authors acknowledge the referee, Richard Henry, for his detailed reading and constructive comments. 
We also thank M. Rodr\'iguez and A. Mampaso for many fruitful discussions and useful suggestions.
G.D.-I., C.M., and G.S. acknowledge support from the following mexican projects CB-2010/153985, 
PAPIIT-IN105511, and PAPIIT-IN112911. G.D.-I. gratefully acknowledges a DGAPA postdoctoral grant 
from the Universidad Nacional Aut\'onoma de M\'exico (UNAM). This work has made use of NASA's Astrophysics 
Data System, and the SIMBAD database operated at CDS, Strasbourg, France.

%%%%%%%%%%%%%%%%%%%%%%%%%%%%%%%%%%%%%%%%%%%%%%%%%%%%%%%%%%%%%%%%%%%%%%%%%%

\label{lastpage}


\begin{thebibliography}{}

\bibitem[Aggarwal \& Keenan (1999)]{Aggarwal_99} 
Aggarwal, K.M. \& Keenan, F. P. 1999,  \apjs,123,311

\bibitem[Alexander \& Balick(1997)]{Alexander_97} 
Alexander, J., \& Balick, B.\ 1997, \aj, 114, 713

\bibitem[Aller \& Czyzak(1983)]{Aller_83} 
Aller, L.~H., \& Czyzak, S.~J.\ 1983, \apjs, 51, 211

\bibitem[Barlow(1987)]{Barlow_87} 
Barlow, M.~J.\ 1987, \mnras, 227, 161

\bibitem[Becker et al. (1989)]{Becker_89}
Becker, S.~R., Butler, K. \& Zeippen, C.~J. 1989, A\&A, 221, 375

\bibitem[Bloecker(1995)]{Bloecker_95} 
Bloecker, T.\ 1995, \aap, 299, 755

\bibitem[Delgado-Inglada \& Rodr\'iguez(2014)]{DI_14} 
Delgado-Inglada, G. \& Rodr\'iguez, M.\ 2014, ApJ, submitted

\bibitem[Esteban et al.(2004)]{Esteban_04} Esteban, C., Peimbert, 
M., Garc{\'{\i}}a-Rojas, J., et al.\ 2004, \mnras, 355, 229

\bibitem[Ferland et al.(1998)]{Ferland_98} 
Ferland, G.~J., Korista, K.~T., Verner, D.~A., et al.\ 1998, \pasp, 110, 761

\bibitem[Galavis et al. (1995)]{Galavis_95} 
Galavis, M.~E., Mendoza, C., \& Zeippen, C.~J. 1995, ADNDT, 66, 1

\bibitem[Galavis et al. (1997)]{Galavis_97} 
Galavis, M.~E., Mendoza, C., \& Zeippen, C.~J. 1997, \aaps, 123, 159

\bibitem[Garc{\'{\i}}a-Rojas et al.(2009)]{GR_09} 
Garc{\'{\i}}a-Rojas, J., Pe{\~n}a, M., \& Peimbert, A.\ 2009, \aap, 496, 139 

\bibitem[Garc{\'{\i}}a-Rojas et al.(2012)]{GR_12} 
Garc{\'{\i}}a-Rojas, J., Pe{\~n}a, M., Morisset, C., Mesa-Delgado, A., \& Ruiz, M.~T.\ 2012, \aap, 538, A54 

\bibitem[Gathier(1987)]{Gathier_87} 
Gathier, R.\ 1987, \aaps, 71, 245 

\bibitem[Giles (1981)]{Giles_81}
Giles, 1981, \mnras, 195, 63

\bibitem[G{\'o}rny et al.(2009)]{Gorny_09} 
G{\'o}rny, S.~K., Chiappini, C., Stasi{\'n}ska, G., \& Cuisinier, F.\ 2009, \aap, 500, 1089 

\bibitem[Griffin \& Badnell(2000)]{Griffin_00} 
Griffin, D.~C. \& Badnell, N.~R. 2000, J. Phys. B: At. Mol. Opt. Phys, 33, 4389

\bibitem[Henry et al.(2004)]{Henry_04} 
Henry, R.~B.~C., Kwitter, K.~B., Balick, B.\ 2004, \aj, 127, 2284 

\bibitem[Henry et al.(2010)]{Henry_10} 
Henry, R.~B.~C., Kwitter, K.~B., Jaskot, A.~E., et al.\ 2010, \apj, 724, 748 

\bibitem[Henry et al.(2012)]{Henry_12} 
Henry, R.~B.~C., Speck, A., Karakas, A.~I., Ferland, G.~J., \& Maguire, M.\ 2012, \apj, 749, 61

\bibitem[{{Karakas} \& {Lattanzio}(2007)}]{Karakas_07}
{Karakas}, A. \& {Lattanzio}, J.~C. 2007, \pasa, 24, 103

\bibitem[Khromov(1989)]{Khromov_89} 
Khromov, G.~S.\ 1989, \ssr, 51, 339

\bibitem[{{Kingsburgh} \& {Barlow}(1994)}]{KB_94}
{Kingsburgh}, R.~L. \& {Barlow}, M.~J. 1994, \mnras, 271, 257

\bibitem[Kisielius et al.(2009)]{Kisielius_09} 
Kisielius, R., Storey, P. J., Ferland, G. J., \& Keenan, F. P.\ 2009, \mnras, 397, 903

\bibitem[Krueger \& Czyzak(1970)]{Krueger_70} 
Krueger, T.~K. \& Czyzak, S.~J. 1970, Proc. R. Soc. Lond. A, 318, 531 

\bibitem[Kwitter \& Henry(2001)]{Kwitter_01} 
Kwitter, K.~B., \& Henry, R.~B.~C.\ 2001, \apj, 562, 804 

\bibitem[Leisy \& Dennefeld(2006)]{Leisy_06} 
Leisy, P., \& Dennefeld, M.\ 2006, \aap, 456, 451 

\bibitem[Liu et al.(2000)]{Liu_00} 
Liu, X.-W., Storey, P.~J., Barlow, M.~J., et al.\ 2000, \mnras, 312, 585 

\bibitem[Liu et~al.(2004)]{Liu_04}
{Liu}, Y., {Liu}, X.-W., {Luo}, S.-G., \& {Barlow}, M.~J. 2004, \mnras, 353, 1231

\bibitem[Lodders(2010)]{Lodders_10} Lodders, K.\ 2010, Principles 
and Perspectives in Cosmochemistry, 379 

\bibitem[Luridiana et al.(2012)]{Luridiana_12} 
Luridiana, V., Morisset, C., \& Shaw, R.~A.\ 2012, IAU Symposium, 283, 422 

\bibitem[Marigo et al.(2001)]{Marigo_01}
Marigo, P., Girardi, L., Groenewegen, M.~A.~T., \& Weiss, A.\ 2001, \aap, 378, 958

\bibitem[McLaughlin \& Bell(2000)]{Mc_00} 
McLaughlin, B. M. \& Bell, K. L.\ 2000, Journal of Physics B Atomic Molecular Physics, 33, 597

\bibitem[Mendoza (1982)]{Mendoza_82}
Mendoza, C. 1982, \jpb, 15, 867

\bibitem[Mendoza \& Zeippen (1982a)]{Mendoza_82a}
Mendoza, C. \& Zeippen, C.~J. 1982a, \mnras, 198, 127

\bibitem[Mendoza \& Zeippen (1982b)]{Mendoza_82b}
Mendoza, C. \& Zeippen, C.~J. 1982b, \mnras, 199, 1025

\bibitem[Mendoza \& Zeippen (1983)]{Mendoza_83}
Mendoza, C. \& Zeippen, C.~J. 1983, \mnras, 202, 981

\bibitem[Milingo et al.(2010)]{Milingo_10} Milingo, J.~B., 
Kwitter, K.~B., Henry, R.~B.~C., \& Souza, S.~P.\ 2010, \apj, 711, 619 

\bibitem[Morisset(2009)]{Morisset_09} Morisset, C.\ 2009, Mem.~S.~A.~It., 80, 397 

\bibitem[Peimbert \& Costero(1969)]{Peimbert_69} 
Peimbert, M., \& Costero, R.\ 1969, Boletin de los Observatorios Tonantzintla y Tacubaya, 5, 3 

\bibitem[Peimbert \& Torres-Peimbert(1971)]{Peimbert_71} 
Peimbert, M., \& Torres-Peimbert, S.\ 1971, \apj, 168, 413 

\bibitem[Peimbert \& Torres-Peimbert(1977)]{Peimbert_77} 
Peimbert, M., \& Torres-Peimbert, S.\ 1977, \mnras, 179, 217 

\bibitem[Peimbert et al.(1992)]{Peimbert_92} 
Peimbert, M., Torres-Peimbert, S., \& Ruiz, M.~T.\ 1992, \rmaa, 24, 155 

\bibitem[Pequignot(1980)]{Pequignot_80} 
Pequignot, D.\ 1980, \aap, 81, 356

\bibitem[Podovedova et al.(2009)]{Podovedova_09} 
Podobedova, L.I., Kelleher, D.E., \& Wiese, W.L.\ 2009, JPCRD, 38, 171

\bibitem[Rauch(2003)]{Rauch_03} Rauch, T.\ 2003, \aap, 403, 709 

\bibitem[Rodr\'iguez \& Delgado-Inglada (2011)]{Rodriguez_11} 
Rodr\'iguez, M. \& Delgado-Inglada, G.\ 2011, ApJ, 733, 50 

\bibitem[Rodr\'iguez \& Delgado-Inglada (2012)]{Rodriguez_12} 
Rodr\'iguez, M. \& Delgado-Inglada, G.\ 2012, IAU Symposium, 283, 488 

\bibitem[Schoenberner(1983)]{Schoenberner_83} 
Schoenberner, D.\ 1983, \apj, 272, 708 

\bibitem[Stasi{\'n}ska(1978)]{Stasinska_78} 
Stasi{\'n}ska, G.\ 1978, \aap, 66, 257 

\bibitem[Stasi{\'n}ska(2002)]{Stasinska_02} 
Stasi{\'n}ska, G.\ 2002, arXiv:astro-ph/0207500 

\bibitem[Stasi{\'n}ska et al.(1991)]{Stasinska_91} 
Stasi{\'n}ska, G., Tylenda, R., Acker, A., \& Stenholm, B.\ 1991, \aap, 247, 173

\bibitem[Stasi{\'n}ska et al.(1997)]{Stasinska_97} 
Stasi{\'n}ska, G., G\'orny, S.~K., \& Tylenda, R.\ 1997, \aap, 327, 736 

\bibitem[Stasi{\'n}ska et al.(1998)]{Stasinska_98} 
Stasi{\'n}ska, G., Richer, M.~G., \& McCall, M.~L.\ 1998, \aap, 336, 667

\bibitem[Storey \& Zeippen(2000)]{Storey_2000} 
Storey, P.~J., \& Zeippen, C.~J.\ 2000, \mnras, 312, 813

\bibitem[Tayal(2011)]{Tayal_11} Tayal, S.~S.\ 2011, \apjs, 195, 12 

\bibitem[Tayal \& Gupta (1999)]{Tayal_99} 
Tayal, S.S. \& Gupta, G.P.\ 1999 \apj 526, 544

\bibitem[Tayal \& Zatsarinny(2010)]{Tayal_10} Tayal, S.~S., \& Zatsarinny, O.\ 2010, \apjs, 188, 32 

\bibitem[Torres-Peimbert \& Peimbert(1977)]{TorresPeimbert_77} 
Torres-Peimbert, S., \& Peimbert, M.,\ 1977, \rmaa, 2, 181 

\bibitem[{{Tsamis} {et~al.}(2003)}]{Tsamis_03}
{Tsamis}, Y.~G., {Barlow}, M.~J., {Liu}, X.-W., {Danziger}, I.~J., \& {Storey}, P.~J. 2003, \mnras, 345, 186
  
\bibitem[Tsamis et al.(2004)]{Tsamis_04}
{Tsamis}, Y.~G., {Barlow}, M.~J., {Liu}, X.-W., {Storey}, P.~J., \& {Danziger}, I.~J.\ 2004, \mnras, 353, 953 

\bibitem[Wang \& Liu(2007)]{Wang_07} 
Wang, W., \& Liu, X.-W.\ 2007, \mnras, 381, 669 

\bibitem[{{Wesson} \& {Liu}(2004)}]{Wesson_04}
{Wesson}, R. \& {Liu}, X.-W. 2004, \mnras, 351, 1026

\bibitem[{Wesson} {et~al.}(2005){Wesson}, {Liu}, \& {Barlow}]{Wesson_05}
{Wesson}, R., {Liu}, X.-W., \& {Barlow}, M.~J. 2005, \mnras, 362, 424 

\bibitem[Wiese et al.(1996)]{Wiese_96} 
Wiese, W.~L., Fuhr, J.~R., \& Deters, T.~M.\ 1996, JPCRD, Monograph 7, Atomic transition probabilities of carbon, 
nitrogen, and oxygen: A critical Data Compilation (Woodbury, NY: AIP Press)

\bibitem[{Zeippen (1982)}]{Zeippen_82}
Zeippen, C.~J. 1982, \mnras, 198, 111

\bibitem[Zhang \& Liu(2003)]{Zhang_03} 
Zhang, Y., \& Liu, X.-W.\ 2003, \aap, 404, 545
\end{thebibliography}
\end{document}